\documentclass[aps,pre,reprint,showpacs,floatfix]{revtex4-1}

\usepackage{subfigure}
\usepackage{graphicx}
\usepackage{epstopdf}
\usepackage{enumitem}
\usepackage{dcolumn}
\usepackage{diagbox}
\usepackage{amssymb,latexsym}
\usepackage{amsmath}

\newtheorem{deff}{Definition}

\newtheorem{lem}[deff]{Lemma}

\providecommand{\cc}{\!:\!}

\begin{document}

\title{Roundness of grains in cellular microstructures}
\date{\today}

\author{F. H. Lutz}
\email{lutz@math.tu-berlin.de}
\affiliation{Technische Universit\"at Berlin, 10623 Berlin, DE.}
\author{J. K. Mason}
\email{jeremy.mason@boun.edu.tr}
\affiliation{Bo\u{g}azi\c{c}i University, Bebek, Istanbul 34342, TR.}
\author{E. A. Lazar}
\email{mlazar@seas.upenn.edu}
\affiliation{University of Pennsylvania, Philadelphia, PA 19104, USA.}
\author{R. D. MacPherson}
\email{rdm@math.ias.edu}
\affiliation{Institute for Advanced Study, Princeton, New Jersey 08540, USA.}

\begin{abstract} 

Many physical systems are composed of polyhedral cells of varying sizes and shapes. These structures are simple in the sense 
that no more than three faces meet at an edge and no more than four edges meet at a vertex. This means that individual cells 
can usually be considered as simple, three-dimensional polyhedra. This paper is concerned with determining the distribution 
of combinatorial types of such polyhedral cells. We introduce the terms \emph{fundamental} and \emph{vertex-truncated} types 
and apply these concepts to the grain growth microstructure as a testing ground. For these microstructures we demonstrate 
that most grains are of particular fundamental types, whereas the frequency of vertex-truncated types decreases exponentially 
with the number of truncations. This can be explained by the evolutionary process through which grain growth structures are formed, 
and in which energetically unfavorable surfaces are quickly eliminated. Furthermore, we observe that these grain types are `round' 
in a combinatorial sense: there are no `short' separating cycles that partition the polyhedra into two parts of similar sizes. 
A particular microstructure derived from the Poisson--Voronoi initial condition is identified as containing an unusually large 
proportion of round grains. This Round microstructure has an average of $14.036$ faces per grain, and is conjectured 
to be more resistant to topological change
than the steady-state grain growth microstructure.

\end{abstract}


\maketitle

\section{Introduction}
\label{sec:introduction}

\begin{table*}
\caption{\label{tbl:glossary}Glossary of terminology appearing throughout the article, listed alphabetically.}
\begin{ruledtabular}
\begin{tabular}{p{0.19\textwidth}p{0.79\textwidth}}
  Term & Description \\
  \hline \noalign{\smallskip}
  $1$-skeleton & The restriction of a triangulation to the vertices and edges. \\
  $3$-polytope & A three-dimensional convex solid with flat polygonal faces. \\
  combinatorially $k$-round & A simple $3$-polytope that is obtained from the tetrahedron or a flag$^*$ simple $3$-polytope by at most $k$ vertex truncations. \\
  complete subgraph & A subgraph for which every pair of vertices is connected by an edge. \\
  degenerate grain & A grain where one of the faces is a $2$-gon or two faces share more than one edge. \\
  dual graph & For a set of faces of a $3$-polytope, the dual graph contains a vertex for every face in the set, and an edge connects vertices whenever the corresponding faces share an edge. \\
  dual polytope & For a polytope $P$, the dual polytope $P^*$ has faces and vertices in an incidence-preserving correspondence with the vertices and faces of $P$. \\
  dual triangulation & For the boundary of a simple $3$-polytope, the dual triangulation is the boundary of the dual polytope. \\
  flag($^*$) & A flag$^*$ simple $3$-polytope has no triangular faces or non-trivial $3$-belts. A flag simplicial $3$-polytope is dual to a flag$^*$ simple $3$-polytope. \\
  fundamental & A simple $3$-polytope that is either the tetrahedron or does not have any triangular faces. \\
  $\ell$-belt & A set of $\ell$ faces of a simple $3$-polytope such that the dual graph is a triangulated circle. \\
  Schlegel diagram & A projection of the boundary of a $3$-polytope onto one of its faces in such a way that vertices not belonging to the face lie inside it and no two edges cross. \\
  severely constricted($^*$) & A severely constricted simple $3$-polytope has no triangular faces but at least one non-trivial $3$-belt. A severely constricted$^*$ simplicial $3$-polytope is dual to a severely constricted simple $3$-polytope. \\
  simple & A $3$-polytope where two faces meet on at most one edge and exactly three faces meet at each vertex. \\
  simplicial & A $3$-polytope where every face is a triangle. The boundary is a triangulation of the sphere $S^2$. \\
  split type & A flag$^*$ simple $3$-polytope has split type $k\!:\!\ell\!:\!m$ if $k$ is the largest integer less than or equal to $m$ such that the polytope has an $\ell$-belt and the removal of this belt leaves two disks of $k$ and $m$ faces. \\
  stacked($^*$) & A stacked$^*$ simple $3$-polytope is obtained from the tetrahedron by a sequence of vertex truncations. A stacked simplicial $3$-polytope is dual to a stacked$^*$ simple $3$-polytope. \\
  trivial $3$-belt & A $3$-belt where the three faces share a unique vertex. \\
  vertex-truncated & A simple $3$-polytope that is not the tetrahedron and has at least one triangular face.
\end{tabular}
\end{ruledtabular}
\end{table*}

\begin{figure}[b]
\includegraphics[width=0.5\columnwidth]{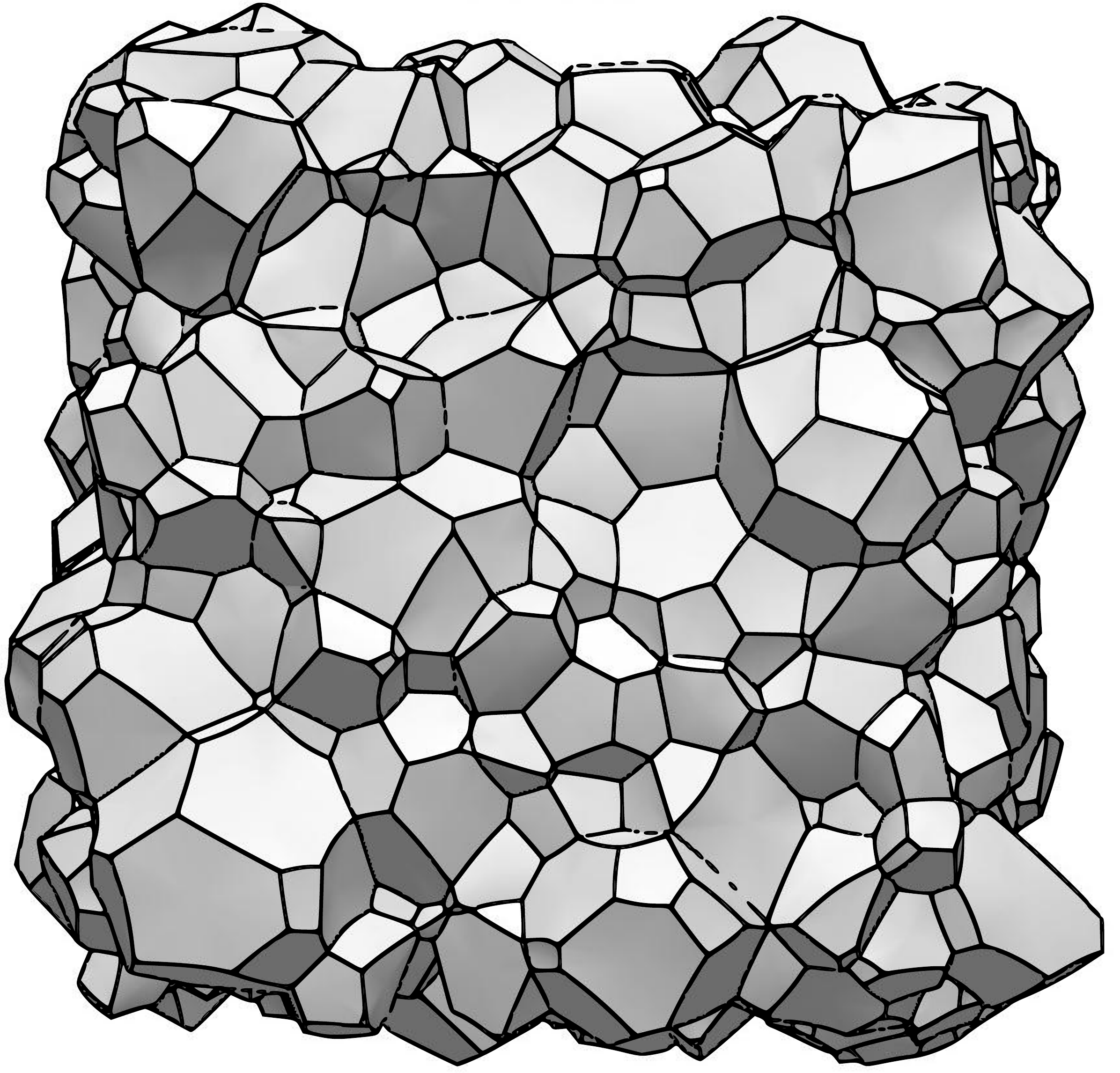}
\caption{A simulated microstructure resulting from a process of grain growth, assuming isotropic grain boundary energies and mobilities.}
\label{fig:3D_microstructure}
\end{figure}

\begin{figure}
\includegraphics[width=0.2\columnwidth]{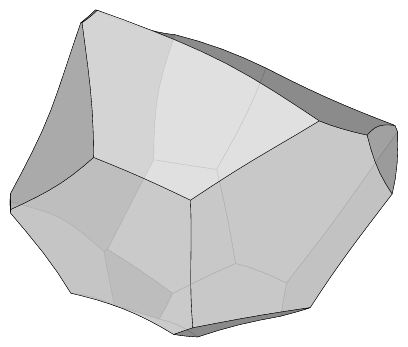}
\includegraphics[width=0.2\columnwidth]{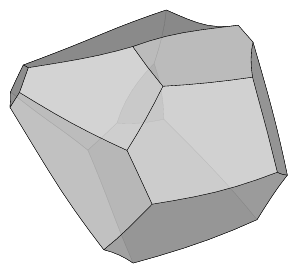}
\includegraphics[width=0.2\columnwidth]{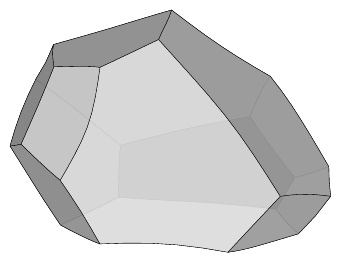}
\includegraphics[width=0.2\columnwidth]{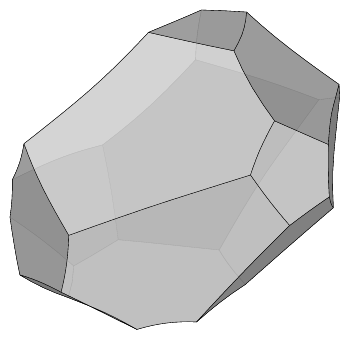}
\caption{A typical set of grains from the grain-growth microstructure.}
\label{fig:single_grains}
\end{figure}

Many physical and biological systems share the property of being composed of irregular polyhedra subject to some form of energy minimization. 
The microstructures of polycrystalline materials in general and of metals in particular \cite{1952burke,1952smith} often serve as motivating examples, 
and can be modeled as resulting from a process of three-dimensional grain growth subject to isotropic boundary energies and mobilities 
\cite{2011lazar, mason2015geometric}. A portion of the resulting simulated microstructure is given in Figure~\ref{fig:3D_microstructure}, 
with some isolated grains displayed in Figure~\ref{fig:single_grains}.

An investigation of these figures reveals several significant features. First, the networks of edges on the surfaces
of the grains are connected in many different ways (i.e., the number of topological types is large). While this could 
be explained as the result of some inherent disorder in a physical system that is constrained to be space-filling, 
the distribution of topological types has been observed to be far from random \cite{2012lazar}. Still, relatively little 
is currently known about why certain types appear more frequently than others. Second, the examples 
in Figure~\ref{fig:single_grains} show that many of the grains appear to be geometrically round, in the sense 
of the surface area to volume ratio. This could be explained as the result of energy minimization, since the energy 
of the model system is proportional to the sum of the surface areas of the grains. A reasonable question would be 
whether this property is related to the observed distribution of topological types (e.g., the topological types 
in a material with equiaxed grains are likely different from those in a material with extruded~grains).

The purpose of this paper is threefold. First, some language and concepts inspired by combinatorial topology 
are introduced to make the discussion and classification of the topological types of grains more precise, 
and to provide a framework to compare the distribution of topological types in different systems. These are 
intended to be generally useful for the analysis of cellular structures. Data from simulations suggest that the dynamics 
of grain growth in real space induce dynamics on the space of topological types, involving the switching of edges 
and the creation and deletion of triangular faces. The second purpose of the paper is to offer a preliminary description 
of these dynamics. Third, a detailed investigation of the evolution of a microstructure from a typical initial condition 
reveals a point where the grains appear to be particularly round in a specific combinatorial sense. This seems to result 
from triangular faces in the initial condition disappearing before triangular faces on tetrahedral grains can appear 
in comparable numbers. The possibility that this microstructure is particularly stable to thermal coarsening suggests a
further investigation into the properties of the corresponding experimental microstructure.

\section{Combinatorial Analysis of Cellular Microstructures}
\label{sec:combinatorial_analysis}

\subsection{Simple and Simplicial Polytopes}
\label{subsec:polytopal}

A \emph{three-dimensional polytope} ($3$-polytope) is defined as a three-dimensional convex hull of finitely many points 
in three-dimensional space. With only a few degenerate exceptions (described in Section~\ref{sec:extremal}), most of the grains 
observed in the simulated grain growth microstructures (or in other space-filling cellular structures, e.g.\ foams
\cite{weaire2001physics}) have the combinatorial types of \emph{simple} $3$-polytopes. These obey the constraints that
two faces meet on at most one edge and exactly three faces meet at each vertex. Similarly, a space-filling three-dimensional
microstructure is \emph{simple} if exactly four cells meet at each vertex, three cells meet at each edge, and two cells meet
on at most one face.

\begin{figure}
\includegraphics[height=50mm]{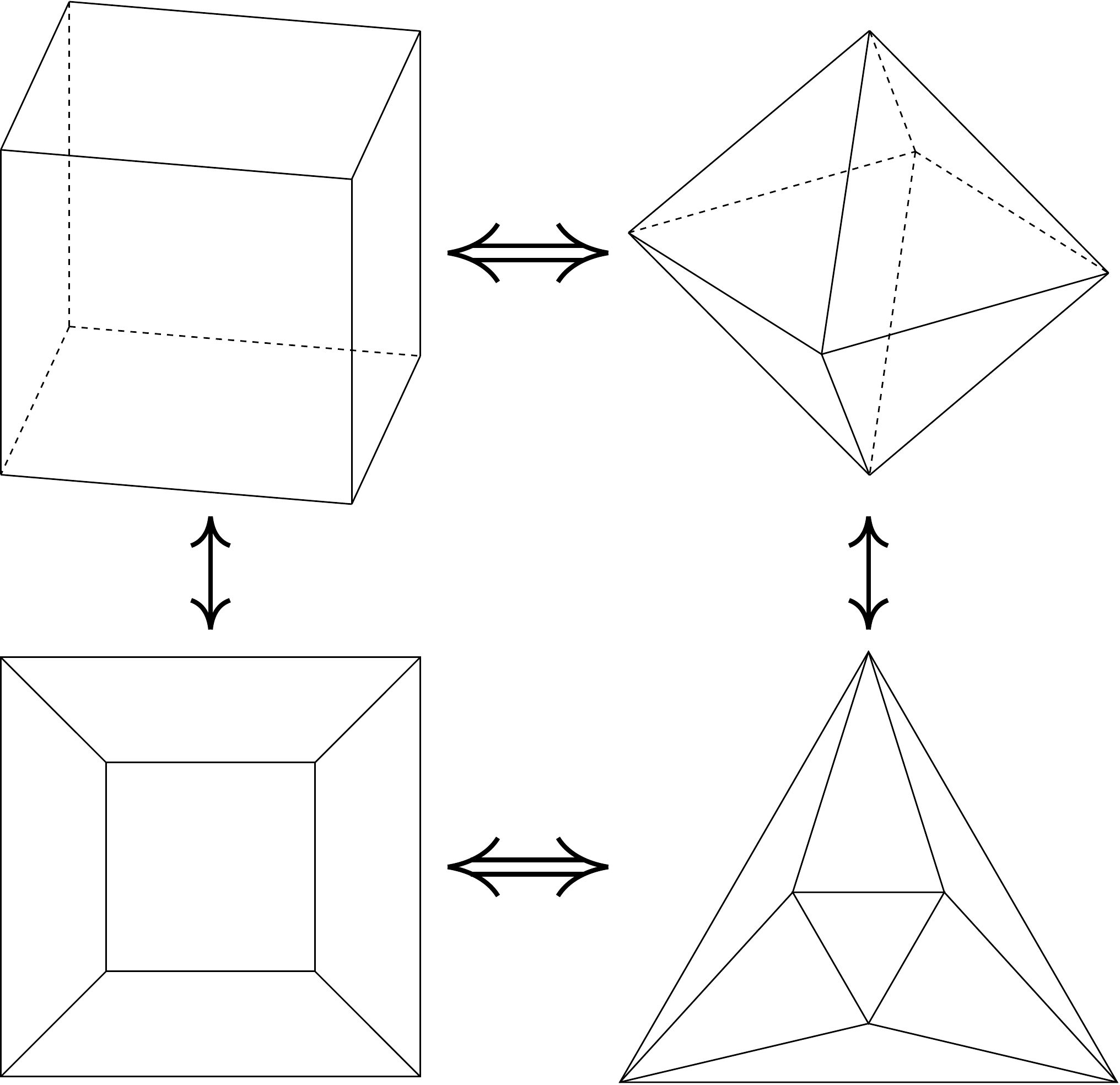}
\caption{The relationship of a simple 3-polytope to the corresponding dual simplicial $3$-polytope (double arrow), and of a polytope 
to the corresponding Schlegel diagram (single arrow).}
\label{fig:schlegel_and_dual}
\end{figure}

Many concepts in this article are more naturally described and algorithmically computed by means of the \emph{dual polytope}. 
For any polytope $P$, there is a dual polytope $P^*$ such that the faces and vertices of $P$ have an incidence-preserving correspondence
with the vertices and faces of $P^*$, respectively. This implies that when $P$ is a simple $3$-polytope (three faces meet at every vertex), 
$P^*$ is a \emph{simplicial} $3$-polytope (every face is a triangle), and conversely. Note that the boundary of a simplicial $3$-polytope 
is always a triangulation of the sphere $S^2$, the \emph{dual triangulation} to the simple boundary of $P$.
As examples, the octahedron is dual to the cube (shown at the top of Figure~\ref{fig:schlegel_and_dual}),
and the icosahedron is dual to the pentagonal dodecahedron. The tetrahedron is the only simple $3$-polytope that is self-dual,
and as such it is both simple and simplicial.

While the combinatorial types of the cube and the octahedron are clear from Figure~\ref{fig:schlegel_and_dual}, the combinatorial types 
of more complicated $3$-polytopes can be difficult to visualize from the embeddings. The \emph{Schlegel diagrams} at the bottom 
of Figure~\ref{fig:schlegel_and_dual} are often more convenient to show the combinatorial type of a $3$-polytope. This diagram 
is constructed by projecting the boundary of a polytope onto one of the faces in such a way that vertices not belonging to the face 
lie inside it, and no two edges cross \cite{schlegel1881theorie,schlegel1886uber}.

All simple $3$-polytopes belong to one of three classes, identified by the following characteristics:
\begin{itemize}[noitemsep]
\item four triangular faces (the tetrahedron),
\item more than four faces and no triangular faces,
\item more than four faces and some triangular faces.
\end{itemize}
Truncating any corner of a simple $3$-polytope in the first or second classes gives a simple $3$-polytope in the third class, 
and any simple $3$-polytope in the third class can be constructed in this way. This suggests the following definition:
\begin{deff}
A simple $3$-polytope is \emph{fundamental} if it cannot be obtained by truncating a corner of a simple $3$-polytope 
with fewer faces, and is \emph{vertex-truncated} otherwise.
\end{deff}

\begin{figure}
\includegraphics[height=28mm]{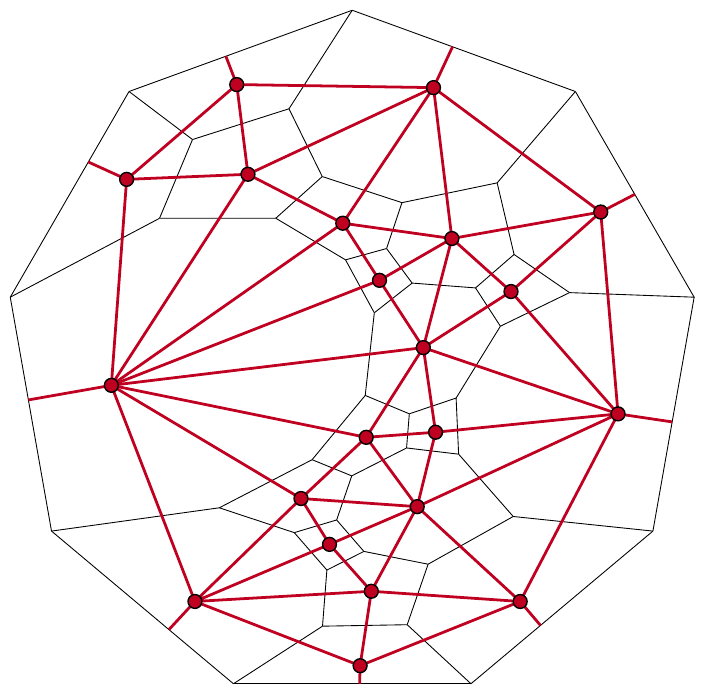}
\caption{The Schlegel diagram of a grain and the dual graph of the set of all faces; the dual graph vertex for the back face is not shown.}
\label{fig:dual_graph}
\end{figure}

\begin{figure}
\includegraphics[height=28mm]{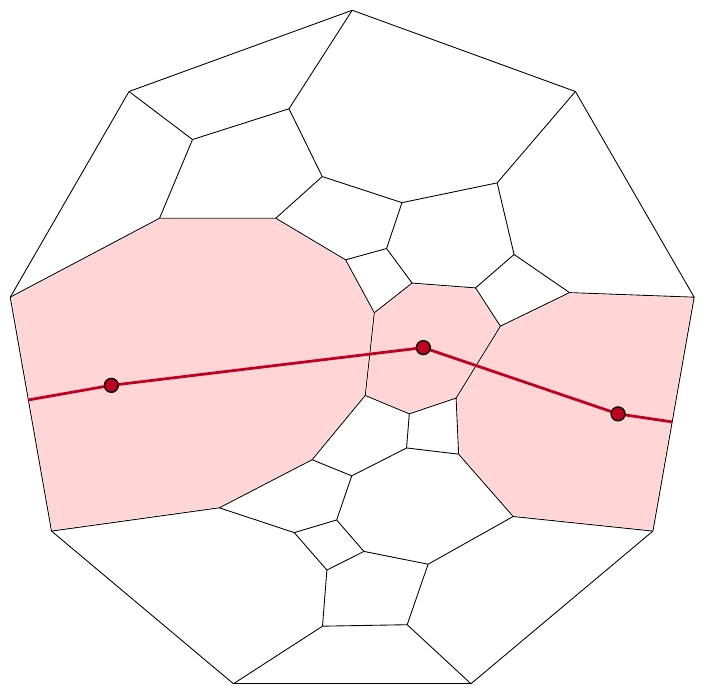}
\hspace{8mm}
\includegraphics[height=28mm]{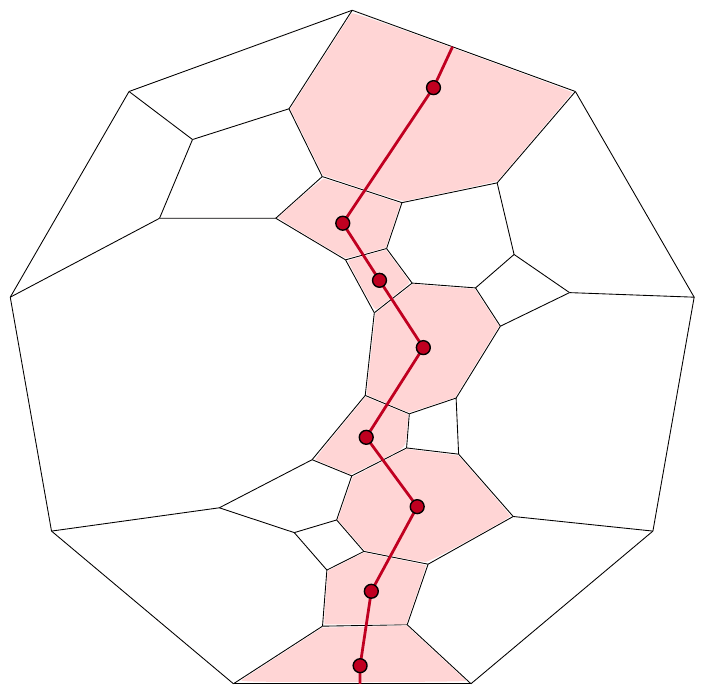}
\caption{A $4$-belt (left) and a $9$-belt (right) of a grain shown on the Schlegel diagram along with their dual graphs. The back face belongs to the belt in both cases.}
\label{fig:schlegel_short_belt}
\end{figure}

As described in Section~\ref{sec:introduction}, the polytopes belonging to physical systems subject to energy minimization often appear 
to be geometrically round. If some characteristic combinatorial feature of these polytopes could be found, then that would motivate 
a corresponding notion of combinatorial roundness. In this vein, a round object can be described as having an absence of constrictions,
i.e., a belt of any orientation around the object would slide off. The combinatorial analogue of such a geometric
constriction uses the notion of a \emph{dual graph}. Given a set of faces, the dual graph contains a vertex
for each face in the set, and an edge connects two vertices whenever the corresponding faces share an edge. The
dual graph of the set of all faces of an example grain is shown in Figure~\ref{fig:dual_graph}. A combinatorial constriction 
can now be defined as follows:
\begin{deff}
An \emph{$\ell$-belt} of a simple $3$-polytope is any set of $\ell$ faces such that the dual graph of the faces 
is a cycle (a triangulated circle). A $3$-belt is \emph{trivial} if the three faces share a unique vertex, and is \emph{non-trivial} otherwise.
\end{deff}
Figure~\ref{fig:schlegel_short_belt} shows a $4$-belt (left) and a $9$-belt (right) of the grain in Figure~\ref{fig:dual_graph}, with the faces of the respective belts shaded in red. One of the faces of each of these belts is the back face. Observe that the dual graph of an $\ell$-belt is a cycle containing $\ell$ vertices that are cyclically connected by $\ell$ edges.

\begin{lem}
The removal of an $\ell$-belt from the boundary of a simple $3$-polytope leaves one (for a trivial $3$-belt) or two connected 
components that are discs.
\end{lem}

\begin{figure}
\includegraphics[height=28mm]{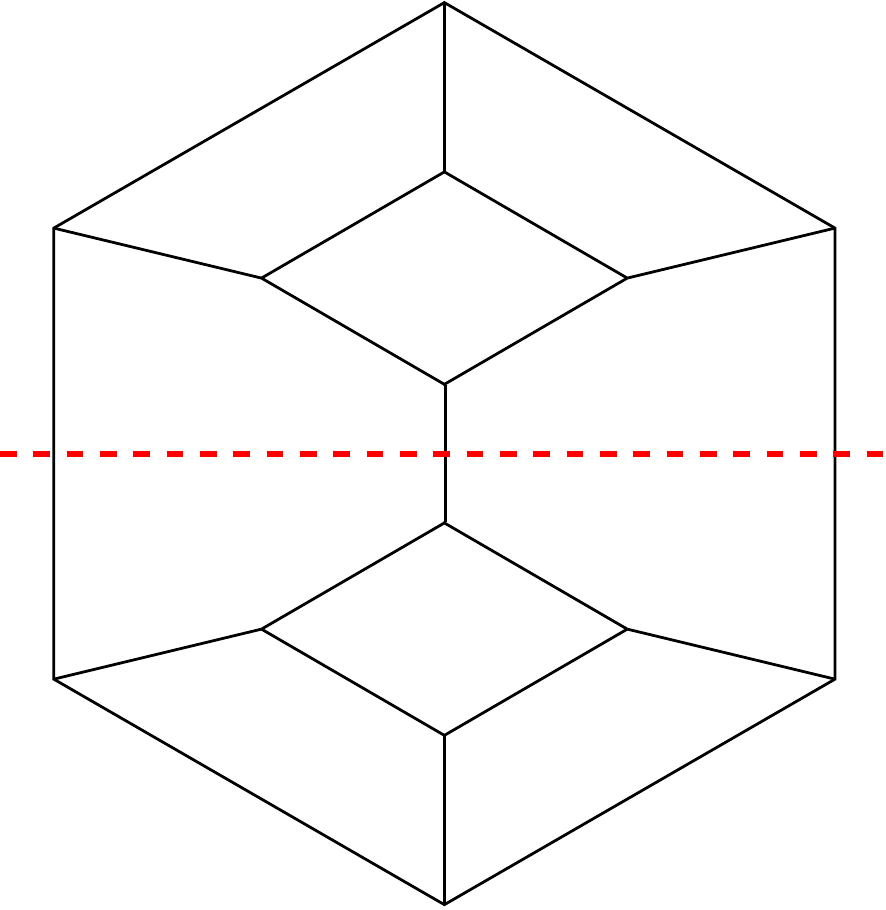}
\hspace{2mm}
\includegraphics[height=28mm]{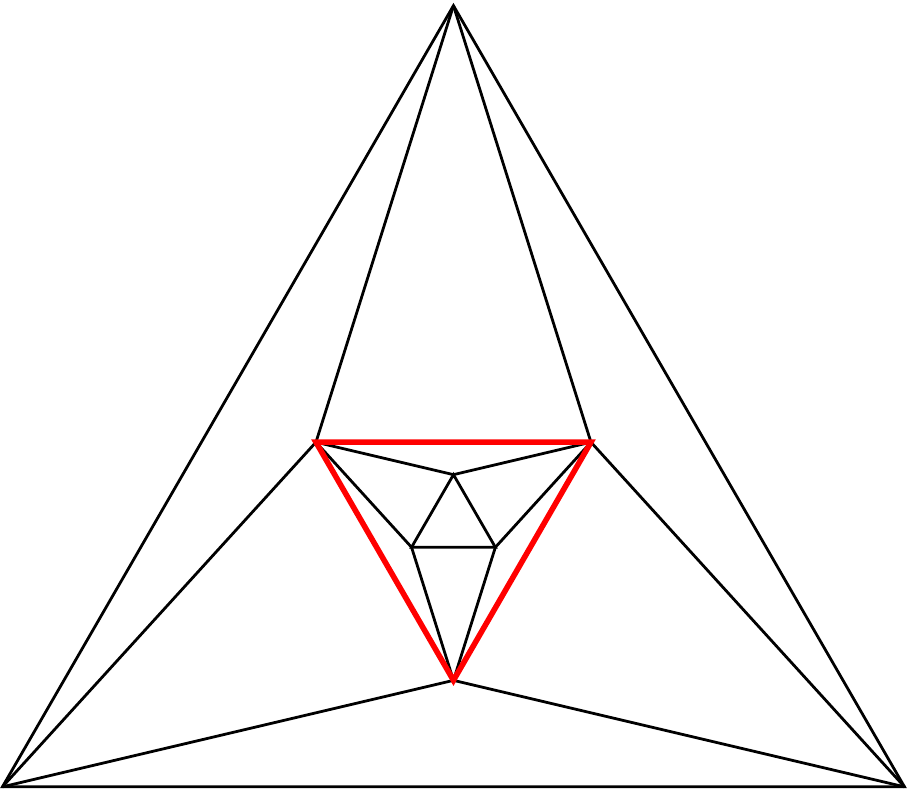}
\hspace{2mm}
\includegraphics[height=28mm]{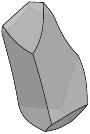}
\caption{The Schlegel diagram of the smallest possible fundamental simple $3$-polytope with a non-trivial $3$-belt (indicated by the dotted red line),
the dual triangulation with an \emph{empty triangle} (solid red line), and an example from a three-dimensional evolved microstructure 
with this combinatorial type.}
\label{fig:smallest_nonflag}
\end{figure}

Any vertex-truncated simple $3$-polytope has at least one triangular face, and the neighboring faces of such a triangle form 
a non-trivial $3$-belt around the triangle. Perhaps more significantly, a fundamental $3$-polytope that is not the tetrahedron 
can have a non-trivial $3$-belt as well, despite not having any triangular faces; Figure~\ref{fig:smallest_nonflag} shows
the smallest (in terms of number of faces) such example. The severe geometric distortion that accompanies the $3$-belt
of this polytope suggests the following definition:
\begin{deff}
A fundamental simple $3$-polytope that is not the tetrahedron (i.e., has no triangular faces) is \emph{flag$^*$} if it has only trivial $3$-belts, 
and is \emph{severely constricted (sc)} otherwise.
\label{def:flag}
\end{deff}

The term \emph{flag} is usually used in combinatorial topology to describe a simplicial complex $K$ where every complete subgraph 
in the $1$-skeleton (edge graph) of $K$ spans a simplex of $K$ \cite{1989brown,2007kozlov}. This means that a flag simplicial complex 
does not have any \emph{empty faces}, and that every flag simplicial complex can be reconstructed from its $1$-skeleton by inserting 
a face in every complete subgraph of the vertices. The full tetrahedron is the smallest example of a three-dimensional flag simplicial complex; 
the $1$-skeleton is a complete graph on $4$ vertices, implying that the complex contains a three-dimensional face (the original tetrahedron). 
The boundary of the full tetrahedron is a two-dimensional simplicial complex that is not flag, since the $1$-skeleton is the same 
but the three-dimensional face is not included. The smallest example of a two-dimensional flag simplicial complex that triangulates $S^2$ 
is the boundary of the full octahedron. 

The existence of a non-trivial $3$-belt of a simple $3$-poly\-tope is closely related to the \emph{connectivity} of the boundary 
of the dual simplicial $3$-polytope. Specifically, the $1$-skeleton of a simplicial $3$-polytope is $4$-connected if, for every 
pair $\{v,w\}$ of distinct vertices $v$ and $w$, the $1$-skeleton contains four independent paths connecting $v$ and $w$. 
For example, the south-pole of the octahedron of Figure~\ref{fig:schlegel_and_dual} can be connected to the north-pole 
by four independent paths. Observe that there can be no empty triangles in a $4$-connected $1$-skeleton of a simplicial $3$-polytope 
since only three non-intersecting paths could be routed via the three vertices of an empty triangle. This allows an alternative definition 
of a flag simplicial $3$-polytope via the following lemma (cf.~\cite{2011Athanasiadis}):
\begin{lem}
The $2$-dimensional boundary complex of a simplicial $3$-polytope is flag if and only if its $1$-skeleton is $4$-connected.
\end{lem}

The motivation for extending the definition of flag to simple polytopes in Definition~\ref{def:flag} is the observation that any flag$^*$ 
simple $3$-polytope $P$ is dual to a flag simplicial $3$-polytope $P^*$. That is, if $P$ (except for the tetrahedron) does not have 
any non-trivial $3$-belts, then the $1$-skeleton of $P^*$ does not contain any empty triangular faces and is $4$-connected.

\subsection{Combinatorial Roundness}
\label{subsec:roundness}

As specified in Definition~\ref{def:flag}, a flag$^*$ simple $3$-polytope $P$ satisfies a number of restrictive requirements. Since $P$ 
is fundamental and not the tetrahedron, $P$ cannot have any triangular faces. Since $P$ only has trivial $3$-belts, the $1$-skeleton 
of the dual simplicial $3$-polytope must be $4$-connected. While there is some motivation to consider flag$^*$ simple $3$-polytopes 
as combinatorially round, using this as a definition is excessively restrictive for physical systems where simple $3$-polytopes 
evolve by energy minimization. The reason is that a flag$^*$ simple $3$-polytope acquires a triangular face and briefly becomes 
vertex-truncated whenever one of the neighboring polytopes is about to disappear as a tetrahedron. More generally, computational results 
suggest that triangular faces frequently appear and disappear as transitional topological events in material microstructures \cite{2011lazar}. 
Ideally, our definition of combinatorial roundness would be robust to this type of transient event.

One approach is to define combinatorial roundness in such a way to include combinatorial types that are a few vertex truncations 
away from being flag$^*$. To make this more precise, observe that repeated truncations are locally tree-like. The first truncated vertex 
can be considered as the root of a tree, and the tree grows whenever one of the newly obtained vertices is truncated. If a different 
original vertex is truncated instead, this begins a different tree. For the purpose of notation, let $W = \{T_1,T_2,\dots,T_t\}$ 
be a collection of $t$ (not necessarily distinct) rooted trees. Since every truncation introduces three new vertices that can be used
for further truncations, the $T_i$ are all ternary trees.

\begin{deff}
A simple $3$-polytope $P$ belongs to the class flag$^*_{\,W}$ or sc$_{\,W}$ if $P$ is derived respectively from a flag$^*$ or sc simple $3$-polytope 
by a sequence of vertex truncations, with the truncation sequence locally represented by the collection of rooted \emph{truncation trees}~$W$.
\end{deff}

\begin{figure}
\includegraphics[height=28mm]{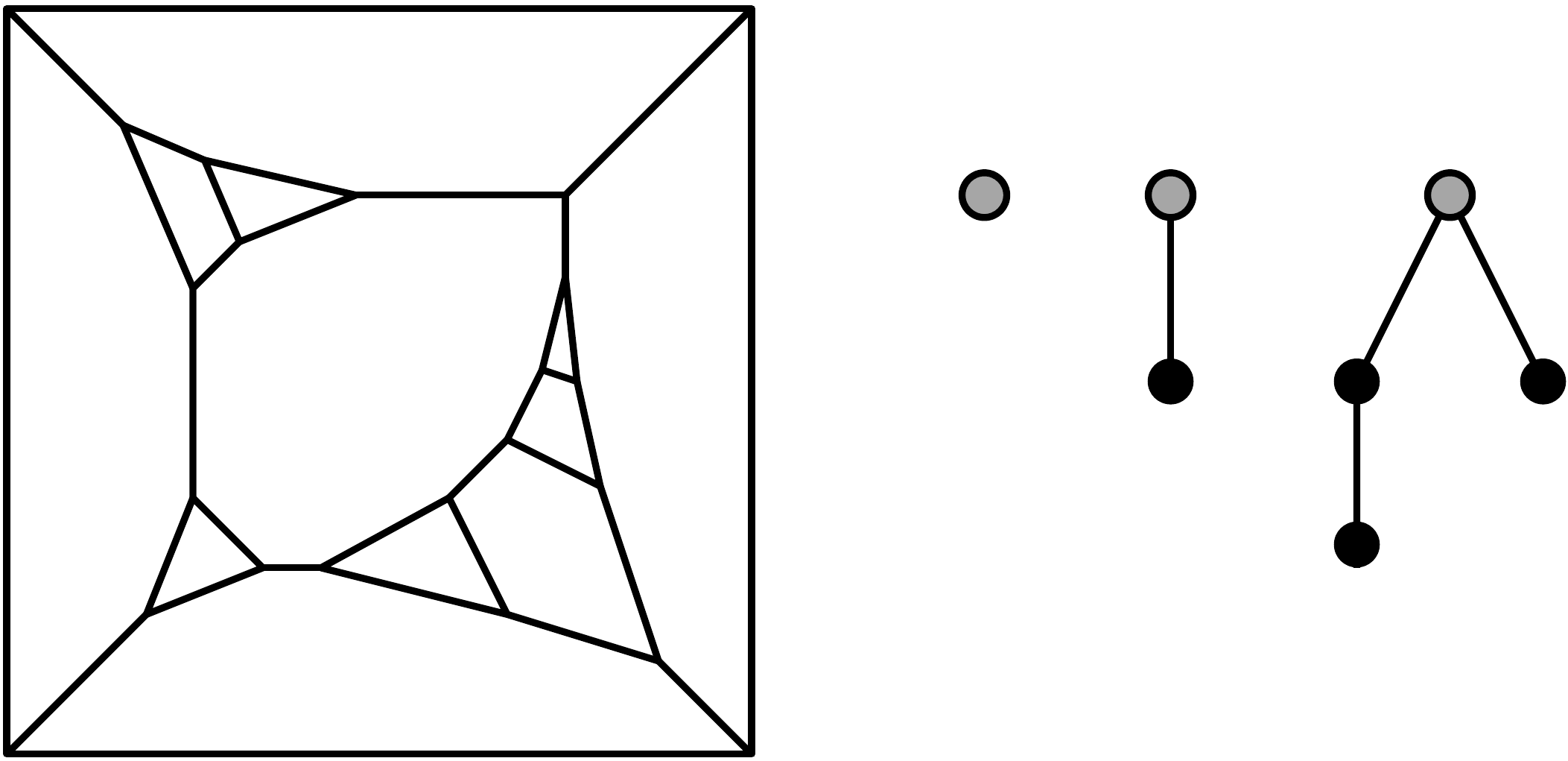}
\caption{A simple $3$-polytope with $13$ faces that is seven truncations away from the cube. The collection of rooted truncation trees 
is on the right, with a truncation depth of three.}
\label{fig:truncated_cube}
\end{figure}

A fundamental $3$-polytope $F$ and a collection of truncation trees $W$ do not uniquely specify the derived simple $3$-polytope $P$, 
since a truncation tree identifies neither the original vertex at which the tree occurs nor the vertex that is truncated when the tree is extended. 
That $W$ is composed of trees does imply that a simple $3$-poly\-tope $P$ is enough to uniquely specify the originating fundamental 
$3$-polytope $F$, though.
\begin{lem}
Any simple $3$-polytope is either fundamental, or can be reduced to a unique fundamental $3$-polytope by reversing truncations in any order.
\end{lem}
For example, the simple $3$-polytope on the left of Figure~\ref{fig:truncated_cube} is derived from the cube by seven truncations. 
This can be verified by repeatedly collapsing any one of the remaining triangular faces until the $3$-polytope is fundamental.

Let $+r_i$ indicate that $T_i$ is a path (i.e., a tree without branches) with $r_i$ vertices. In particular, the class flag$^*_{+1}$ contains 
all simple $3$-polytopes that are derived from flag$^*$ $3$-polytopes by a single truncation. Simple $3$-polytopes that are derived 
from flag$^*$ $3$-polytopes by exactly two truncations fall into two classes, flag$^*_{+1+1}$ and flag$^*_{+2}$. Examples in the first class 
have two triangular faces (obtained by truncating two distinct original vertices), while examples in the second class have only one triangular face 
(obtained by truncating one original vertex and one of the newly introduced vertices). More generally, examples in the class flag$^*_{+r}$ 
for any positive integer $r$ have only one triangular face. This reinforces that the collection $\{T_1,T_2,\dots,T_t\}$ of truncation trees 
provides a more detailed description of the polytope than merely the number of triangular faces \cite{1982rhines,2013patterson}.

Other significant classes include flag$^*_{\,\leq k}$, flag$^*_{\,\geq k}$, flag$^*_{\,k}$, and flag$^*_{\,\geq}$, which are derived from 
flag$^*$ simple $3$-polytopes by \emph{at most} $k$, \emph{at least} $k$, \emph{exactly} $k$, and \emph{any number of} truncations, 
respectively. The classes sc$_{\,\leq k}$, sc$_{\,\geq k}$, sc$_{\,k}$, and sc$_{\,\geq}$ for severely constricted 
simple $3$-polytopes are defined similarly. Finally, the \emph{truncation depth} of a simple $3$-polytope 
of type flag$^*_{\,W}$ or sc$_{\,W}$ is the number of vertices of a longest rooted path in one of the rooted trees in $W$. 
For example, the simple $3$-polytope in Figure~\ref{fig:truncated_cube} has a truncation depth of three.

The dual relationship of a simple $3$-polytope to a simplicial $3$-polytope implies that operations on simple $3$-polytopes 
have corresponding operations on simplicial $3$-polytopes. Specifically, a vertex truncation of a simple $3$-polytope induces 
a \emph{stacking operation} on the dual simplicial $3$-polytope, where a triangle is subdivided by placing a new vertex slightly `above' 
the center of the triangle. Let a \emph{stacking tree} be defined such that for any simple $3$-polytope $P$ and dual simplicial 
$3$-polytope $P^*$, the collection of truncation trees $\{T_1,T_2,\dots,T_t\}$ of $P$ is identical to the collection of stacking trees 
$\{S_1,S_2,\dots,S_t\}$ of $P^*$. Then the truncation depth of $P$ and the \emph{stacking depth} of $P^*$ are always the same.

Since the tetrahedron is the unique fundamental simple $3$-polytope with triangular faces, some specific terminology is introduced 
to identify the $3$-polytopes derived from the tetrahedron.
\begin{deff}
A simple (simplicial) $3$-polytope is \emph{stacked$^*$} (\emph{stacked}) if it can be obtained from a tetrahedron 
by a sequence of vertex truncations (stacking operations).
\end{deff}
The flag octahedron with six vertices is the smallest simplicial $3$-polytope that is not stacked. There is always a vertex of degree three
on the boundary of a stacked simplicial $3$-polytope (e.g., the vertex most recently introduced by a stacking operation), and all vertices 
of the octahedron have degree four.

A simple $3$-polytope that is at most $k$ truncations away from a flag$^*$ simple $3$-polytope is said to be \emph{$k$-close to flag$^*$}, 
and a simple $3$-polytope that is at most $k$ truncations away from the tetrahedron is said to be \emph{$k$-close to tetrahedral}. 
This allows a combinatorial roundness to be defined for simple $3$-polytopes as follows:
\begin{deff}
A simple $3$-polytope is \emph{combinatorially $k$-round} and belongs to the class \emph{round$_{\,\leq k}$} if it is $k$-close to flag$^*$ 
or $k$-close to tetrahedral.
\end{deff}
There is an analogous definition of the class round$_{\,\leq k}$ for simplicial $3$-polytopes, where $k$-close to flag and $k$-close 
to tetrahedral bound the number of stacking operations rather than the number of vertex truncations. Our intention is to provide 
preliminary evidence that the class round$_{\,\leq k}$ for small values of $k$ connects the combinatorial structure of a grain 
in a material microstructure to its geometric shape (being round in the sense of small surface area), and thereby to the frequency 
of its combinatorial type in energy-minimizing structures.

Finally, triangulations of the boundary $2$-spheres of simplicial $3$-polytopes with up to $23$ vertices have been enumerated 
by Brinkmann and McKay \cite{plantri}, as well as all examples with minimal vertex-degree at least four and all $4$-connected 
flag examples with up to $27$ vertices (Table~\ref{tbl:few_vertices_I}). Explicit lists of the triangulations with up to $14$ vertices 
can be found online \cite{Lutz_PAGE}. 
These will be used in Section~\ref{subsec:microstructure_evolution} to show that round$_{\,\leq k}$ is a small subset of all triangulations 
with a given number of vertices, particularly for small values of $k$.

\section{Computational Results}

\subsection{Microstructure Evolution}
\label{subsec:microstructure_evolution}

Grain growth simulations generally assume that the grain boundary energy and mobility are constants, and that capillary pressure 
drives grain boundary motion. For this model, a generic geo\-metric initial condition converges to a statistical steady state where 
the faces are curved and boundary motion continues but dimensionless statistical quantities are constant \cite{mason2015geometric}. 
The average number of faces per grain in this steady state is $13.769 \pm 0.016$ \cite{2011lazar, mason2015geometric}, 
though a precise justification of this value is not known.

One of the main purposes of this paper is to more closely analyze the grain growth process and the induced dynamics on the space 
of topological types using the terminology introduced in previous sections. The specific numerical model used in the following 
is explained in detail elsewhere \cite{2011lazar}. Time is always expressed in dimensionless units of $1 / m \gamma$, where $m$ 
and $\gamma$ are the constant grain boundary mobility and energy per unit area, respectively.

\begin{figure}
\includegraphics[width=1.\columnwidth]{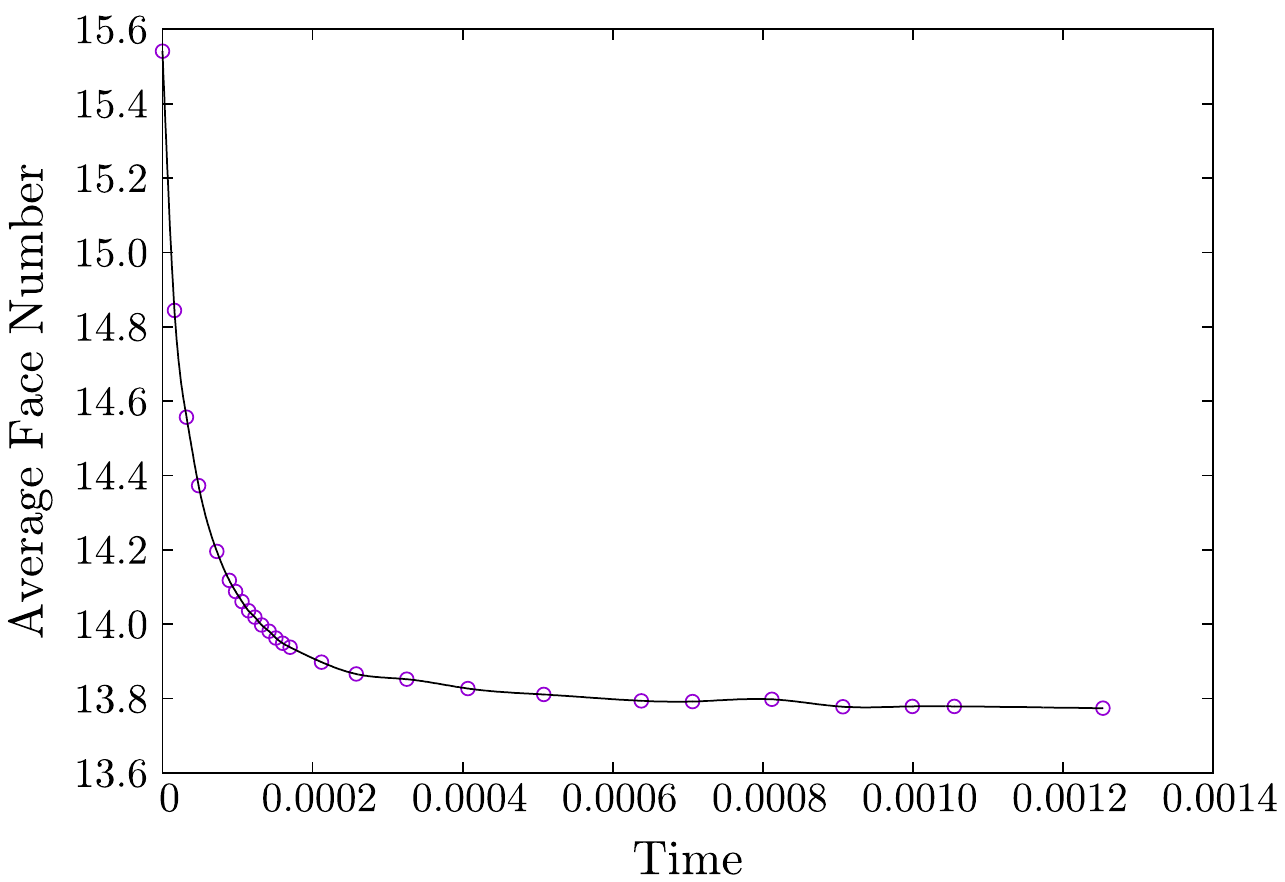}
\caption{Evolution of the average number of faces per grain in time from a Poisson--Voronoi initial condition with constant 
grain boundary energy and mobility.}
\label{fig:face_over_time}
\end{figure}

\begin{figure}
\includegraphics[width=1.\columnwidth]{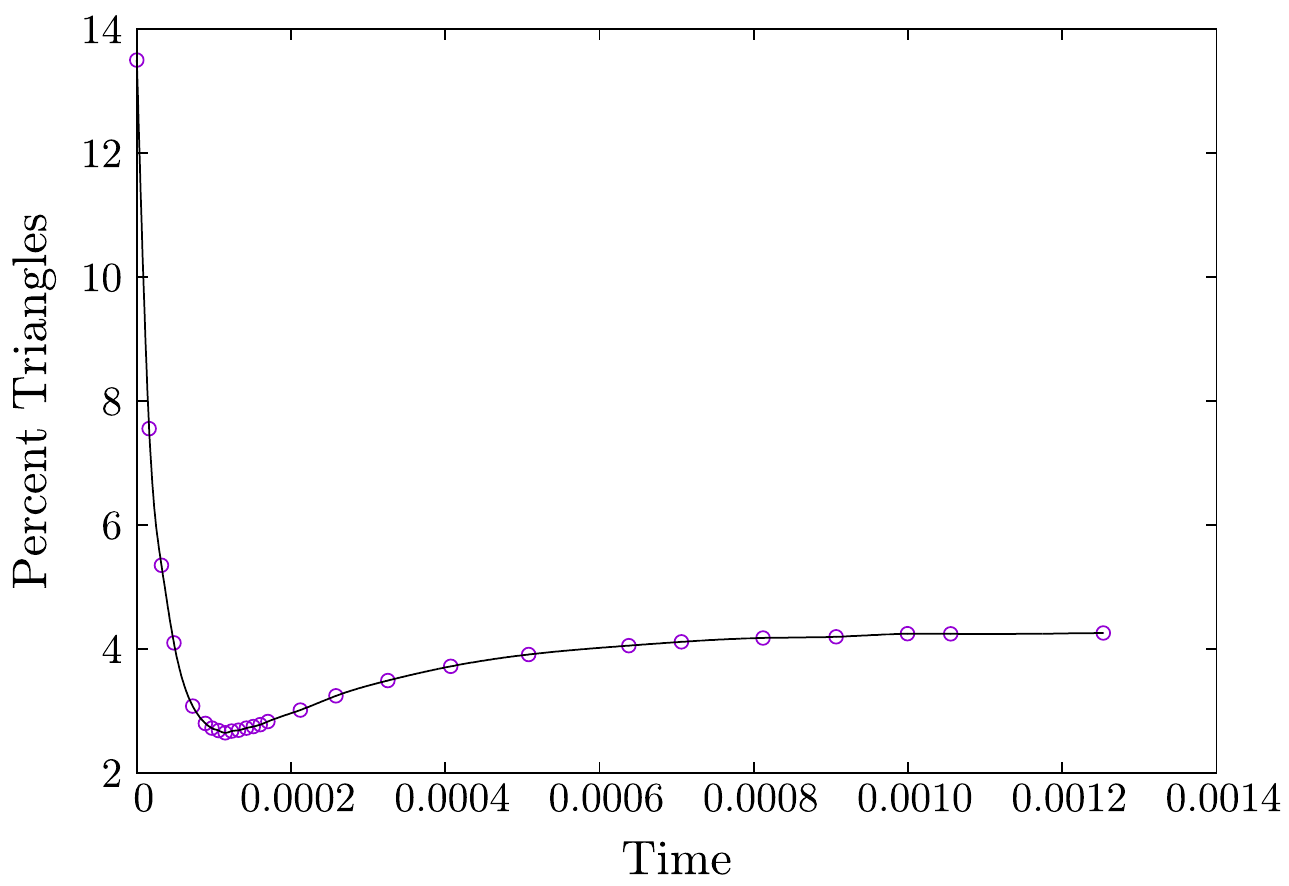}
\caption{Evolution of the percent of triangular faces in time from a Poisson--Voronoi initial condition with constant 
grain boundary energy and mobility.}
\label{fig:triangles_over_time}
\end{figure}

\begin{figure}
\includegraphics[width=1.\columnwidth]{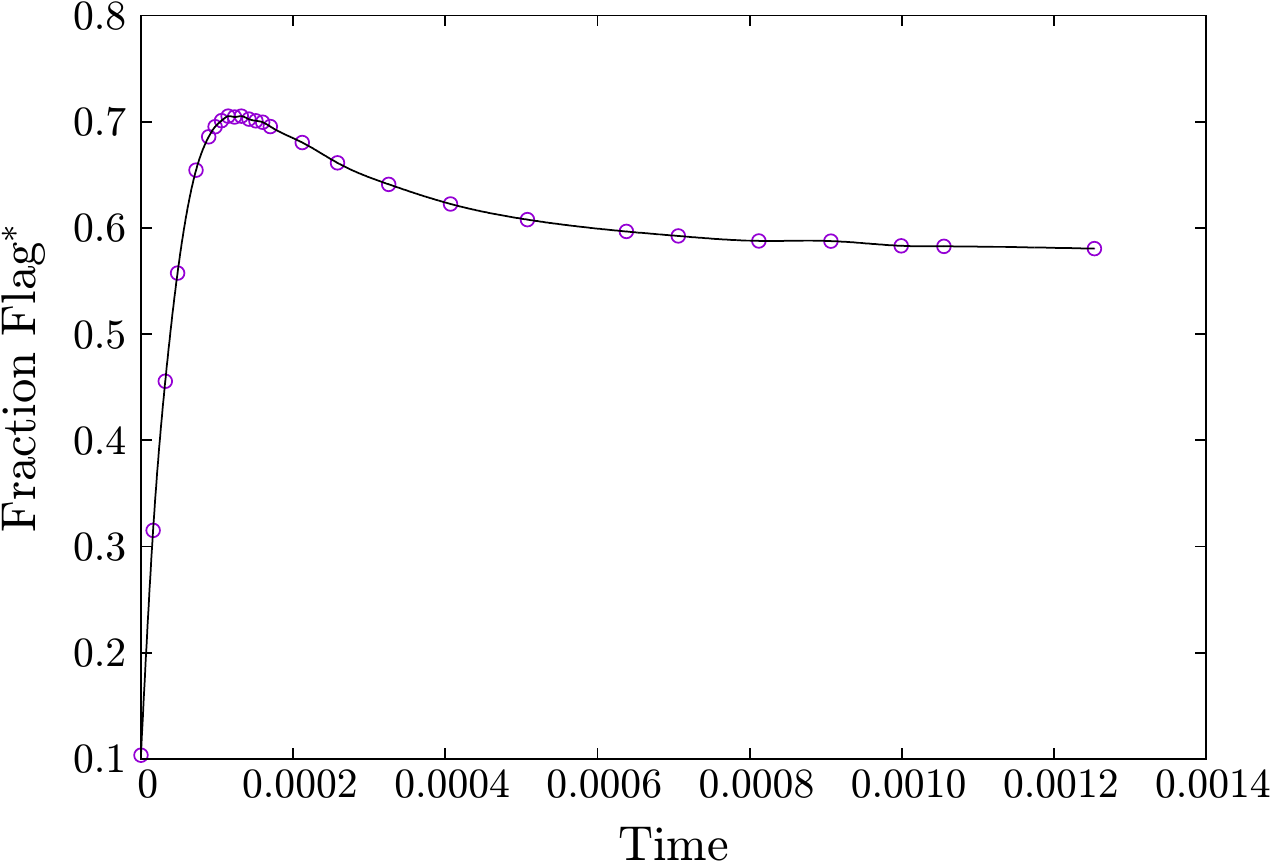}
\caption{Evolution of the fraction of grains that are flag$^*$ in time from a Poisson--Voronoi initial condition with constant 
grain boundary energy and mobility.}
\label{fig:flag_over_time}
\end{figure}

Consider a grain growth simulation initialized with a Poisson--Voronoi microstructure that has an average of about $15.535$ faces 
per grain \cite{1953meijering}. Figure~\ref{fig:face_over_time} shows that the average number of faces per grain 
is (effectively) monotone decreasing in time, and converges to the steady state value of $13.769$ \cite{2011lazar}. However, the percentage 
of faces that are triangles (Figure~\ref{fig:triangles_over_time}) and the fraction of grains that are flag$^*$ (Figure~\ref{fig:flag_over_time}) 
show significant initial transients before converging to the steady state values. The minimum in Figure~\ref{fig:triangles_over_time} 
and the maximum in Figure~\ref{fig:flag_over_time} occur at the same time of $0.000115$. That these should coincide is reasonable since grains 
without triangles are fundamental (flag$^*$ or severely constricted), and the population of severely constricted grains is expected to be small.

Let the initial microstructure be called the \emph{Poisson--Voronoi} condition, the microstructure in the statistical steady state 
be called the \emph{Evolved} condition, and the microstructure at time $0.000115$ be called the \emph{Round} condition. 
The measurements in the following sections are made using three data sets, one for each condition:
\begin{itemize}[noitemsep]
\item The Poisson--Voronoi data set (V-268402) was constructed from 268,402 seeds in the unit cube with periodic boundary conditions.
\item The Evolved data set (E-268402) was obtained by taking $25$ Poisson--Voronoi microstructures, each containing 100,000 grains, 
for a total of 2,500,000 initial grains. These were allowed to evolve by normal grain growth until a total of 269,555 grains 
remained. The data set contains the 268,402 grains that were polytopal (the other $1153$ were degenerate, 
as described in Section~\ref{sec:extremal}).
\item The Round data set (R-268402) was constructed from the same simulation as the Evolved data set by sampling 
268,402 polytopal grains from the microstructure at time $0.000115$.
\end{itemize}

As justification for the behavior in Figures~\ref{fig:triangles_over_time} and \ref{fig:flag_over_time}, observe that the
average grain size increases during the process of grain growth, requiring some grains to shrink and disappear. Since the microstructure 
is roughly equiaxed, reducing the volume of a shrinking grain reduces the edge lengths and face areas of the grain as well.
This induces a sequence of topological operations necessary to maintain the connectivity of the microstructure. From the standpoint 
of a shrinking grain, a vanishing edge is replaced by a different edge that exchanges the neighbors of the bounding vertices (an edge flip), 
and a vanishing face is replaced by a single vertex (a reverse truncation). From the standpoint of the surrounding grains, the same operations 
can involve edge flips, truncations, or reverse truncations. This makes the fraction of triangular faces depend on the relative rates 
of the interacting topological operations in a complicated way. That said, every shrinking grain that remains polytopal must pass through
the tetrahedron before vanishing, and the transient population of tetrahedral grains places a lower bound on the fraction of triangular 
faces present in the microstructure.

For the Poisson--Voronoi data set, $13.503\%$ of the faces are triangles, none of the grains are tetrahedral, $0.01\%$ 
of the grains are stacked$^*$, and $10.36\%$ of the grains are flag$^*$ (Table~\ref{tbl:distribution_V_data}). 
For the Round data set, $2.647\%$ of the faces are triangles, $0.05\%$ of the grains are tetrahedral, 
$0.25\%$ of the grains are stacked$^*$, and $70.52\%$ of the grains are flag$^*$ (Table~\ref{tbl:distribution_R_data}). 
This is interpreted to mean that the capillary pressure characteristic of grain growth drives the grains to be more geometrically round 
and reduces the initially high fraction of triangular faces, while the relative absence of vanishing grains does not bound the fraction 
of triangular faces from below. For the Evolved data set, $4.259\%$ of the faces are triangles, $0.14\%$ of the grains are tetrahedral, 
$0.69\%$ of the grains are stacked$^*$, and $58.05\%$ of the grains are flag$^*$ (Table~\ref{tbl:distribution_E_data}). 
The population of shrinking and vanishing grains has increased relative to the Round data set, eventually reaching a condition 
where the number of triangular faces generated by shrinking grains is equal to the number of triangular faces consumed 
by the action of capillary pressure.

This suggests that the Round microstructure is characterized by grains with unusually stable topology, appearing after the relaxation 
of the Poisson--Voronoi microstructure and before the appearance of a substantial population of vanishing grains. The Round microstructure 
is further notable for having an average of $14.036$ faces per grain, only slightly above the $14$ faces per grain in Kelvin's equal volume foam
\cite{thomson1887division}. While Kelvin's foam was long thought to have the least area of any equal volume foam, 
the Weaire--Phelan foam is more economical with an average of only $13.5$ faces per grain \cite{weaire1994counter}. 
Kusner and Sullivan still conjecture that Kelvin's foam is optimal as an equal volume equal pressure foam \cite{kusner1996comparing}, though.

Real microstructures, for example steels, depart from the above idealized grain growth model in a number of respects. 
The presence of multiple phases, concentration gradients, plastic strain around martensitic lathes, and orientation- and concentration-dependent 
grain boundary properties are all known to affect the evolution of the microstructure, though our understanding of these changes is vague 
due to experimental and theoretical limitations. Fabricating a microstructure in the Round condition would further require 
that the initial microstructure be in the Poisson--Voronoi condition, and this is not always a reasonable approximation 
for samples solidified from the liquid. Nevertheless, an approximate initial condition could be fabricated by dispersing fine insoluble particles 
in the liquid to induce uniform heterogeneous nucleation of the solid phase, resulting in a roughly equiaxed microstructure 
with approximately planar grain boundaries. This should have the same properties as the Poisson--Voronoi data set considered here, 
namely, an excess of triangular faces and a relative absence of tetrahedral grains. A short anneal would still presumably reduce 
the average number of faces per grain, despite the complications mentioned above. An interesting experiment would be to quench 
the sample when the average number of faces per grain is close to $14$ (as an approximation to the Round microstructure), 
both to investigate the topological properties of the grains and whether this topology imparts any enhanced mechanical properties.

\begin{figure}
\includegraphics[width=1.\columnwidth]{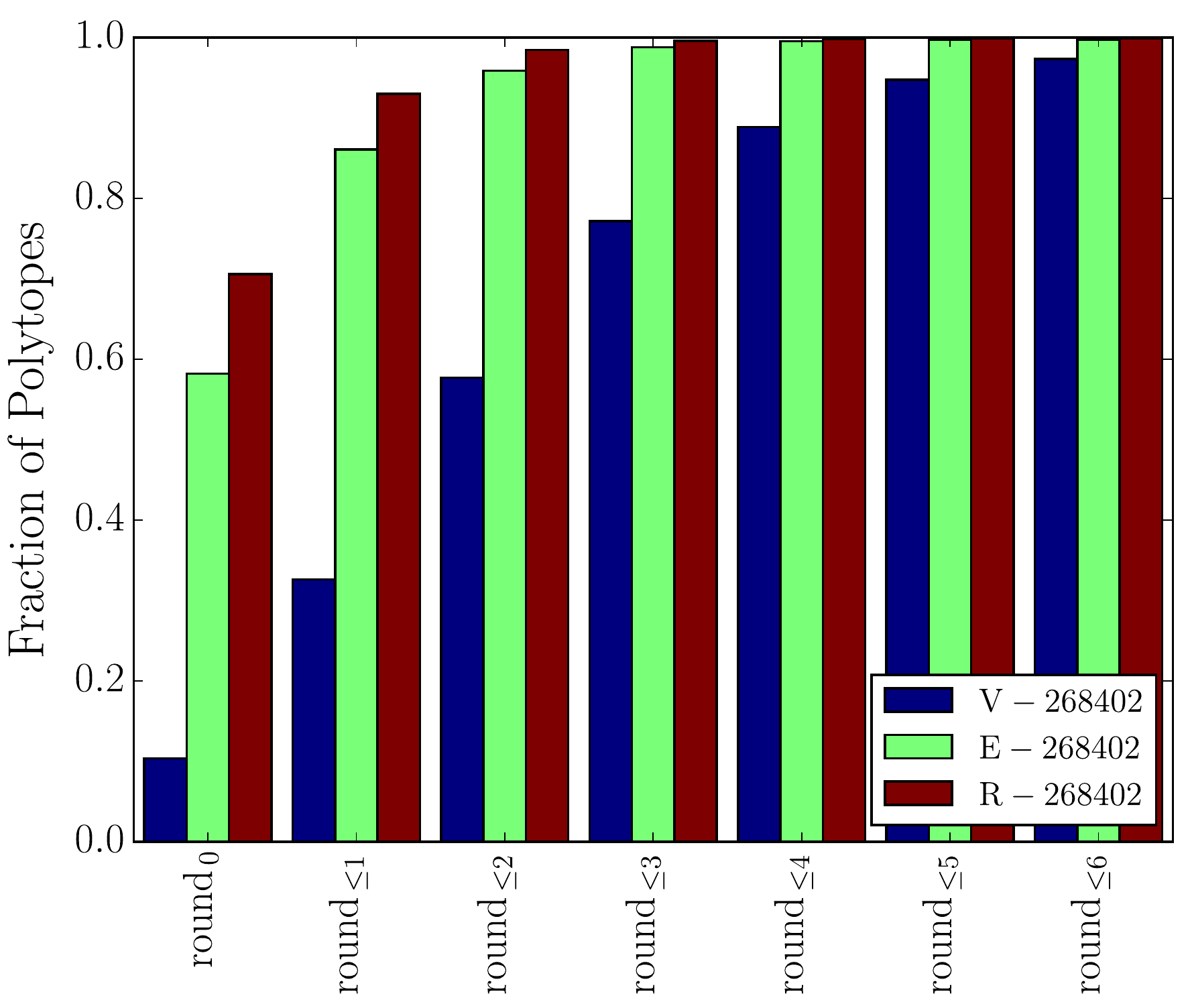}
\caption{Fraction of round$_{\,\leq k}$ grains in the Poisson--Voronoi, Evolved and Round microstructures.}
\label{fig:cumulative_all}
\end{figure}

\begin{figure}
\includegraphics[width=1.\columnwidth]{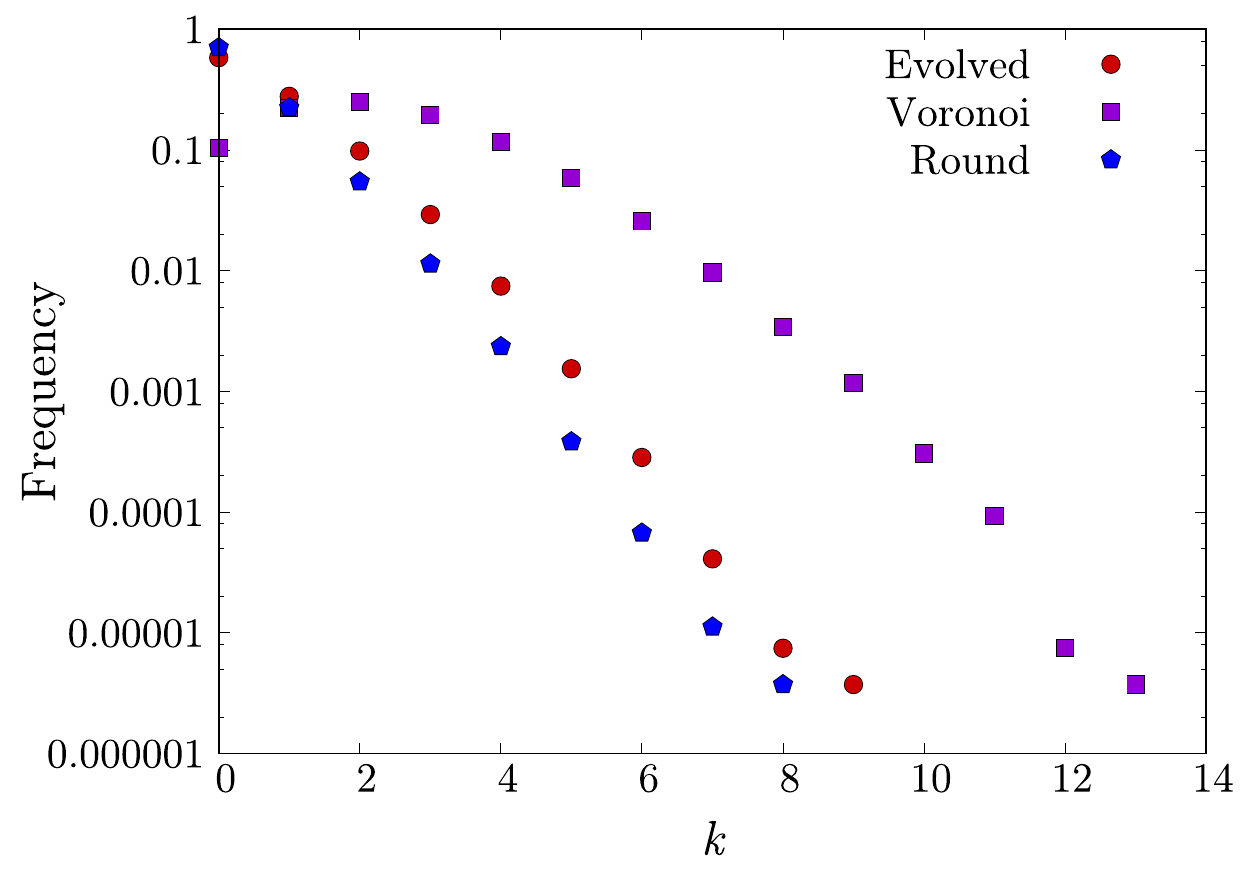}
\caption{The frequencies of round$_k$ grains for the Poisson--Voronoi, Round and Evolved data structures.}
\label{fig:round_k_cuts}
\end{figure}

Whereas Figure~\ref{fig:flag_over_time} shows the fraction of grains that are flag$^*$ as a function of time, the class
round$_{\leq k}$ of all combinatorially $k$-round grains is more robust to the appearance of transient triangular faces.
Table~\ref{tbl:distribution} and Figure~\ref{fig:cumulative_all} give the percentages of $k$-round grains in the three
data sets for $0 \leq k \leq 6$. This is already sufficient to include almost all of the grains since $97.32\%$, $99.86\%$ and $99.72\%$ 
of the polytopal grains in the Poisson--Voronoi, Evolved and Round data sets, respectively, are $6$-round. 
Figure~\ref{fig:round_k_cuts} shows that the percentage of round$_k$ grains
decays roughly exponentially for higher values of $k$, though the reason for this rate of decay is unknown.

\begin{figure}
\includegraphics[width=1.\columnwidth]{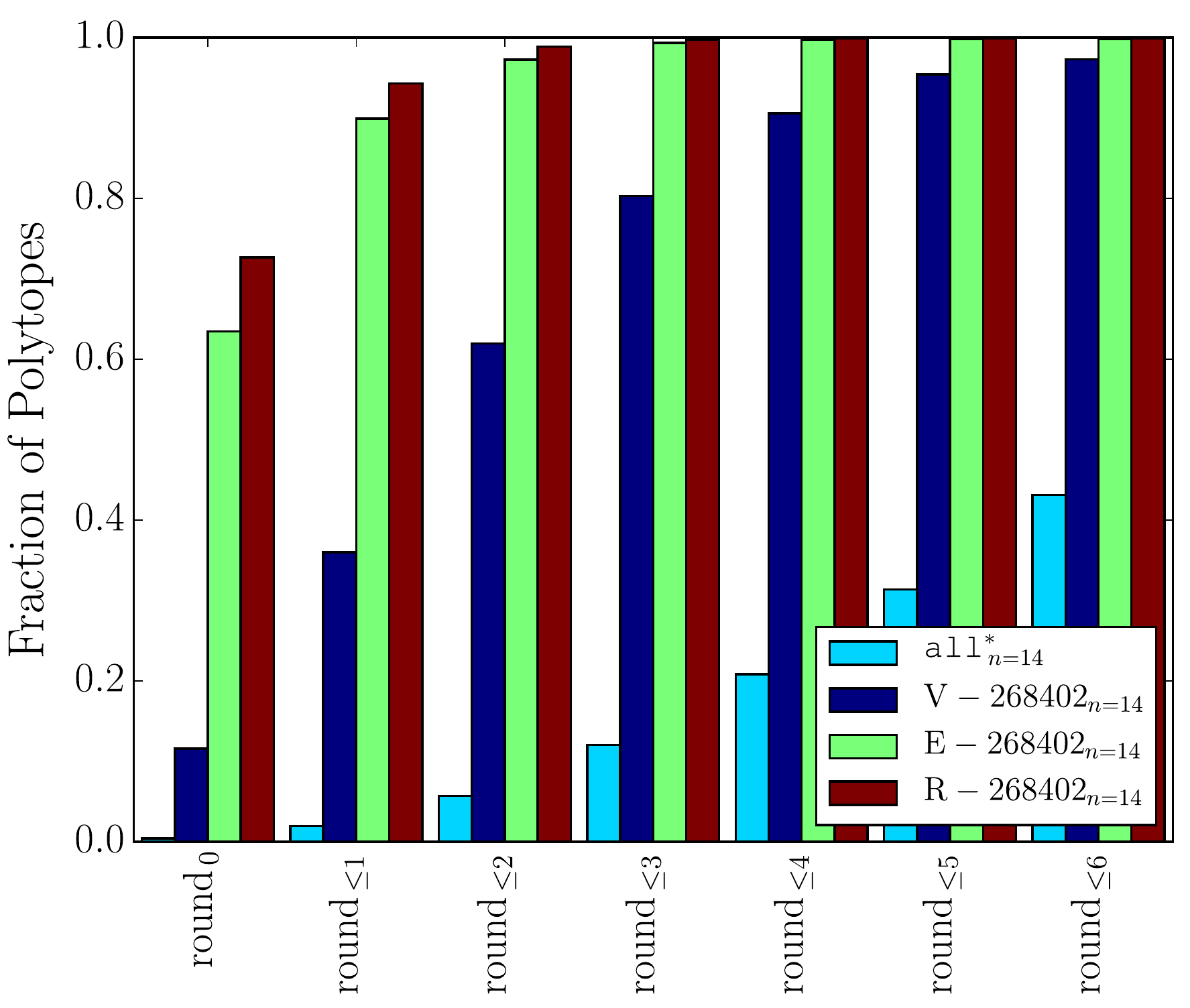}
\caption{Fraction of round$_{\,\leq k}$ grains in the Poisson--Voronoi, Evolved and Round microstructures,
restricted to $n=14$ faces, and compared to all possible grains with $n=14$ faces.}
\label{fig:cumulative_14}
\end{figure}

That the Voronoi data set consistently has fewer $k$-round grains for $0 \leq k \leq 6$ suggests that normal grain growth
strongly supports the formation of combinatorially round grains. Further evidence for this claim is found by comparing
the three data sets to the set of all simple $3$-polytopes with few faces. As mentioned in Section~\ref{subsec:roundness},
triangulations of the $2$-dimensional sphere with up to $23$ vertices have been enumerated by Brinkmann and McKay
\cite{plantri}; refer to Table~\ref{tbl:few_vertices_I} for select statistics. Triangulations with up to $14$ vertices
are further classified in Table~\ref{tbl:few_vertices_II} as stacked, flag, flag$_{+1}$, flag$_{+1+1}$, flag$_{+2}$, flag$_{\,\geq 3}$, 
sc$^*$, sc$^*_{+1}$ or sc$^*_{\,\geq 2}$. This table indicates that the fraction of flag examples decreases with the number of vertices $n$, 
and similarly that the fraction of examples in the class flag$_{\,\leq k}$ decreases with $n$ for a given value of $k$.

Since the Poisson--Voronoi, Round and Evolved data sets all have face averages close to $14$, we compare these with 
the set \texttt{all}$^*_{\,n = 14}$ of simple polytopes that are dual to the set \texttt{all}$_{\,n = 14}$ of all
triangulations of the $2$-sphere with $14$ vertices. The latter includes 339,722 distinct simplicial
$3$-polytopes, only $0.40\%$ of which are flag. Figure~\ref{fig:cumulative_14} shows the fractions of $k$-round 
simple polytopes in \texttt{all}$^*_{\,n = 14}$ 
and the corresponding fractions of $k$-round grains with $14$ faces in the Poisson--Voronoi, Round and Evolved data sets. 
We infer that although $k$-round grains are rare among all possible grains for small~$k$, the Poisson--Voronoi,
Round and Evolved microstructures consist almost entirely of these grains.

Alternatively, consider the grains that do not belong to $k$-round for small $k$. From Table~\ref{tbl:distribution},
$43.11\%$ of all possible grains with $14$ faces are $6$-round, and the $56.89\%$ remaining necessarily belong to the types
flag$^*_{\,\geq 7}$ or sc$_{\,\geq}$. By comparison, only $2.68\%$, $0.28\%$ and $0.14\%$ of the grains with $14$ faces
in the Poisson--Voronoi, Evolved and Round data sets, respectively, are flag$^*_{\,\geq 7}$ or sc$_{\,\geq}$.
That is, grains of the types flag$^*_{\,\geq 7}$ or sc$_{\,\geq}$ are rare in the space-filling Poisson--Voronoi, Evolved and
Round microstructures; we conjecture that this is true of generic space-filling microstructures as well.

\subsection{Geometric Roundness}
\label{subsec:geometric}

The motivation for our study of the combinatorial types of grains is that there should exist some combinatorial property 
that is positively correlated with the geometric roundness of a grain. That the combinatorial roundness as defined 
in Section~\ref{subsec:roundness} should satisfy this criterion is perhaps not obvious. To the contrary, starting with 
a convex polyhedron and repeatedly truncating the vertex with the highest angular deficit would appear to make 
the polyhedron more spherical. That is, round$_k$ for any $k > 0$ could contain grains that are \emph{more} geometrically round 
than those in round$_0$, not less as suggested in Section~\ref{subsec:microstructure_evolution}.

This reasoning does not account for the effect of the vertex truncations on the surrounding microstructure, though. 
Specifically, for an isotropic boundary energy the system evolves to reduce the total boundary area, with the consequence 
that the tangents to the edges at a vertex always meet at the tetrahedral angle. Introducing vertices on the surface 
of a grain by vertex truncation can then increase the grain's surface area to volume ratio if the adjoining faces acquire 
a negative curvature to satisfy the edge constraint.

Of course, the natural recourse to discern the relationship of combinatorial roundness to geometric roundness is direct measurement. 
A standard measure for geometric roundness of a $3$-dimensional object is the dimensionless \emph{isoperimetric quotient} 
\begin{equation}
Q = 36 \pi V^2 / S^3,
\end{equation}
where $V$ is the volume and $S$ is the surface area of the object. The isoperimetric quotient is confined to the interval $0 \leq Q \leq 1$, 
with values of $1$ for the sphere, $0.755$ for the dodecahedron, $0.524$ for the cube, and $0.302$ for the tetrahedron (all regular). 
That is, the value increases with geometric roundness.

\begin{figure}
\includegraphics[width=1.\columnwidth]{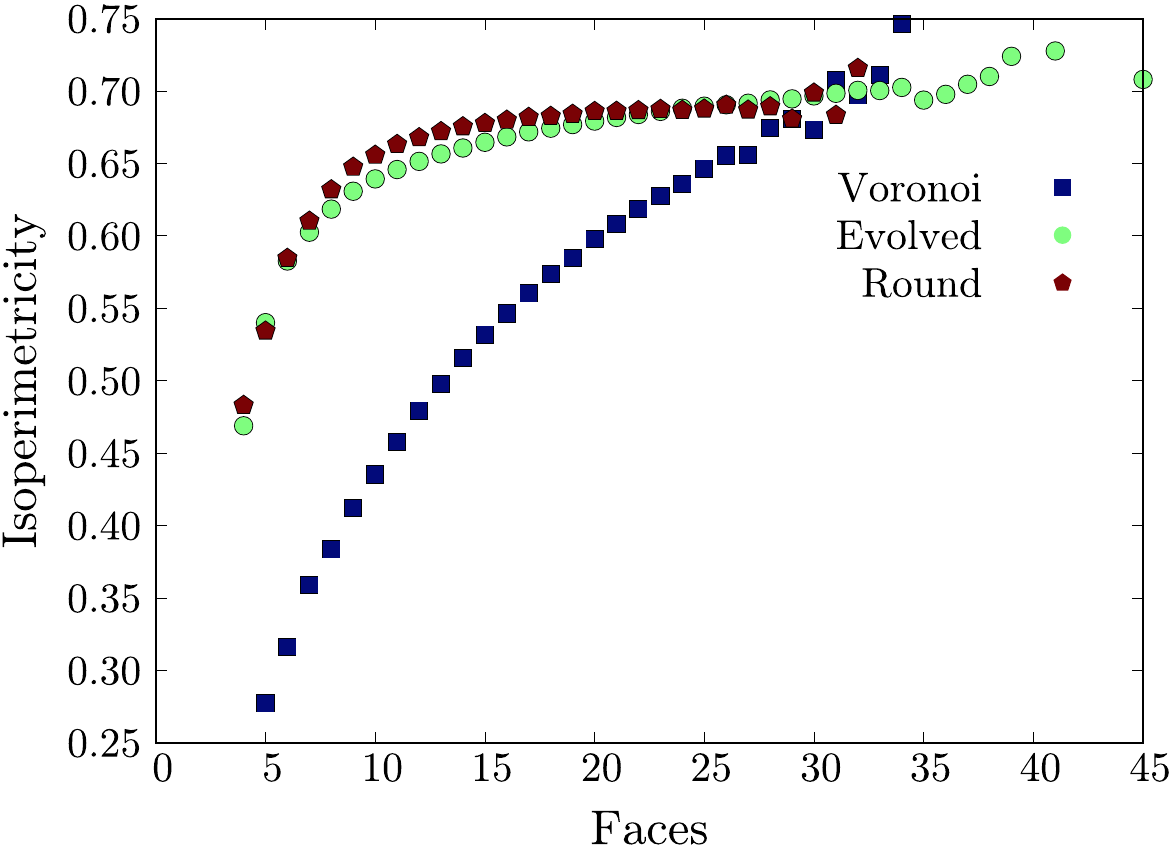}
\caption{The average isoperimetricity of grains with a given number of faces for the Poisson--Voronoi, Round and Evolved microstructures.}
\label{fig:faces_vs_isop}
\end{figure}

\begin{figure}
\includegraphics[width=1.\columnwidth]{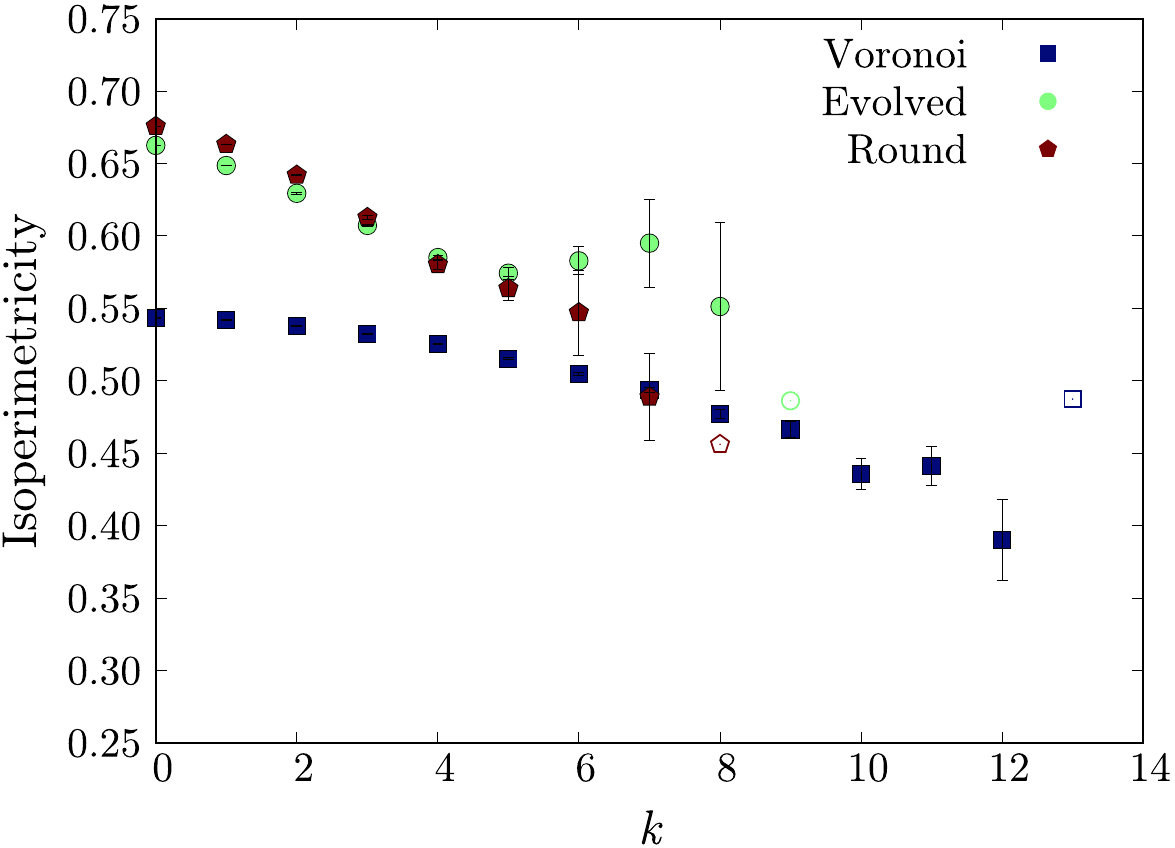}
\caption{The average isoperimetricity of round$_k$ grains for the Poisson--Voronoi, Round and Evolved data structures. 
Error bars are the standard error of the mean, and classes with a single grain are indicated by an open marker.}
\label{fig:cuts_vs_isop}
\end{figure}

Figure~\ref{fig:faces_vs_isop} shows the average isoperimetric quotient of the grains as a function of the number of faces 
for the Poisson--Voronoi, Round and Evolved microstructures. The average reliably increases with the number of faces, 
with the plateau for the Round and Evolved microstructures possibly related to an upper bound for uniform space-filling cellular structures. 
Grains with fewer than $23$ faces (the vast majority) consistently have the highest average isoperimetric quotients 
in the Round microstructure and the lowest in the Voronoi microstructure, consistent with the fractions of round$_{\leq k}$ grains 
in Figure~\ref{fig:cumulative_all}. The scatter for the highest number of faces is a consequence of sampling error, 
though the Evolved microstructure does contain a population of grains with many faces and high isoperimetric quotient 
that is absent from the other microstructures.


Figure~\ref{fig:cuts_vs_isop} more directly addresses the central concern of this section, namely, the correlation 
of the average isoperimetric quotient of round$_k$ grains with $k$. That the isoperimetric quotient generally decreases 
with $k$ indicates that the combinatorial roundness and geometric roundness are indeed positively correlated, 
as hypothesized in Section~\ref{sec:combinatorial_analysis}. The obvious exception is the increasing average isoperimetric quotient 
for grains in the Evolved microstructure that are $5 \leq k \leq 8$ vertex truncations away a fundamental grain. 
A more detailed investigation reveals that this population mainly consists of two groups, a group of stacked$^*$ grains 
with $n \leq 13$ faces and of a group of truncated flag$^*$ grains with $n \geq 20$ faces. The first group 
has a much smaller average isoperimetricity ($0.491 \pm 0.065$) than the second ($0.665 \pm 0.037$), 
which along with the small population size explains the increase in the standard error of the mean.

An explanation for the existence of the group with $n \geq 20$ faces and the shift in the average isoperimetic quotient begins 
with the observation that the Evolved microstructure contains some grains that are significantly larger than the average grain, 
and that often continue to grow and consume the surrounding structure. Since the microstructure has a well-defined average edge length, 
these grains are expected to have more faces than the average grain. Since the surrounding grains are shrinking and transitioning through 
tetrahedra, more of the faces are expected to be triangles than for the average grain. That is, the largest grains are overrepresented 
in the Evolved microstructure for $5 \leq k \leq 8$ in Figure~\ref{fig:cuts_vs_isop}, increasing the average isoperimetric quotient. 
This explanation is supported by the intimation of a similar deviation for the Round microstructure where the population 
of largest grains is not yet fully developed.

\subsection{Grain Classes and Error Analysis}
\label{subsec:grain_classes}

\begin{figure}
\includegraphics[width=1.\columnwidth]{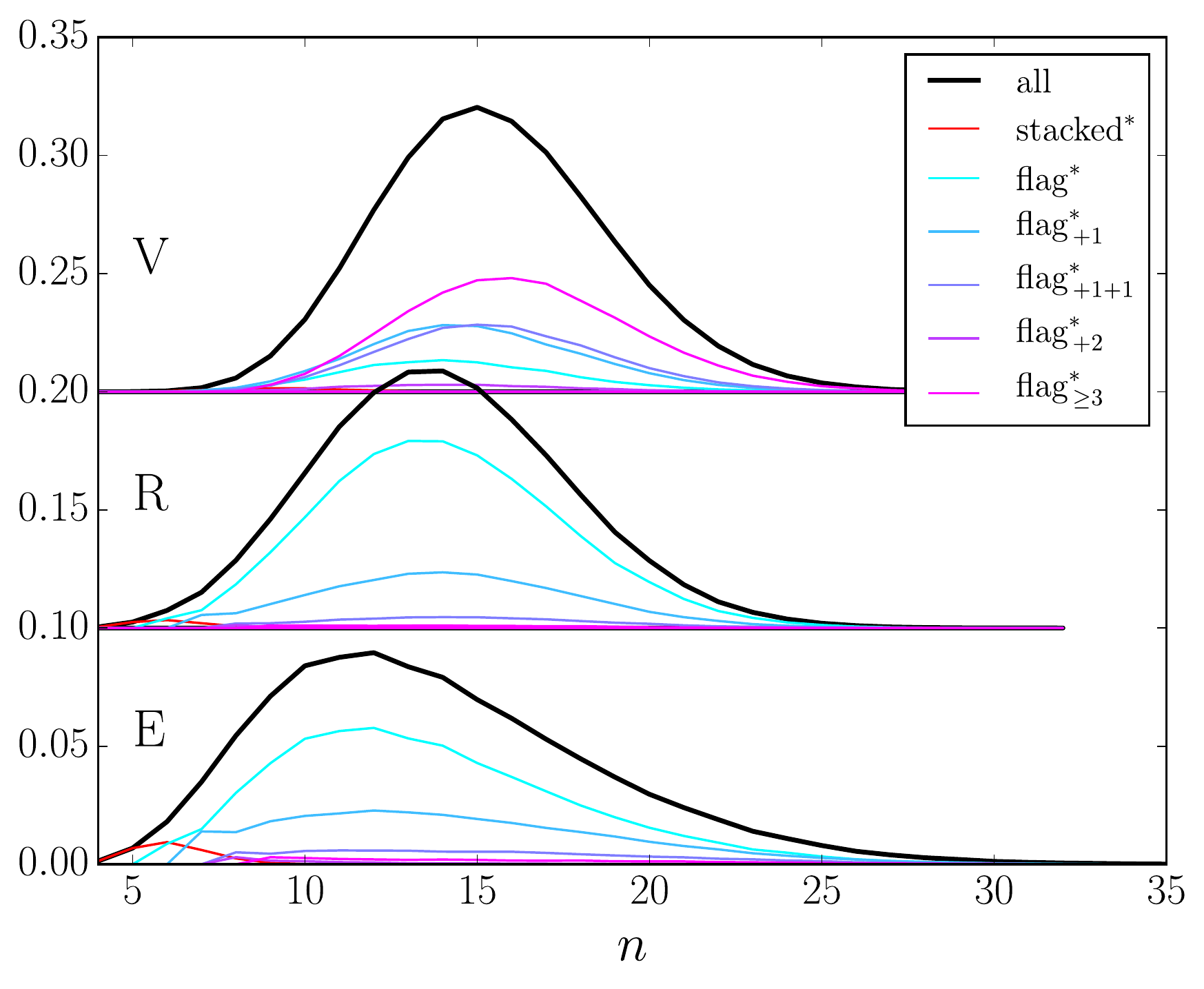}
\caption{Frequencies of various grain classes in the Poisson--Voronoi (V), Round (R), and Evolved (E) microstructures, 
where $n$ is the number of faces. Values for the Poisson--Voronoi and Round microstructures are vertically offset 
by $0.20$ and $0.10$, respectively.}
\label{fig:aggregate}
\end{figure}

The dependence of the frequencies of the simple polytope classes defined in Section~\ref{subsec:roundness} on the system
evolution is considered further in Figure~\ref{fig:aggregate}, which shows the frequencies of several simple
polytope classes as functions of the number of faces~$n$ in the Poisson--Voronoi, Round, and Evolved microstructures.
The black curves are the distributions of the number of faces, and show that the system evolution on average
reduces the number of grains with many faces and increases the number of grains with few faces. This is 
consistent with the average number of faces per grain being $15.535$, $14.036$, and $13.769$ for the Poisson--Voronoi,
Round, and Evolved microstructures, respectively.

Figure~\ref{fig:aggregate} further shows a noticeable decrease in the flag$^*_{+1+1}$ and flag$^*_{\geq 3}$ grain populations
from the Poisson--Voronoi to the Round microstructure, with a corresponding increase in the population of flag$^*$
grains. Since a $0$-round grain is necessarily either tetrahedral or belongs to the class flag$^*$,
the increase in the population of flag$^*$ grains is consistent with the Round microstructure
having more $0$-round
grains than the Poisson--Voronoi microstructure in Figure~\ref{fig:cumulative_all}. Note also that grains of type
flag$^*_{+1+1}$ are consistently more frequent than grains of type flag$^*_{+2}$
(Tables~\ref{tbl:distribution_V_data}, \ref{tbl:distribution_R_data} and \ref{tbl:distribution_E_data}).
The reason for this is likely combinatorial. Given a flag$^*$ grain, a grain of type flag$^*_{+1+1}$ is obtained
by truncating any two of the original vertices. A grain of type flag$^*_{+2}$ is only obtained if the second
truncated vertex is one of the three vertices newly introduced by the first truncation.

There are three further significant changes in the frequencies of simple polytope classes as the system evolves 
from the Round to the Evolved microstructure. First, non-negligible populations of stacked$^*$ and flag$^*_{+1}$ grains 
with small numbers of faces appear. These grains are likely in the process of losing faces before vanishing, and correspond to
a transient population that is not substantially present in either the Poisson--Voronoi or Round microstructures.
Second, the frequency of flag$^*$ grains is somewhat reduced overall, with the distribution skewed toward small
$n$. The significance of the skewness is not known, but the reduced frequency of flag$^*$ grains is consistent with
the Evolved microstructure having fewer $0$-round grains than the Round microstructure in Figure~\ref{fig:cumulative_all}.
Third, the Evolved microstructure has more grains (and in particular flag$^*$ grains) with $n\geq 22$ faces than the 
Round microstructure. These grains represent heterogeneities that could degrade certain mechanical properties, and
provide further motivation to investigate the properties of materials that approximate the Round microstructure.

\begin{figure}
\includegraphics[width=0.3\columnwidth]{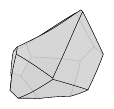}
\caption{Dodecahedron from the Evolved data set with several short edges.}
\label{fig:short_edge}
\end{figure}

The preceding analysis of combinatorial types presupposes that the topology of a grain is well-defined. The example
grain in Figure~\ref{fig:short_edge} suggests that arbitrarily small perturbations to a grain's geometry could induce
topological transitions that change the grain's combinatorial type without changing the number of faces, though. 
This raises a reasonable question about the independence of the statistics reported in this section to errors introduced 
by implementation details and finite precision arithmetic.

\begin{figure}
\includegraphics[width=0.4\columnwidth]{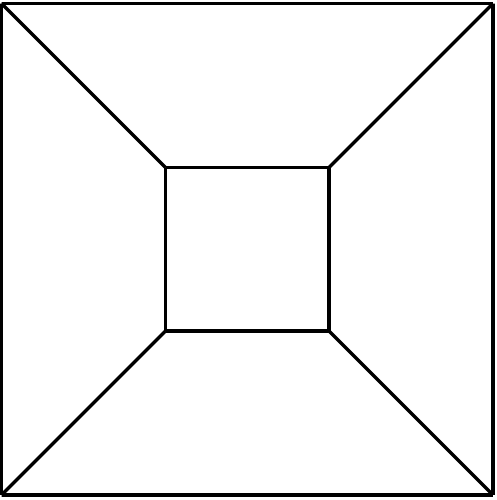}\hspace{5mm}
\includegraphics[width=0.46\columnwidth]{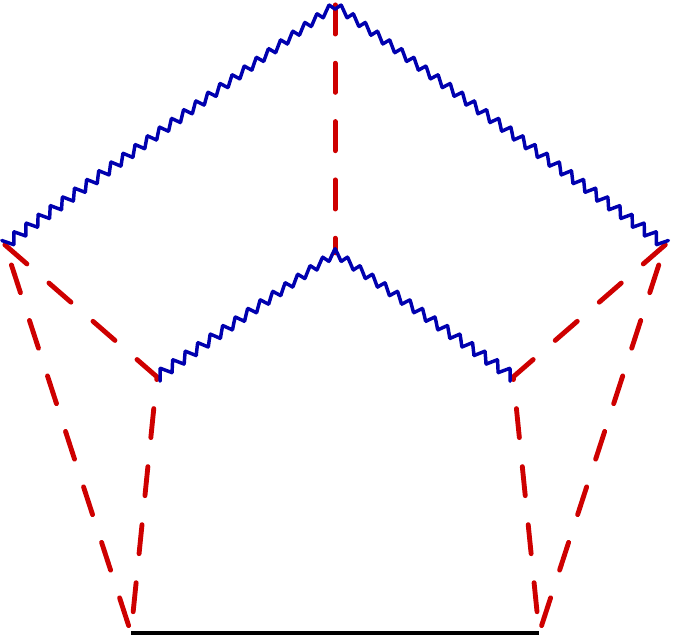}
\caption{Schlegel diagrams of the two polytopes with six faces. The flip of a bold edge gives the other polytope, 
the flip of a zigzag edge gives the same polytope, and the dashed edges are non-admissible.}
\label{fig:grains_with_6_faces}
\end{figure}

Consider the effect of a fixed rate of topological edge errors per grain on the frequency of round$_k$ grains.
Starting from the Evolved data set with 268,402 grains, every grain is subjected to a fixed number of edge flips
selected uniformly at random from the set of admissible edges (ones that do not give a degenerated grain). Note
that the presence of combinatorial symmetries means that distinct flips can sometimes give the same
combinatorial type, or even leave the combinatorial type unchanged. For example, the only two simple $3$-polytopes
with six faces are the cube with $p$-vector $(0,6,0)$ and the unique simple $3$-polytope with $p$-vector $(2,2,2)$,
where the $p$-vector is described in Appendix~\ref{sec:constraints}. The Schlegel diagrams of these polytopes are
given in Figure~\ref{fig:grains_with_6_faces}. The flip of any edge of the cube gives the second polytope. For
the second polytope, the flip of any of four edges does not change the polytope, the flip of one edge gives the
cube, and the remaining seven edges are non-admissible. There is generally more variability for grains with more faces.

\begin{figure}
\includegraphics[width=1.0\columnwidth]{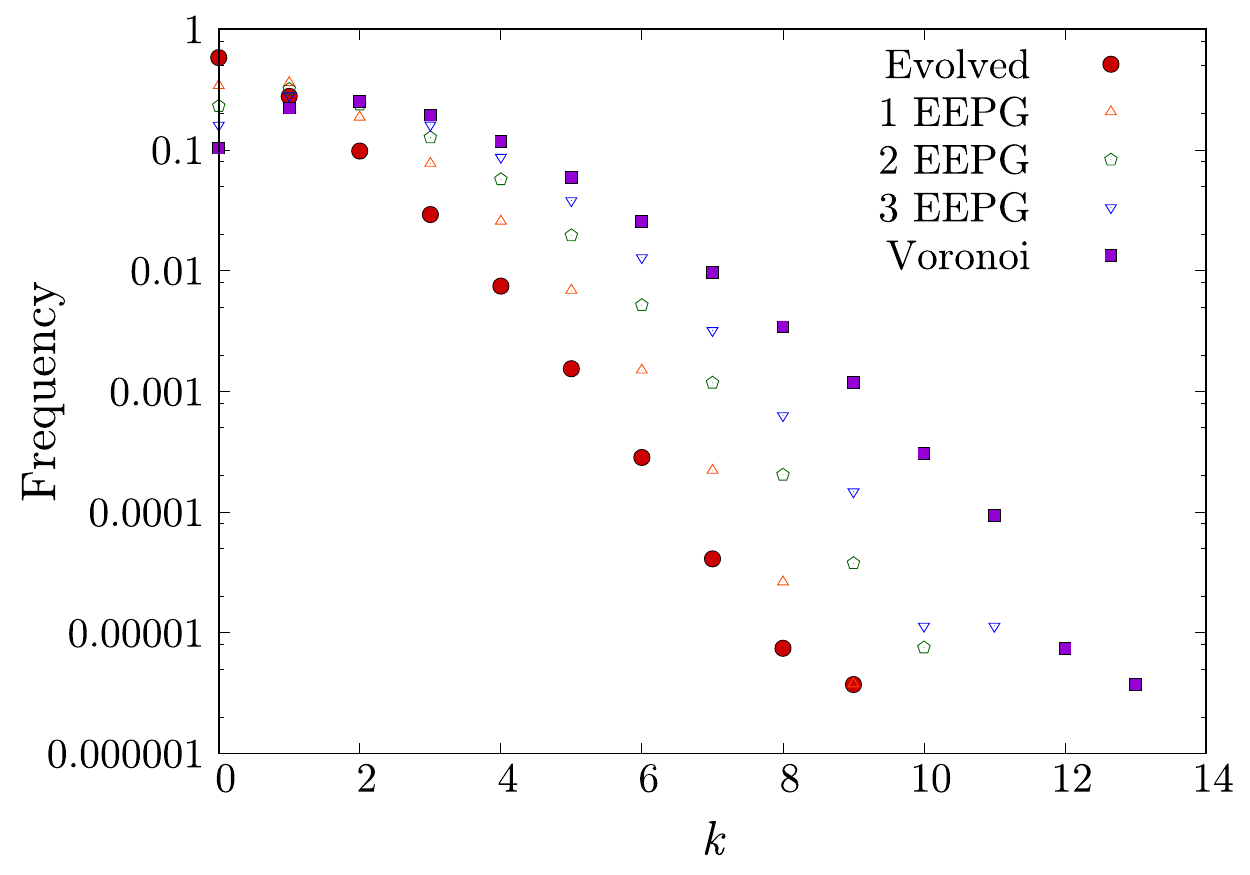}
\caption{The frequencies of round$_k$ grains in the Poisson--Voronoi and Evolved data sets, and for grains 
in the Evolved data set after flipping a fixed number of edges per grain. EEPG stands for edge errors per grain.}
\label{fig:combined_cut_types}
\end{figure}

Figure~\ref{fig:combined_cut_types} shows the effect of the above procedure on the frequencies of round$_k$ grains in the Evolved data set,
along with the corresponding frequencies for the Poisson--Voronoi and Evolved data sets. Observe that
the frequencies do not change substantially for one edge error per grain. Since the fraction of edges in the
Evolved data set with less than one-tenth the average length is about $0.027$ and a grain in the
Evolved microstructure has about $35$ edges \cite{mason2015geometric}, there is roughly one error candidate per
grain. That said, this is a very conservative estimate, the actual average number of edge errors per grain is
likely much less than one, and this source of error is unlikely to significantly change the reported statistics. 
We take this opportunity to further observe that increasing the number of edge errors per grain to two or three 
causes the frequencies of round$_k$ grains to move in the direction of those of the Poisson--Voronoi data set
--- which is still far from randomly picking grains from the set of all grains with few faces.

\subsection{Flag$^*$ Grains}
\label{subsec:split_types}

The above sections define several combinatorial classes of simple polytopes and consider the frequencies of grains belonging 
to those classes in a space-filling microstructure evolving by normal grain growth. This section is instead concerned 
with a refined analysis of the class of flag$^*$ simple $3$-polytopes and the extent to which the frequency 
of a particular combinatorial type can be predicted from combinatorial information alone. The present proposal is restricted to flag$^*$ simple 
$3$-polytopes since this is the main class considered in the previous sections, and the topo\-logy of these polytopes is relatively well constrained.

As explained in Section~\ref{subsec:polytopal}, one of the characteristics of a flag$^*$ simple $3$-polytope is the absence 
of non-trivial $3$-belts. Let the notation $k\!:\!\ell\!:\!m$ indicate that a polytope with $k + \ell + m$ faces has an $\ell$-belt, 
and that the removal of this belt leaves two disks with $k$ and $m$ faces where \mbox{$k \leq m$}. Then the property 
of flag$^*$ simple $3$-polytopes mentioned above is equivalent to simple 3-polytopes
with $n\geq 6$ faces not having any $3$-belts except for ones of the type $0\!:\!3\!:\!(n - 3)$.

Intuitively, a belt of type $k\!:\!\ell\!:\!m$ for fixed $\ell$ imposes a constriction on the polytope that becomes increasingly severe 
with increasing $k$. For example, a cube has a $4$-belt of type $1\!:\!4\!:\!1$ and is reasonably described as unconstricted, 
while a simple $3$-polytope with a belt of type $6\!:\!4\!:\!6$ is likely constricted similarly to the grain in Figure~\ref{fig:smallest_nonflag}. 
This motivates the following definition:
\begin{deff}
A flag$^*$ simple $3$-polytope $P$ with $n$ faces has \emph{split type} $k\!:\!\ell\!:\!m$ for a given $\ell\geq 4$ if $k$ 
is the largest integer (with $k\leq m$) such that $P$ has an $\ell$-belt of type $k\!:\!\ell\!:\!m$. If $P$ does not have an $\ell$-belt, 
it is said to have split type $0\!:\!\ell\!:\!(n - \ell)$.
\label{def:split_type}
\end{deff}
This definition induces a corresponding notion of split-type for the dual simplicial $3$-polytope $P^*$. Specifically, $P^*$ 
has split type $k\!:\!\ell\!:\!m$ if $k$ is the largest integer such that $P^*$ has a cycle of length $\ell$ that, when removed, 
separates the triangulation into two components with $k$ and $m$ vertices.

A simple $3$-polytope with a non-trivial $3$-belt is either vertex-truncated or severely constricted, and is excluded from the present analysis. 
The most severe possible constriction of a flag$^*$ simple $3$-polytope is therefore from a non-trivial $4$-belt. The present analysis considers 
only $4$-belts for simplicity, though a more complete analysis would likely include information concerning $\ell$-belts for arbitrary $\ell$
(where the existence of a non-trivial $\ell$-belt with $\ell\leq 5$ is guaranteed by Equation~\ref{eq:valence} in Appendix~\ref{sec:constraints}).
Given this restriction, the notation $k\!:\!4\!:\!m$ is abbreviated to $k\!:\!m$ and the phrase `split type' refers to $4$-belts in the following.

Table~\ref{tbl:flag_few_vertices} lists the split types of all flag triangulations with up to $n = 14$ vertices. The corresponding numbers 
of grains with every observed split type in the Poisson--Voronoi, Round and Evolved data sets are reported 
in Tables~\ref{tbl:flag_distribution_V_data}--\ref{tbl:flag_distribution_E_data}, respectively. The notable absence 
of split types $1\!:\!3$ and $0\!:\!9$ in Table~\ref{tbl:flag_few_vertices} implies that flag$^*$ grains with these split types are forbidden. 
For example, a flag$^*$ grain with split type $0\!:\!9$ would necessarily have $13$ faces, none of which could be $3$-gons or $4$-gons, 
and would have the $p$-vector $(0,0,12,1)$ that is forbidden by Equation~\ref{eq:eberhard} \cite[p.~271]{Gruenbaum1967}.

Among all flag triangulations with $14$ vertices, the split types $5\!:\!5$, $4\!:\!6$, $3\!:\!7$, $2\!:\!8$, and $1\!:\!9$ occur 
with the strictly decreasing frequencies $40.83\%$, $19.23\%$, $17.32\%$, $15.40\%$, and $7.15\%$, respectively. 
This is contrary to the trend observed for our space-filling structures in Tables~\ref{tbl:flag_distribution_V_data}--\ref{tbl:flag_distribution_E_data}, 
where the frequency of grains with $14$ faces and split type $k\!:\!m$ generally decreases with increasing $k$ 
(consistent with the motivation for Definition \ref{def:split_type}). For example, the frequencies of flag$^*$ grains 
with $14$ faces and split type $5\!:\!5$ are only $8.02\%$, $2.38\%$ and $2.06\%$ for the Voronoi, Round, 
and Evolved microstructures, respectively.

A flag$^*$ simple $3$-polytope does not have any truncated vertices by definition. A flag$^*$ simple $3$-polytope with a $4$-gon face 
admits a $4$-belt around that face, and therefore has split type $k\!:\!m$ with $1 \leq k$. As a consequence:
\begin{lem} 
A flag$^*$ simple $3$-polytope of split type $0\!:\!m$ has no $3$-gon or $4$-gon faces.
\end{lem} 

By Equation~\ref{eq:valence}, a flag$^*$ simple $3$-polytope with split type $0\!:\!m$ has at least $12$ faces; the boundary of the dodeca\-hedron 
is the unique flag$^*$ simple $3$-polytope with split type $0\!:\!8$ and $12$ faces. 
The next smallest example is the unique flag$^*$ simple $3$-polytope with split type \mbox{$0\!:\!10$} and $14$ faces. 
Since the average number of faces per grain in the Poison--Voronoi, Round and Evolved data sets is less than or equal to $15.535$
(and since initially the fraction of flag$^*$ examples is small in the Poison--Voronoi data set), 
flag$^*$ simple $3$-poly\-topes with split type $0\!:\!m$ and more than~$14$ faces should be infrequent in these or any other generic microstructure.
A notable example of a non-generic microstructure is the Weaire--Phelan foam, which consists of the two cell types with split types
$0\!:\!8$ and $0\!:\!10$ and has an average of $13.5$ faces per grain~\cite{kusner1996comparing}.
Note that it is not possible to tile space with only the dodeca\-hedron --- this would yield the notorious 120-cell with positive curvature.
Decompositions of non-isotropic spaces $\Sigma^2\times S^1$ (with $\Sigma^2$ any closed surface) are discussed 
elsewhere~\cite{JMSullivan2000}.

\begin{figure}
\includegraphics[width=1.\columnwidth]{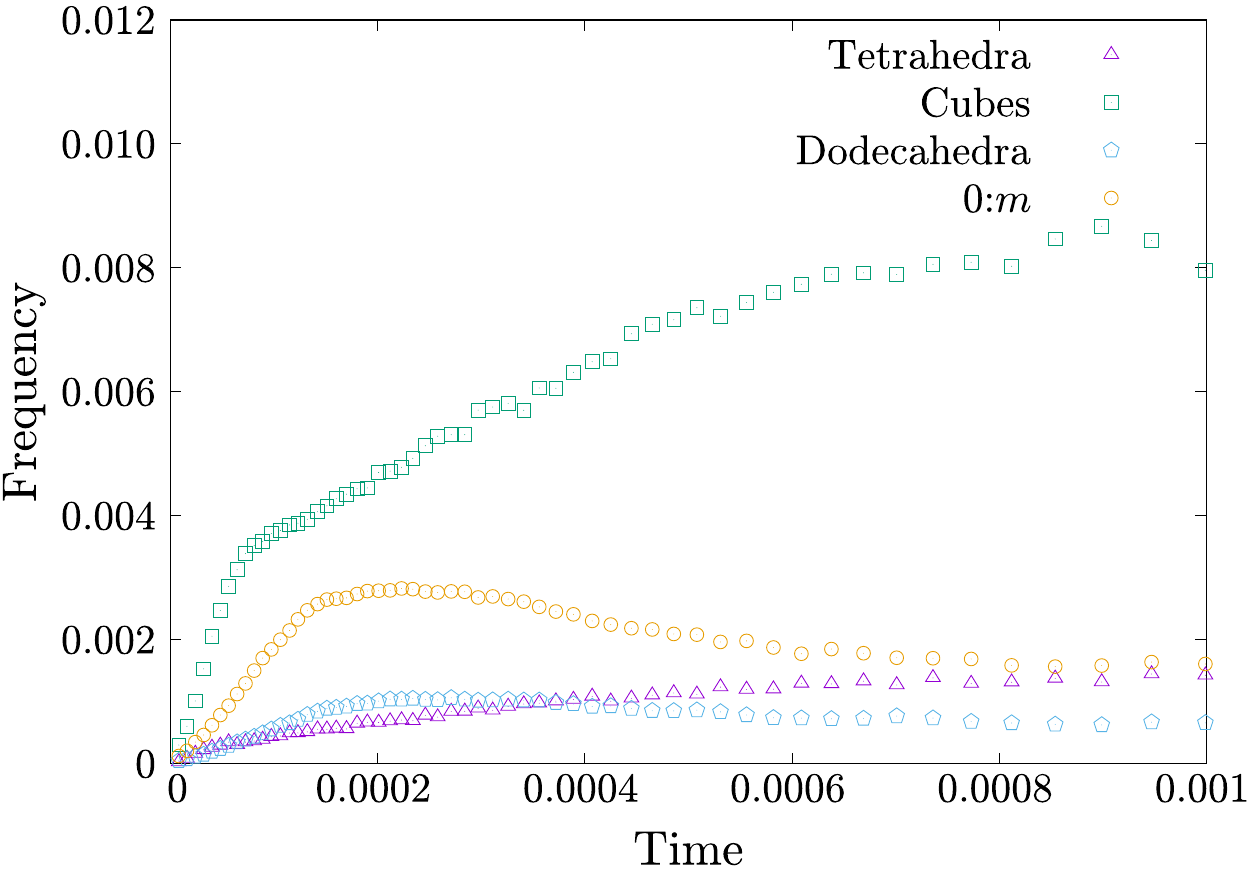}
\caption{Percentage of tetrahedra, cubes, dodecahedra, and grains of type 0:$m$ over time.}
\label{fig:combinatorial_parameters}
\end{figure}

We now consider four prominent species of fundamental grains, the tetrahedron (the smallest fundamental grain),
the cube (the smallest flag$^*$ grain), the dodecahedron (the smallest flag$^*$ grain of split type $0\!:\!m$)
and the set of all flag$^*$ grains of split type $0\!:\!m$.
The fractions of these four species are shown in Figure~\ref{fig:combinatorial_parameters} 
as functions of time, starting from the Poisson--Voronoi microstructure and ending with the Evolved microstructure. 
The gradual increase in the fraction of tetrahedra is caused by a higher initial rate of grains losing faces and transitioning 
to tetrehedra than of tetrahedra losing faces and vanishing. 
The fraction of tetrahedra stabilizes in the Evolved microstructure since these two processes necessarily occur at the same rate 
in the steady state. The fractions of dodecahedra and grains with split type $0\!:\!m$ peak at time $0.000212$, 
with $0.108\%$ and $0.287\%$ of grains being of these types, respectively. 
Note that this is significantly after the point where the fraction of flag$^*$ grains peaks and the Round microstructure is identified. 
The cause of the abrupt shoulder in the fraction of cubes is unknown.


The above examples indicate that combinatorial information alone is not generally sufficient to predict the frequency of
a particular flag$^*$ simple $3$-polytope; Figure~\ref{fig:combinatorial_parameters} shows
that the frequency of cubes changes significantly from the Poisson--Voronoi to the Evolved microstructure. More extreme
examples are provided by the Kelvin foam and the Weaire--Phelan foam. Kelvin's foam consists of a single cell
(the truncated octahedron) with six square faces and eight hexagonal faces. The Weaire--Phelan foam contains two types
of cells with significant numbers of pentagonal faces. Since the distributions of cell types for these two foams are
disjoint, there cannot be any function that precisely describes the frequency of combinatorial types in \emph{all}
space-filling cellular structures.

\begin{figure}
\includegraphics[width=1.\columnwidth]{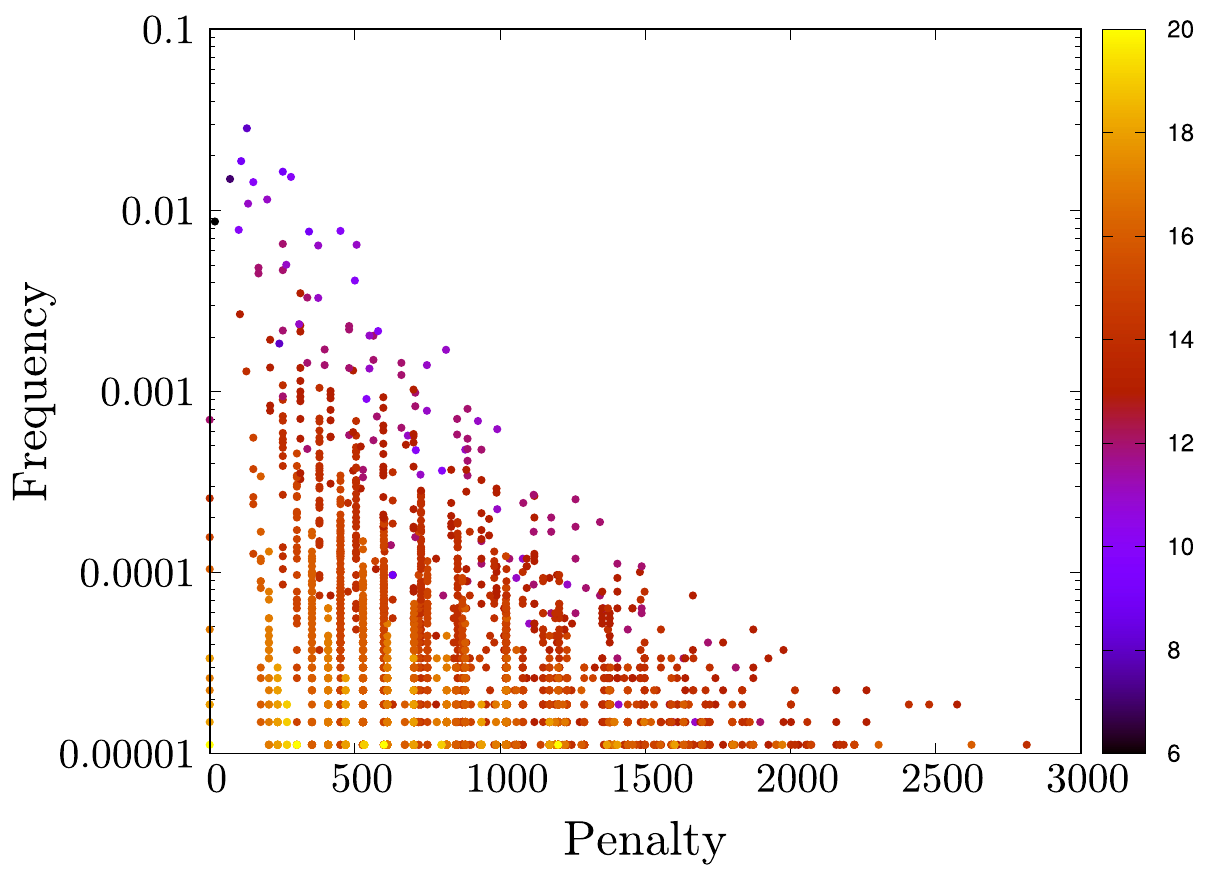}
\caption{Flag$^*$ grains with frequencies of at least $0.00001$ in the Evolved microstructure as a function of penalty value. 
Color indiates the number of faces.}
\label{fig:penalty}
\end{figure}

Instead, we propose a penalty function that effectively bounds the frequency of a given flag$^*$ simple $3$-polytope
from above. Intuitively, the penalty should increase with the number of faces, with the severity of the constriction by
$4$-belts, and with the number of $4$-belts. One function with these properties follows.
\begin{deff}
The \emph{penalty value} of a flag$^*$ grain with $n$ faces is
\begin{equation}
{\cal P} = n \displaystyle\sum_k x_k k (n-k-4),
\end{equation}
where $x_k$ is the number of $4$-belts of type $k\!:\!(n-k-4)$.
\end{deff}
The penalty function sums over all $4$-belt types of a flag$^*$ grain, with the sum weighted by the numbers of faces.
Figure~\ref{fig:penalty} shows flag$^*$ grains with frequencies of at least $0.00001$ in the Evolved microstructure as a function of
their penalty value. Flag$^*$ simple $3$-poly\-topes with split type $0\!:\!m$ have penalty value $0$,
but are not frequent. The largest value for $p$ that we observed in the Evolved
microstructure is 21,514 for a grain with $31$ vertices. The occurrence of such a grain
is most likely due to a \emph{long tail effect} of the statistics --- while any individual grain
with many vertices and high penalty value will be very infrequent, there are many such possible flag$^*$ grains
from which some individual grains are sampled.
As intended though, the penalty function does appear to give a reasonable upper bound on the frequency
of a given flag$^*$ simple $3$-polytope, at least for this particular microstructure.
A more accurate and general penalty function would likely use additional information concerning belts of other lengths,
the number of vertex-truncations the polytope is away from a fundamental polytope, and potentially the order of the symmetry group.

With regard to the symmetry group, grains in the Poisson--Voronoi, Evolved, and intermediate microstructures generally have 
no \emph{geometric} (reflectional or rotational) symmetries at all. For the Pois\-son--Voronoi case this immediately
follows from a general position argument for the (random) normal vectors of the faces of the individual  simple
$3$-polytopes, forcing all edges of the polytopes to have different lengths and therefore destroying geometric
symmetries. For the Evolved and intermediate cases, the edges and the faces of grains (in simulations and experiments) are curved,
presenting a further obstacle to geometric symmetries.

That said, it has been observed ``that the most frequent grain \emph{topologies} [emphasis added] in grain growth microstructures 
are substantially more symmetric than the corresponding ones for Poisson--Voronoi microstructures'' \cite{2012lazar}. 
This follows from grains in the Evolved microstructure having on average fewer
faces and fewer triangular faces than grains in the Poisson--Voronoi microstructure. By Equation~\ref{eq:valence},
$\frac{1}{n}(p_3+p_4+\dots)\,=\,6-\frac{12}{n}$ for any simple $3$-polytope with $n$ faces and $p$-vector $(p_3,p_4,\dots)$.
If $n$ is small and $p_3=0$, then $p_k\neq 0$ for large $k$  is not possible. Hence, the most frequent grains
will naturally have only $4$-gons, $5$-gons, and $6$-gons, giving them a higher chance of being combinatorially symmetric;
see~\cite{Lutz2008b} for a complete list of the examples with $n\leq 10$ along with their symmetry groups.

\begin{figure}
\includegraphics[width=0.3\columnwidth]{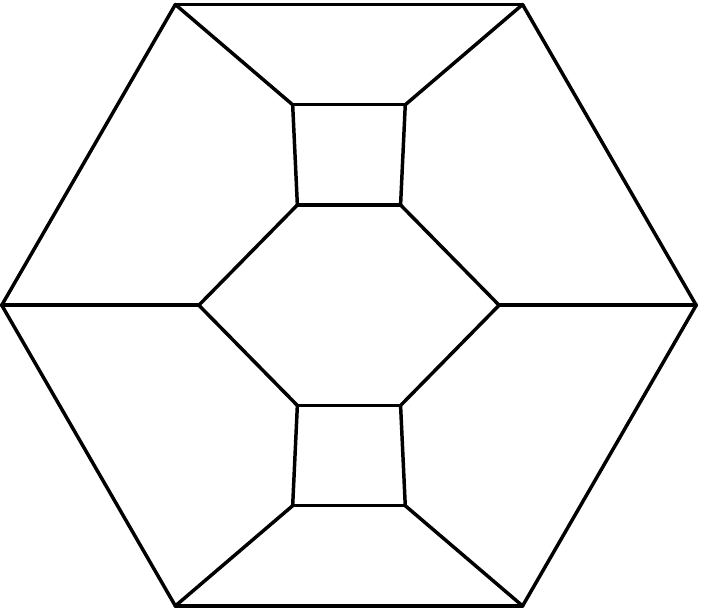}\hspace{2mm}
\includegraphics[width=0.3\columnwidth]{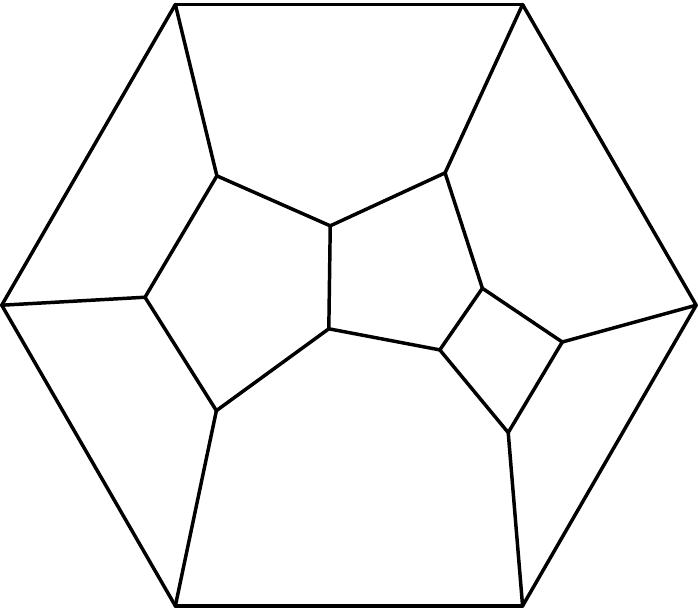}\hspace{2mm}
\includegraphics[width=0.3\columnwidth]{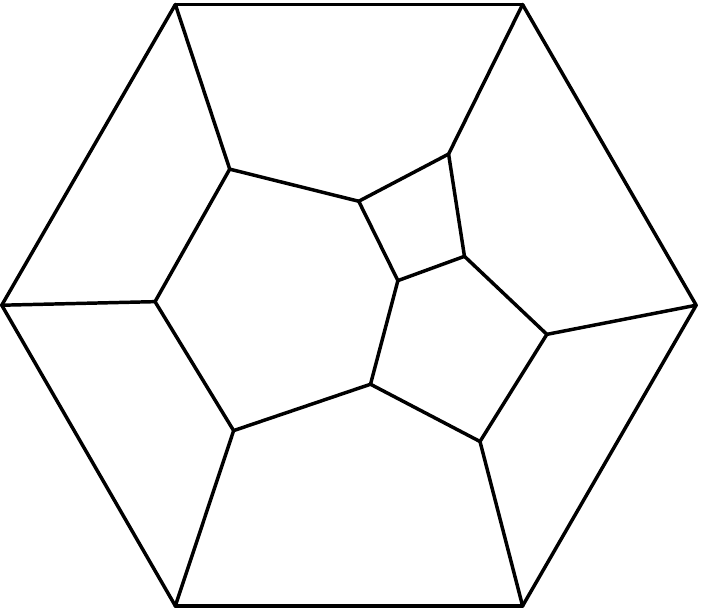}\\
A\hspace{26mm}B\hspace{26mm}C
\caption{Three grains A, B, and C with the same $p$-vector $(0,4,4,2)$.}
\label{fig:three_grains}
\end{figure}

\begin{figure}
\includegraphics[width=0.95\columnwidth]{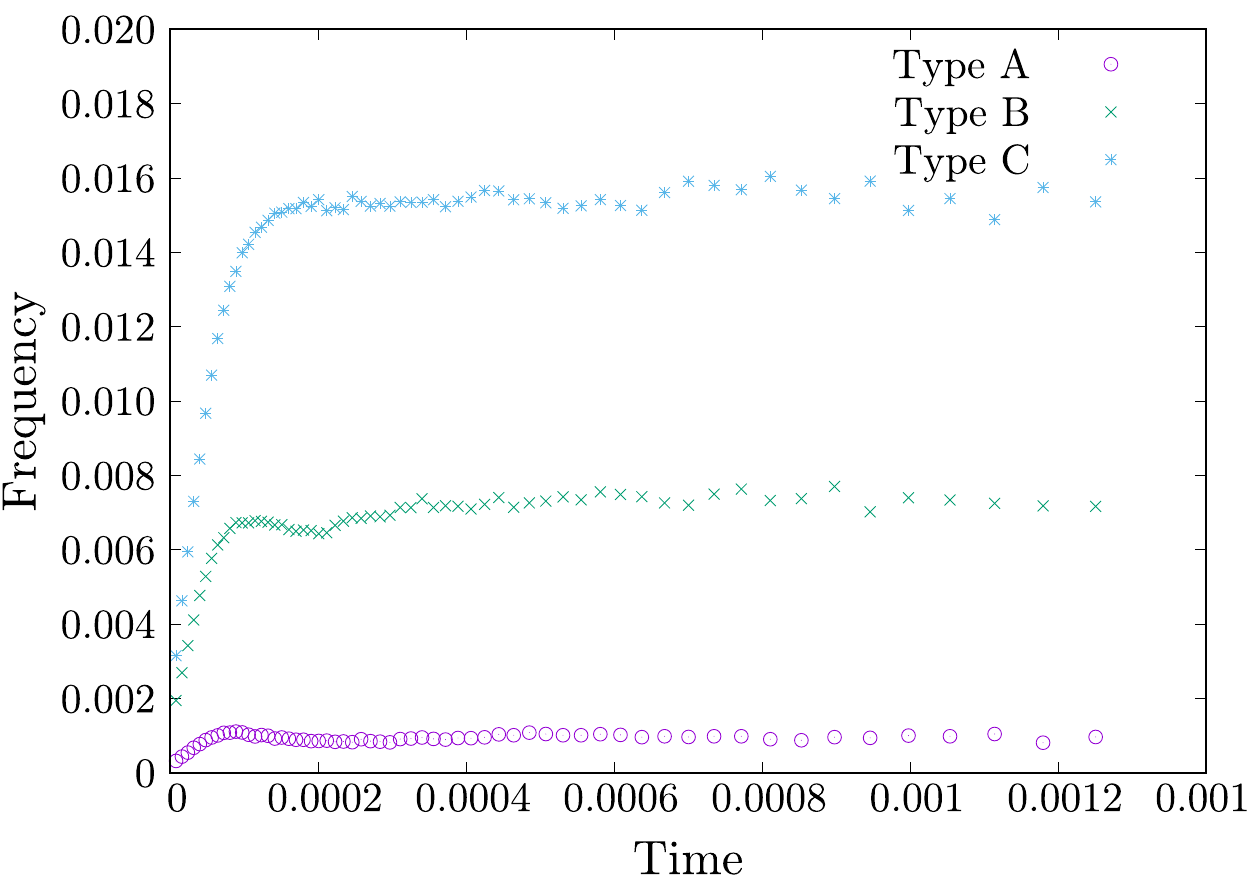}
\caption{Fractions of the three grains with $p$-vector $(0,4,4,2)$ over time.}
\label{fig:three_grains_freq}
\end{figure}

As one set of examples, consider the three simple 3-polytopes A, B, and C in Figure~\ref{fig:three_grains},
each with 10 faces and $p$-vector (0,4,4,2). Figure~\ref{fig:three_grains_freq} shows that at all times during
the evolution, type A is the least frequent and type C is roughly twice as frequent as type B; in fact, type~C 
is the most frequent type of all grains with 10 faces in the Evolved microstructure. Type A has a symmetry
group of order 8, while types B and C have symmetry groups of order 2. That is, the symmetry group seems to be related
to frequency of the type, but is not sufficient to distinguish types B and C.

These differences can instead be explained by the measures of combinatorial roundness considered above. 
Type~C has no belts of type $3\!:\!4\!:\!3$, is the most round of the three examples,
and is the most frequent one. Type~A has two (orthogonal) belts of type $3\!:\!4\!:\!3$, while B has only one. This makes
A the most constricted and the least frequent of the three, despite being the most symmetric. Note that this
situation is more succinctly described in terms of the penalty values of the three types. The penalty values of types A, B,
and C are 380, 290, and 280 respectively. The larger the penalty value, the lower the frequency of the type.

\begin{figure}
\includegraphics[width=0.3\columnwidth]{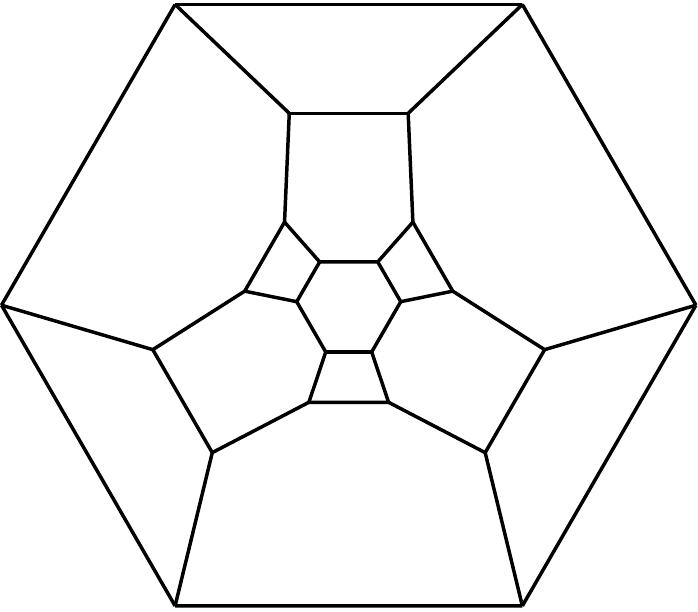}\hspace{2mm}
\includegraphics[width=0.3\columnwidth]{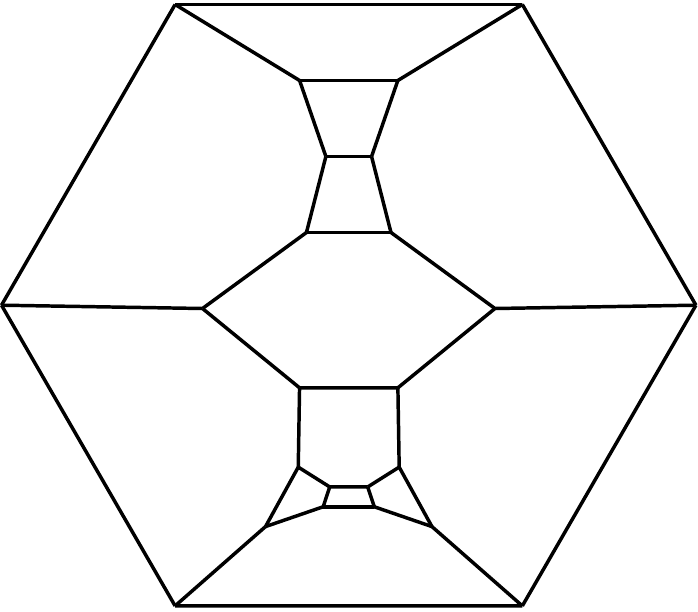}\hspace{2mm}
\includegraphics[width=0.3\columnwidth]{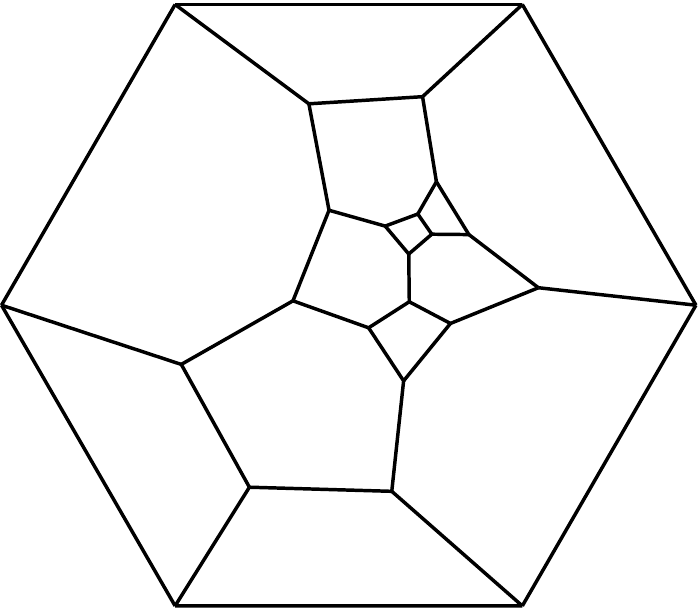}\\
K\hspace{26mm}L\hspace{26mm}M
\caption{Three grains K, L, and M with the same $p$-vector $(0,6,0,8)$.}
\label{fig:three_grains0608}
\end{figure}

Observe that predicting the frequency of a combinatorial type instead of an upper bound on the frequency would require
information about the compatibility of that type with the surrounding microstructure. Consider the three
simple $3$-polytopes K, L, and M illustrated in Figure~\ref{fig:three_grains0608}, each with $14$ faces and $p$-vector $(0,6,0,8)$. 
Note that K is the Kelvin cell, which can be used to tile space.
None of these types appears often at any point during grain growth. This can be understood as a consequence of the
absence of $5$-gons, which are the most frequent face type. Since every face is shared by two grains, a simple
$3$-polytope without $5$-gons is unlikely to be compatible with the faces presented by the surrounding microstructure;
indeed, the most frequent grain with 14 faces in the Evolved microstructure has $p$-vector (0,1,10,3).
Similarly, the disorder of the microstructure makes combinatorial types with only $5$-gons (dodecahedra) rare, despite
these types being highly symmetric and combinatorially round.
The preference of combinatorially round grains that approximate the average $p$-vector
is another property that could be encoded in a generalized penalty function.

\section{Degenerate and Extremal Grains}
\label{sec:extremal}

\begin{figure}[b]
\center
\subfigure[]{%
	\label{subfig:trihedron}{%
		\includegraphics[height=0.3\columnwidth]{%
			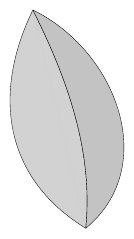}}} \hspace{15mm}
\subfigure[]{%
	\label{subfig:pillow}{%
		\includegraphics[height=0.27\columnwidth]{%
			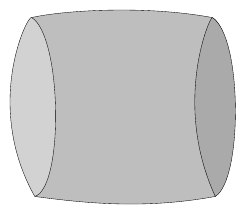}}}
\caption{Examples of degenerate grains with $2$-gon faces. (a) The trihedron, and (b) the pillow.}
\label{fig:trihedron_pillow}
\end{figure}

\begin{figure}
\includegraphics[width=0.95\columnwidth]{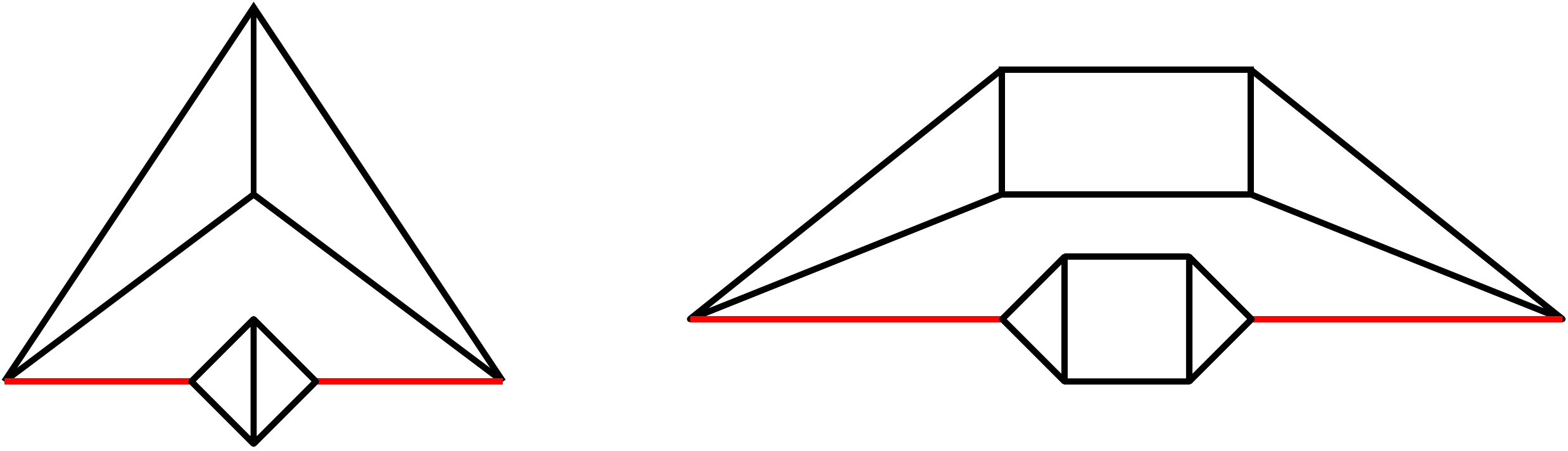}
\caption{Examples of degenerate grains with two faces sharing multiple edges.}
\label{fig:multiple_edges}
\end{figure}

Not all of the grains in grain growth microstructures are polytopal, i.e., a small fraction of them cannot be
realized as the boundaries of three-dimensional polytopes with straight edges and planar faces. Such grains
are said to be \emph{degenerate}. Degenerate grains occur when one of the faces is a $2$-gon, as for the grains in
Figure~\ref{fig:trihedron_pillow}, or when two faces share more than one edge, as for the grains in Figure~\ref{fig:multiple_edges}. 
Note that, with the exception of the trihedron in Figure~\ref{subfig:trihedron}, the presence of a $2$-gon requires
the neighboring faces to share more than one edge. The alternative, that faces share more than one edge in the
absence of a $2$-gon (like the grains in Figure~\ref{fig:multiple_edges}), is extremely rare.

\begin{figure}
\includegraphics[width=0.95\columnwidth]{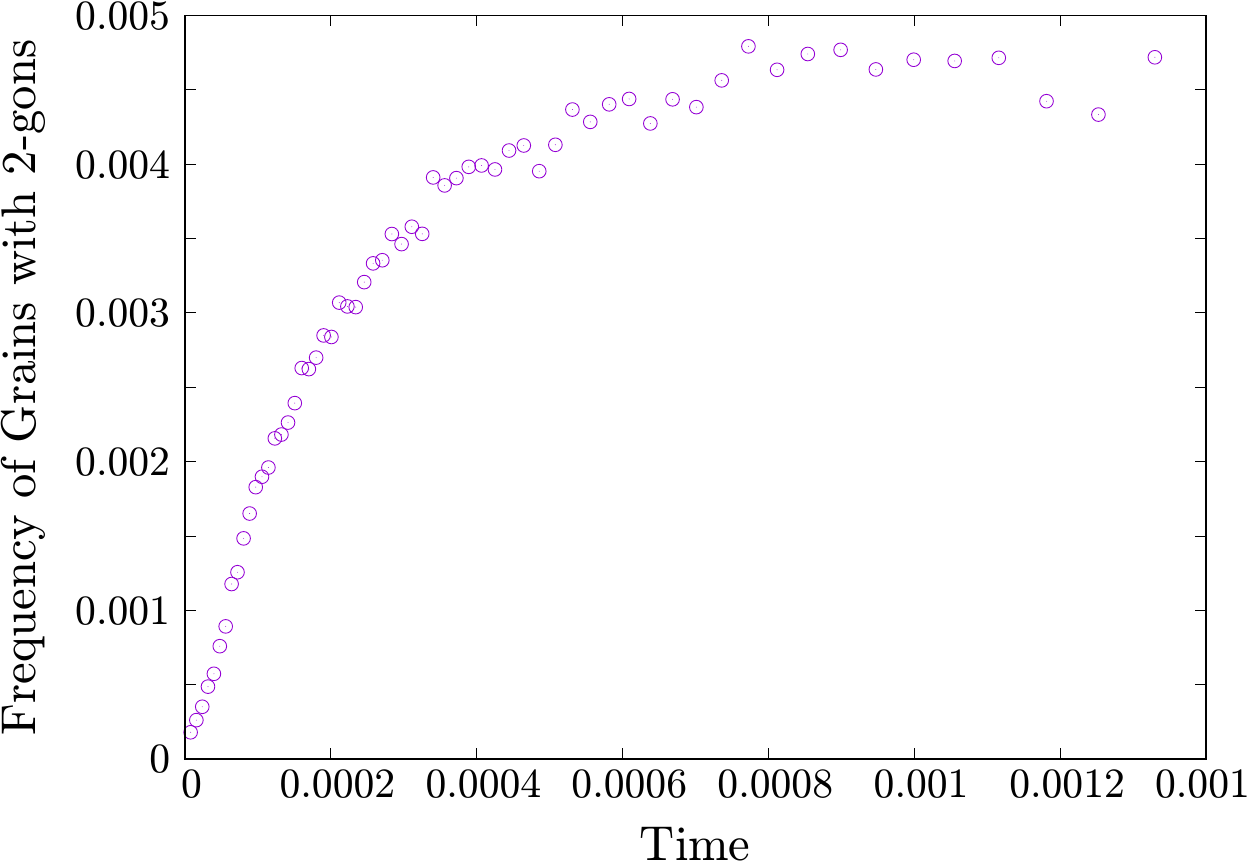}
\caption{Fraction of grains with a $2$-gon face over time.}
\label{fig:digon_grains}
\end{figure}

While degenerate grains occur as a natural consequence of grain growth, they are not very frequent. Figure~\ref{fig:digon_grains}
shows that degenerate grains with $2$-gon faces constitute less than $0.5\%$ of all grains throughout our simulation. The
initial increase in the number of these grains is a consequence of the possibility of generating a $2$-gon when a $3$-gon
participates in a topological transition. The fraction of such grains saturates because the action of capillary
pressure causes $2$-gons to quickly disappear, with the result depending on the grain type.

Most of the degenerate grains in the Evolved microstructure have exactly one $2$-gon (97\%), 
and the disappearance of the $2$-gon makes these grains polytopal. The 0.6\% 
of the degenerate grains that are the trihedron shown in Figure~\ref{fig:trihedron_pillow}(a)
disappear entirely following the disappearance of one of the $2$-gons since the result has a forbidden
topology. The 1.6\% of the degenerate grains with two $2$-gons 
and the remaining 0.6\% (except for the trihedron) with three or more $2$-gons 
generally undergo further transitions. 

\begin{figure}
\includegraphics[height=28mm]{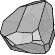}
\includegraphics[height=28mm]{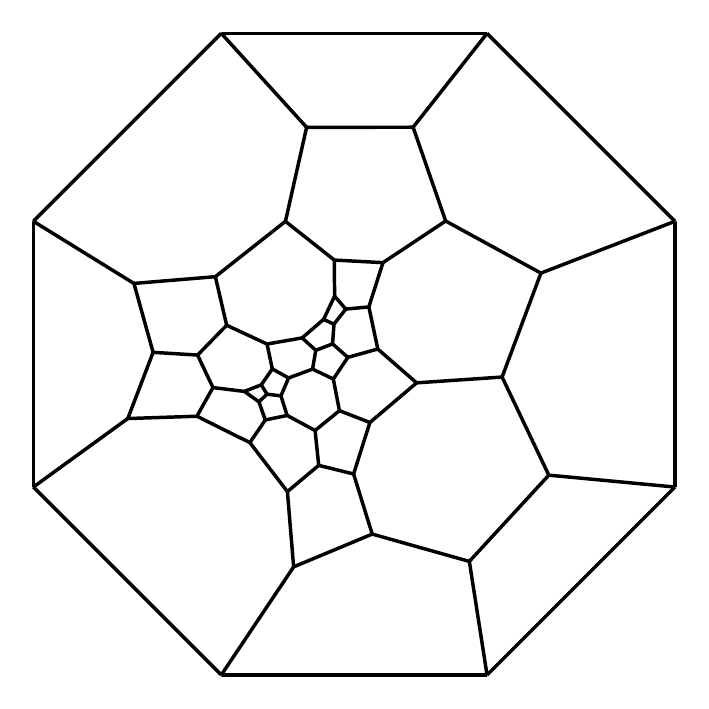}
\caption{Extremal grain with $31$ faces, type $1\!:\!4\!:\!26$, and $p$-vector $(0,4,13,7,5,2)$ from the Poisson--Voronoi microstructure.}
\label{type0413752}
\end{figure}

\begin{figure}
\includegraphics[width=0.3\columnwidth]{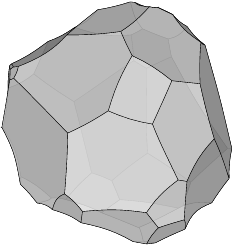}\quad\quad
\includegraphics[width=0.3\columnwidth]{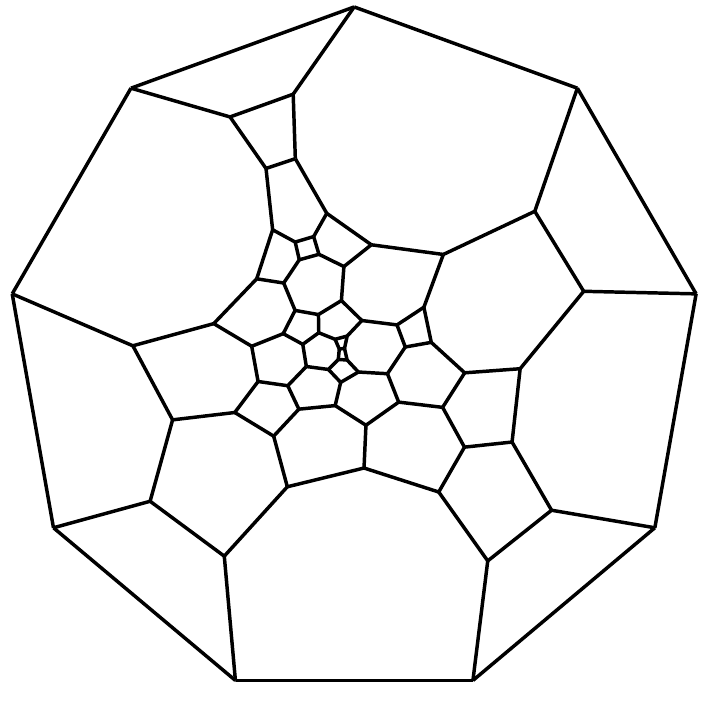}\\
\includegraphics[width=0.3\columnwidth]{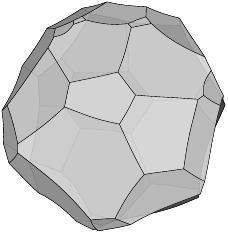}\quad\quad
\includegraphics[width=0.3\columnwidth]{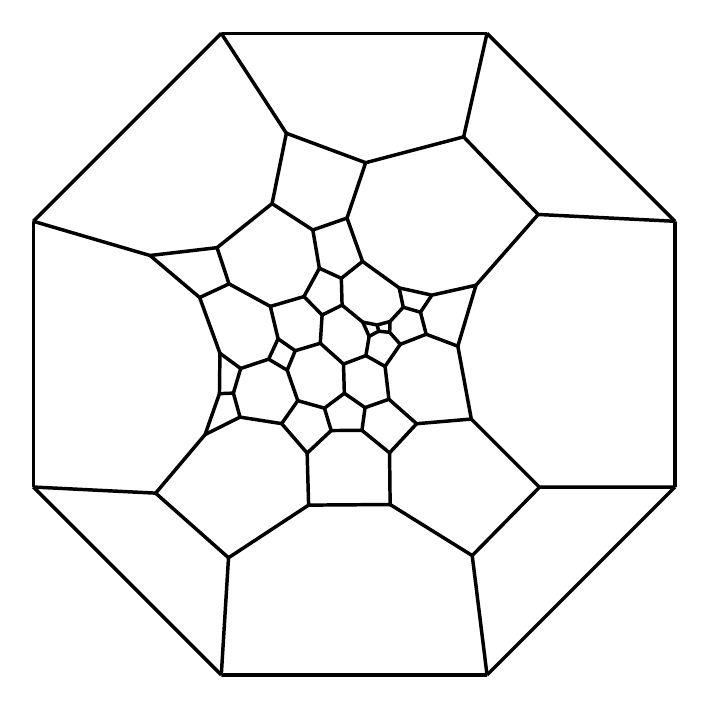}
\caption{Extremal grains with $38$ faces, type $2\!:\!4\!:\!32$, and $p$-vectors $(0,9,8, 12, 5,3,1)$ and $(0,10,8,8,8,4)$ 
from the Evolved microstructure.}
\label{type38faces2_32}
\end{figure}

The distributions of the number of faces in Figure~\ref{fig:aggregate} imply the existence of flag$^*$ grains in the
Poisson--Voronoi and Evolved microstructures with some maximum number of faces. For the Poisson--Voronoi microstructure,
the unique largest flag$^*$ grain with $31$ faces is shown in Figure~\ref{type0413752} and has split type
$1\!:\!4\!:\!26$. For the Evolved microstructure, the two largest grains with $38$ faces are shown in Figure~\ref{type38faces2_32} 
and have split type $2\!:\!4\!:\!32$. These examples show that, as one might expect, grains with the maximum number of faces 
generally have little combinatorial constriction.

\begin{figure}
\includegraphics[height=30mm]{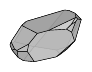}
\includegraphics[height=30mm]{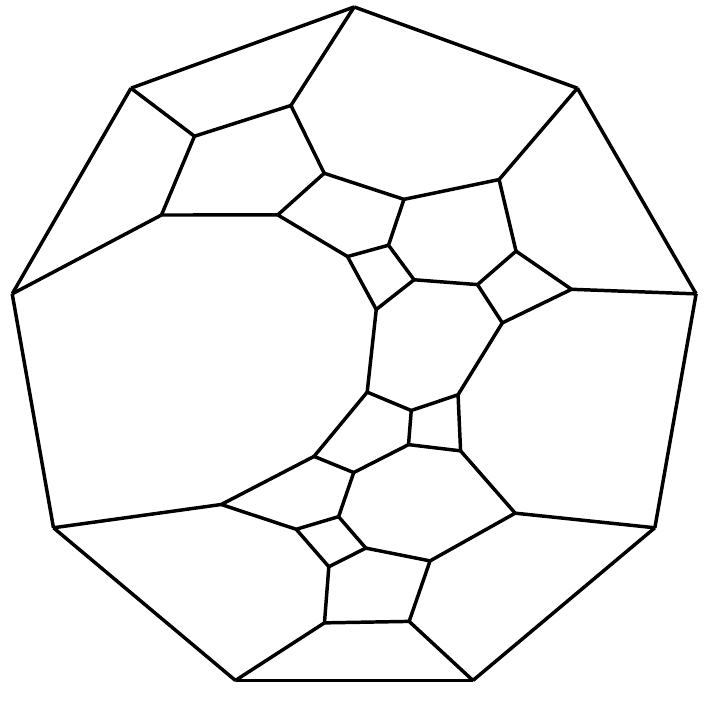}
\caption{Extremal grain with $22$ faces, type $9\!:\!4\!:\!9$, and $p$-vector $(0,7,7,3,3,0,2)$ from the Poisson--Voronoi microstructure.}
\label{type0007733020}
\end{figure}

\begin{figure}
\includegraphics[width=25mm, angle=90]{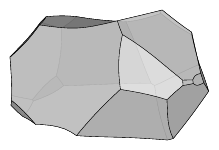}
\includegraphics[height=25mm]{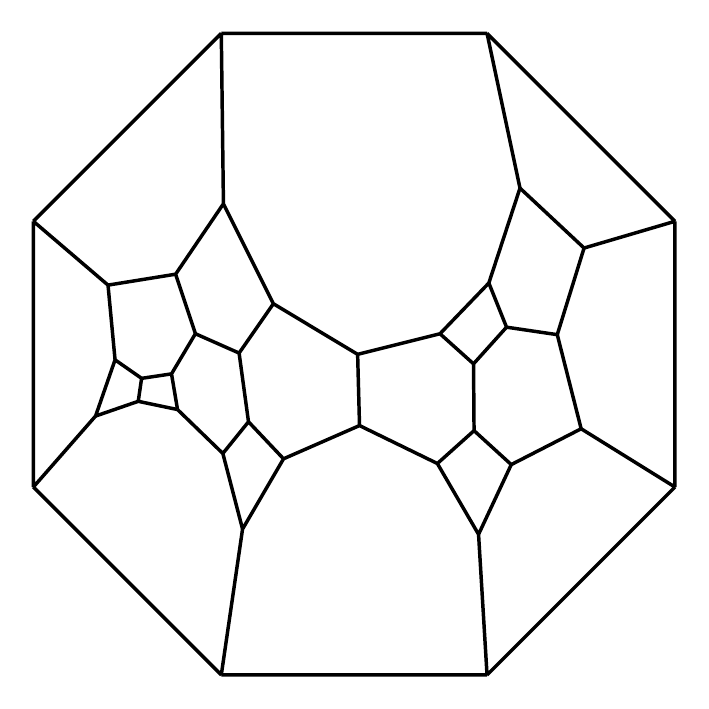}
\caption{Extremal grain with $21$ faces, type $8\!:\!4\!:\!9$, and $p$-vector $(0,6,6,5,2,2)$ from the Evolved microstructure.}
\label{type66522}
\end{figure}

The grains with the most combinatorial constriction (in the sense of the penalty function) generally have severe
geometrical constriction as well. The example from the Poisson--Voronoi microstructure in Figure~\ref{type0007733020}
is notably extended along the axis perpendicular to the $4$-belt, and the example from the Evolved microstructure in
Figure~\ref{type66522} visibly curves in along the $4$-belt.

\section{Conclusion}

The grain growth microstructure is an unusual physical system in the sense that the governing equation is relatively
simple, and yet reasonable explanations for quantities as basic as the average number of faces continue to elude
researchers. Our main purpose here is not to resolve this situation, but rather to use the grain growth microstructure
as motivation to develop novel topological and combinatorial concepts.

Grain growth dynamics strongly encourages the collapse of triangles once they are formed. This suggests classifying
simple $3$-polytopes as \emph{fundamental} polytopes without triangular faces (except for the tetrahedron), or
\emph{vertex-truncated} polytopes with triangular faces. Vertex-truncated polytopes are less stable with respect to
topological transitions, and every vertex-truncated polytope can be reduced to a unique fundamental polytope by
reversing vertex truncations.

The fundamental simple $3$-polytopes therefore provide a useful point to begin investigating the combinatorial
types of polytopes in the grain growth microstructure. Not all fundamental polytopes are equally frequent, though.
Those that have a non-trivial belt of three faces passing around the polytope are said to be \emph{severely constricted},
and appear only infrequently in the grain growth microstructure. The remaining fundamental polytopes 
(except for the tetrahedron) are said to be \emph{flag$^*$}.

That the examples of severely constricted polytopes in our simulated microstructure are all severely geometrically
constricted as well suggests that grain growth dynamics favors grains that are round --- both geometrically and
combinatorially. More specifically, a simple $3$-polytope is \emph{combinatorially $k$-round} if it is at most $k$
truncations away from a non-severely constricted fundamental polytope. Our simulations provide extensive evidence 
that $k$-round grains for small values of $k$ appear vastly more often in space-filling microstructures than could be
expected from the distribution of the combinatorial types of all simple $3$-polytopes. Moreover, the combinatorial
roundness is positively correlated with the geometric roundness (as expressed by the isoperimetric quotient)
for all three
simulated microstructures considered here. That is, the definition of combinatorial roundness proposed 
in Section~\ref{subsec:roundness} seems to be natural in the sense of capturing a meaningful 
and non-trivial aspect of the physical system.

Although the combinatorial roundness of a simple $3$-polytope is related to the frequency of that polytope in a
generic microstructure, this relationship is not precise. In fact, the examples of the Kelvin foam and the Weaire--Phelan
foam show that the frequency of a polytope cannot, in general, be a function of combinatorial information alone.
Instead, a penalty function is proposed, where the penalty increases with the number of faces and with the constriction
as quantified by belts of faces around the polytope. This is intended to assign a value to the difficulty of
accommodating the polytope in a space-filling microstructure, and (per Figure~\ref{fig:penalty}) seems to function
as an upper bound for the frequency of the polytope.

Having developed this mathematical machinery, a particular microstructure derived from the Poisson--Voronoi initial
condition is identified as containing an unusually large proportion of round grains. This Round microstructure has an
average of $14.036$ faces per grain (suggestively close to the $14$ faces per grain of the Kelvin foam), and is
conjectured to be more resistant to topological change than the steady-state grain growth microstructure. The
investigation of the mechanical properties of the Round microstructure is proposed as a subject for further research.

\begin{acknowledgments}

We are grateful to Junichi Nakagawa, Boris Springborn and John M.~Sullivan for helpful discussions
and suggestions. We also thank the anonymous referees for remarks that helped to improve the presentation of the paper.
The first author was partially supported by the DFG Research Group ``Polyhedral Surfaces'', 
by the DFG Coll.\ Research Center TRR 109 ``Discretization in Geometry and Dynamics'',
by \textsc{VILLUM FONDEN} through the Experimental Mathematics Network 
and by the Danish National Research Foundation (DNRF) through the Centre for Symmetry and Deformation.
The third author acknowledges support of the NSF Division of Materials Research through Award 1507013.

\end{acknowledgments}

\appendix

\section{Combinatorial Constraints}
\label{sec:constraints}

The \emph{face vector} of a $3$-dimensional polytope is the triple $f = ( f_0, f_1, f_2 )$ of the numbers of vertices, edges, 
and faces of the polytope, respectively. A simple $3$-polytope $P$ with $n$ faces, $e$ edges, and $t$ vertices 
has the face vector $f_P = ( t, e, n )$, while the dual simplicial $3$-polytope $P^*$ has the face vector $f_{P^*} = ( n, e, t )$. 
The Euler characteristic of $P^*$ requires that $n - e + t = 2$, and equating the number of edge-triangle incidences 
and the number of triangle-edge incidences of $P^*$ gives $2 e = 3 t$. 
Taken together, these imply that $f_{P^*} = ( n, 3n-6, 2n-4 )$, and by duality, that $f_{P} = ( 2n-4, 3n-6, n )$.

For any vertex $v_i$ of $P^*$, the \emph{degree} ${\rm deg}(v_i)$ of the vertex is the number of adjoining edges. Equating the number 
of vertex-edge incidences and the number of edge-vertex incidences of $P^*$ gives $\sum_{i = 1, \dots, n}{\rm deg}(v_i) = 2e = 6 n - 12$. 
The average vertex-degree of $P^*$ is then
\begin{equation}
\overline{{\rm deg}} = \frac{1}{n} \sum_{i = 1, \dots, n} {\rm deg}(v_i) = 6 - \frac{12}{n} < 6.
\label{eq:valence}
\end{equation}
One of the implications of Equation \ref{eq:valence} is that every simplicial $3$-polytope has at least one vertex $v$ with degree ${\rm deg}(v) < 6$. 
A second implication comes from the observation that since the degree of every vertex of a flag simplicial $3$-polytope is at least $4$, 
the average vertex-degree must be at least $4$ as well. Equation \ref{eq:valence} then requires that $n \geq 6$ for all flag simplicial $3$-polytopes. 
The octahedron is the unique example with $n = 6$ vertices.

The \emph{$p$-vector} of a $3$-dimensional polytope is the tuple $p=(p_3,p_4,p_5,\dots,p_{\rm max})$ of the numbers 
of $3$-gons (triangles), $4$-gons, $5$-gons, etc.\ of the polytope, with $p_{\rm max}$ being the number of polygons
with the maximum number of sides. Eberhard's theorem \cite{Eberhard1891} states that for every sequence 
$\hat{p} = ( p_3, p_4, p_5; p_7, p_8, \dots, p_{\rm max} )$ of nonnegative integers satisfying 
\begin{equation}
3 p_3 + 2 p_4 + p_5 = 12 + \sum_{7 \leq k \leq{\rm max}} ( k - 6 ) p_k
\label{eq:eberhard}
\end{equation}
and some nonnegative integer $p_6\geq 0$, there is a simple $3$-polytope with $p$-vector $p = ( p_3, p_4, p_5, p_6, p_7, \dots, p_{\rm max} )$. 
The allowed values of $p_6$ are known in several cases. 
For example, the partial $p$-vector $(0, 0, 12)$ allows simple $3$-polytopes for all $p_6 \neq 1$ \cite[p.~271]{Gruenbaum1967}; 
cf.\ also \cite{izmestiev2013there}. However, there is no $13$-faced simple $3$-polytope with $p=(0,0,12,1)$.

Three-dimensional grain growth simulations generally use periodic boundary conditions, and represent the microstructure 
as a simple decomposition of the $3$-dimensional torus $T^3$ into simple $3$-polytopes. Let the dual triangulation $T$ 
of any simple decomposition of $T^3$ have face vector $f_T = ( f_0, f_1, f_2, f_3 )$. The Euler characteristic of $T$ requires 
that $f_0 - f_1 + f_2 - f_3 = 0$, and equating the number of triangle-tetrahedra incidences and the number of tetrahedra-triangle 
incidences gives $2 f_2 = 4 f_3$. These imply that $f_T = ( f_0, f_1, 2 f_1 - 2 f_0, f_1 - f_0 )$, or that the face vector is completely 
determined by $f_0$ and $f_1$.

The number of edges of any triangulation of a closed $3$-manifold $M$ is bounded above by the number of distinct pairs 
of vertices of the triangulation, or $f_1 \leq \binom{f_0}{2}$. 
A triangulation of $M$ is called \emph{neighborly} if $f_1$ achieves this bound. According to Walkup~\cite{Walkup1970}, 
there is a largest integer $\gamma (M)$ such that for every triangulation of $M$ with $f_0$ vertices and $f_1$ edges 
the following inequality is satisfied:
\begin{equation}
\label{eq:Walkup1}
f_1\geq 4f_0 + \gamma(M).
\end{equation}
Moreover, for any $3$-manifold $M$ there is a smallest integer $\gamma^*(M) \geq \gamma (M)$ such that there is a triangulation of $M$ 
for every pair of integers $( f_0, f_1 )$ provided that $f_0 \geq 0$ and  that the following inequalities are satisfied:
\begin{equation}
\label{eq:Walkup2}
\binom{f_0}{2} \geq f_1 \geq 4 f_0 + \gamma^*(M).
\end{equation}
Specifically for the $3$-torus $T^3$, the best known upper bounds are $\gamma^*(T^3) \leq 45$ and $\gamma (T^3) \leq 44$, 
and the best known lower bounds are $\gamma^*(M) \geq \gamma (M) \geq 11$~\cite{LutzSulankeSwartz2009}. 
Equation \ref{eq:Walkup2} then implies that there are neighborly triangulations of $T^3$ for any $f_0\geq 15$.

Analogously to a single simplicial $3$-polytope, the average vertex-degree of a triangulation of the $3$-dimensional torus $T^3$ is
\begin{equation}
\label{eq:average_3d}
\overline{{\rm deg}} = \frac{1}{f_0}\sum_{i = 1, \dots, n}{\rm deg}(v_i) = \frac{2 f_1}{f_0}.
\end{equation}
This implies that when $f_1 = 4 f_0 + \gamma(T^3)$ as in Equation~\ref{eq:Walkup1}, the average degree is $8 + 2 \gamma(T^3) / f_0$, 
and that the average degree of a neighborly triangulation is $f_0 - 1$. Since there is no upper bound on $f_0$, the average degree 
satisfies \mbox{$8 < \overline{{\rm deg}} < \infty$}. That is, there are simple decompositions of $T^3$ into grains with the combinatorial types 
of simple $3$-polytopes where the average number of faces per grain is as few as $8$ or as many as the number of $3$-polytopes 
in the decomposition minus one.

\begin{figure}[t]
\includegraphics[height=26mm]{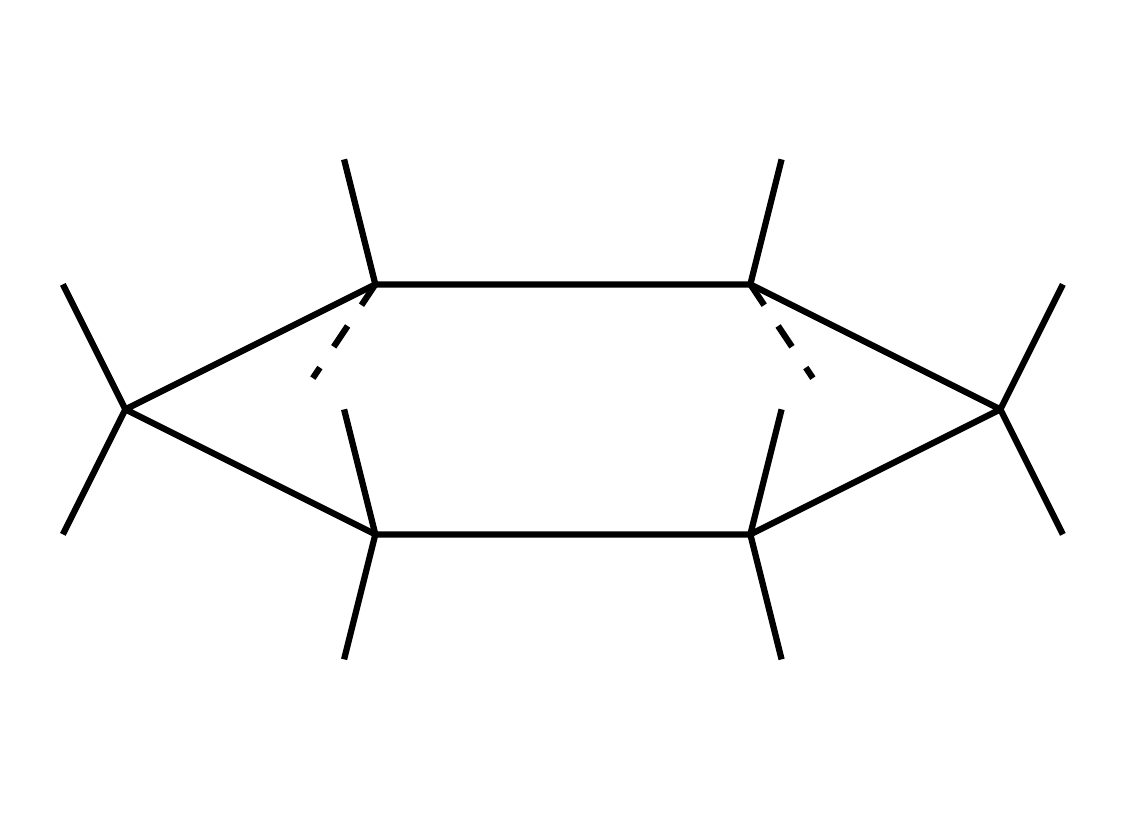}
\hspace{6mm}
\includegraphics[height=26mm]{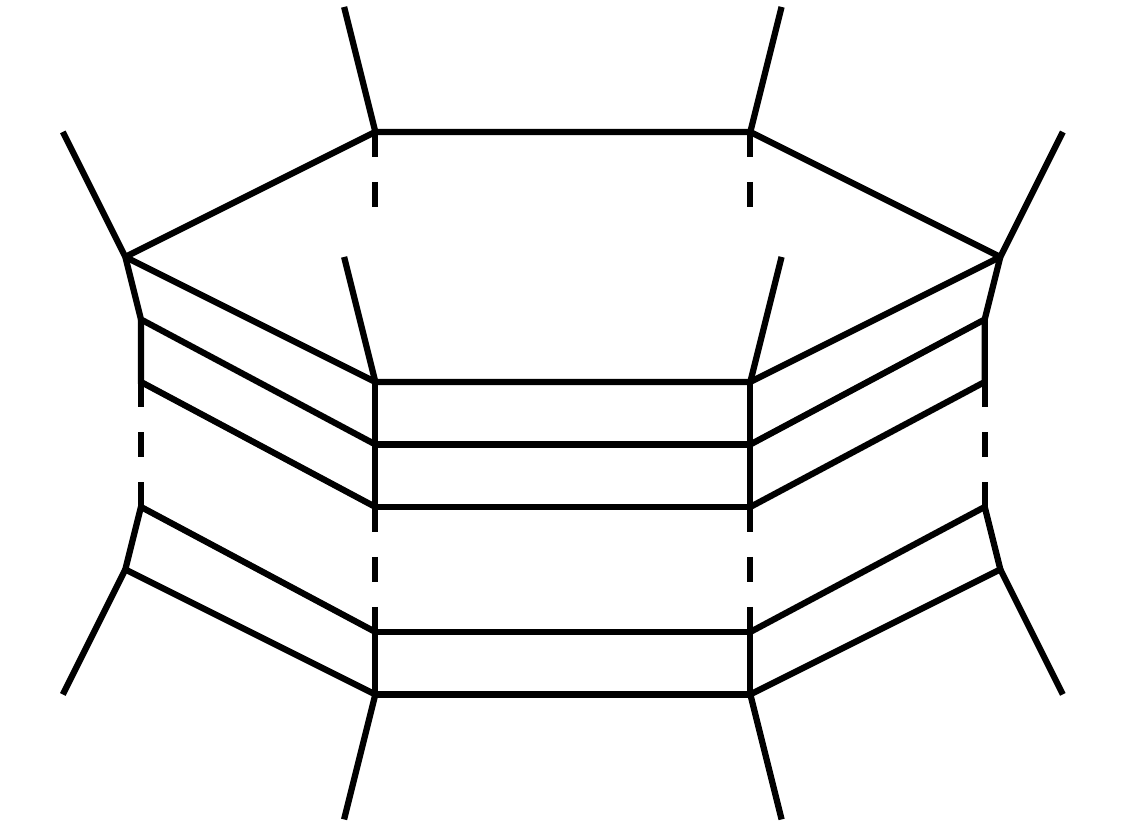}
\caption{A $6$-gon in a simple cellular microstructure (left) and its expansion into prisms over the $6$-gon (right).}
\label{fig:prism_over_6gon}
\end{figure}

The arguments above do not require all grains to be proper convex $3$-polytopes (not just combinatorial types)
with planar faces and straight edges, though. Let this condition be called \emph{geometric}. Poisson--Voronoi decompositions 
have an average of $48 \pi^2 / 35 + 2 \simeq 15.535$ faces per grain \cite{1953meijering}, and are clearly geometric. 
The lower bound on the average number of faces can be achieved by replacing some of the vertices of any initial
geometric decomposition with (small) tetrahedra. Repeating this procedure produces a geometric decomposition with an average number
of faces per grain arbitrarily close to~$8$. The upper bound can be achieved by observing that a $q$-gon in a geometric
decomposition can always be expanded into a prism over the $q$-gon. Repeating this procedure produces a geo\-metric simple
decomposition with an average number of faces per grain arbitrarily close to~$2q+2$;
see Figure~\ref{fig:prism_over_6gon} for this construction over the $6$-gon.
Since there is some geometric decomposition
that contains a $q$-gon for any choice of~$q$, the average number of faces per grain of a geometric decomposition can be arbitrary
large (we are grateful to John~M.\ Sullivan for pointing this out). Note that this construction is dual to applying a sequence
of stellar subdivisions to an edge of degree $q$ as described by Luo and Stong~\cite{LuoStong1993}.
\begin{lem}
\label{lem:space_average}
There are geometric simple decompositions with average number of faces per grain ranging over
\begin{equation}
\label{eq:average_range}
8 < \frac{2 f_1}{f_0} < \infty,
\end{equation}
where $f_0$ and $f_1$ are the numbers of grains and of faces (corresponding to $f_0$ vertices and $f_1$ edges of the dual triangulations), respectively.
\end{lem}
The lemma notwithstanding, examples of generic random decompositions with an average number of faces per grain substantially higher
than that of the Poisson--Voronoi case ($15.535$) are not known.

Several analytical models predict the average number of faces per grain for particular simple decompositions.
Coxeter~\cite{1958coxeter} gives a value of $13.564$, whereas Kusner \cite{1982rivier,1992kusner,kusner1996comparing} 
proved a lower bound of $13.397$ on the average number of faces for a cell in an equal volume equal pressure foam.
It would be of strong interest to provide further insight into the reasons why simple material microstructures seem 
to be narrowly confined to an average of between 12 and 16 faces per grain though, the more so since topology 
and geometry alone do not seem to enforce any restrictions.

\section{Algorithmic Aspects}

The following is a brief sketch on how to algorithmically determine
combinatorial information (stacked, flag, severely constricted, split type)
for a given triangulation $T$ of the $2$-dimensional sphere $S^2$.

Let $T$ have $n$ vertices. If $T$ has at least one vertex of degree $3$,
then $T$ is either the boundary of the tetrahedron,
in which case the combinatorial type of $T$ is fully determined,
or $T$ has at least $5$ vertices.

\begin{figure}
\includegraphics[width=.5\columnwidth]{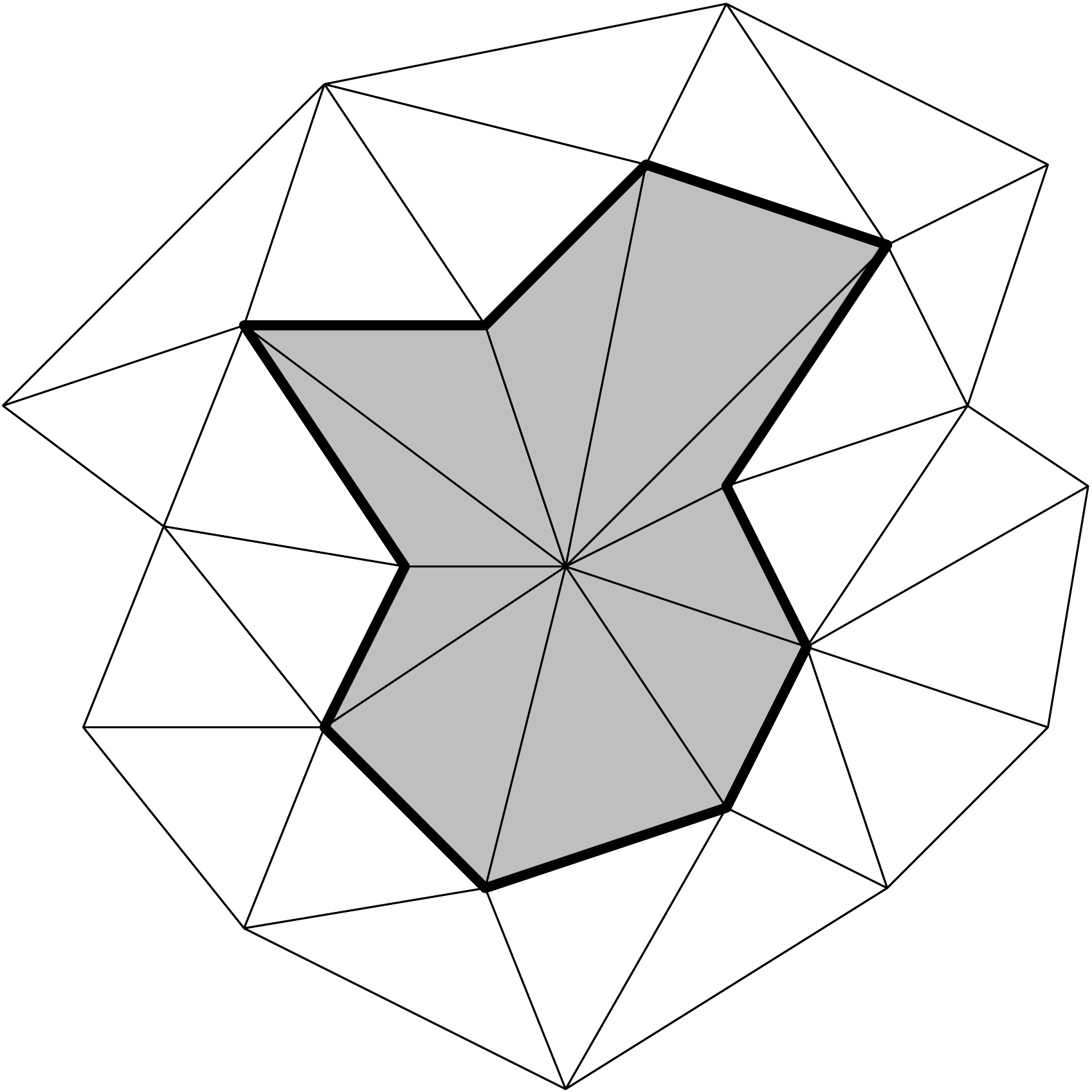}
\caption{The star (in grey) and the link (in bold) of a vertex.}
\label{fig:star_link}
\end{figure}

\begin{deff}
The \emph{star} of a vertex $v$ in a triangulation $T$ of $S^2$ is written as ${\rm star}_vT$, and is the
collection of faces of $T$ that contain the vertex $v$ and all of their subfaces (shown in Figure~\ref{fig:star_link}).
The \emph{link} of $v$ in $T$ is written as ${\rm link}_vT$, and is the collection of all faces in ${\rm star}_vT$
that do not contain~$v$. 
\end{deff}

If ${\rm star}_vT$ consists of $m$ triangles (and their subfaces), then ${\rm link}_vT$ is a cycle of length $m$.
In particular, if $v$ is a vertex of degree $3$, then ${\rm link}_vT$
is a cycle with three vertices $a$, $b$ and~$c$, and edges $ab$, $ac$, and $bc$. If the three triangles $abv$, $acv$, and $bcv$ 
of ${\rm star}_vT$ are removed and replaced by the single triangle $abc$, then this is a local modification that undoes 
a stacking operation, and yields a triangulation of $S^2$ with $n-1$ vertices.

Given the triangulation $T$, we iteratively remove vertices of degree~$3$ (in any order) until the result is either
\begin{itemize}[noitemsep]
\item the boundary of the tetrahedron
\item or a triangulation with no vertex of degree $3$ (flag or severely constricted).
\end{itemize}
If this process results in the boundary of the tetrahedron, then $T$ is a stacked triangulation that is $n-4$ stackings away 
from the boundary of the tetrahedron. If a triangulation $T'$ without vertices of degree $3$ is reached, we search 
for empty traingles to identify whether the resulting triangulation is flag or severely constricted. To do so, for every set 
of three vertices $e$, $f$ and $g$ of $T'$, we check whether $ef$, $eg$ and $fg$ are edges of $T'$. If they are, 
we check whether or not  $efg$ is a triangle of the triangulation $T'$. If it is not, then $efg$ is an empty triangle, 
and $T'$ is severely constricted. If $T'$ has no empty triangle, then it is flag.

To determine the split type of a flag triangulation $T$ with $n$ vertices, we
generate all cycles of length four in the $1$-skeleton of $T$.
To do so, we consider all subsets of four vertices $a$, $b$, $c$, and $d$ of the vertices of $T$.
For each subset, we check if one of the cycles
$a$--$b$--$c$--$d$--$a$, $a$--$b$--$d$--$c$--$a$, or $a$--$c$--$b$--$d$--$a$
is contained in the $1$-skeleton of~$T$. For every $4$-cycle in the $1$-skeleton of~$T$,
we compute the split of $T$ with respect to this $4$-cycle.

The split of the $4$-cycle $a$--$b$--$c$--$d$--$a$ in the $1$-skeleton of $T$
is found by splitting ${\rm star}_aT$ into two parts along $a$--$b$--$c$--$d$--$a$.
The two parts can be extended by a greedy search into the two components of the overall split.
Finally, the numbers of vertices on each side are found (excluding the four
vertices $a$, $b$, $c$ and~$d$ on the splitting cycle). The largest $k\leq \frac{n-4}{2}$
for which there is a split $k\!:\!4\!:\!(n-4-k)$ determines the split type of $T$.



%

\pagebreak

\begin{table*}
\caption{\label{tbl:few_vertices_I}The number of all possible triangulations of the $2$-sphere for small numbers of vertices $n$ \cite{plantri}. The degree of a vertex is the number of adjoining edges.}
\begin{ruledtabular}
\begin{tabular}{*{4}{r}}
  $n$ & all & min.\ deg. $\geq 4$ & flag \\
  \hline \noalign{\smallskip}
  4 &              1 &              0 &             0 \\
  5 &              1 &              0 &             0 \\
  6 &              2 &              1 &             1 \\
  7 &              5 &              1 &             1 \\
  8 &             14 &              2 &             2 \\
  9 &             50 &              5 &             4 \\
 10 &            233 &             12 &            10 \\
 11 &           1249 &             34 &            25 \\
 12 &           7595 &            130 &            87 \\
 13 &          49566 &            525 &           313 \\
 14 &         339722 &           2472 &          1357 \\
 15 &        2406841 &          12400 &          6244 \\
 16 &       17490241 &          65619 &         30926 \\
 17 &      129664753 &         357504 &        158428 \\
 18 &      977526957 &        1992985 &        836749 \\
 19 &     7475907149 &       11284042 &       4504607 \\
 20 &    57896349553 &       64719885 &      24649284 \\
 21 &   453382272049 &      375126827 &     136610879 \\
 22 &  3585853662949 &     2194439398 &     765598927 \\
 23 & 28615703421545 &    12941995397 &    4332047595 \\
 24 &                &    76890024027 &   24724362117 \\
 25 &                &   459873914230 &  142205424580 \\
 26 &                &  2767364341936 &  823687567019 \\
 27 &                & 16747182732792 & 4801749063379
\end{tabular}
\end{ruledtabular}
\end{table*}

\begin{table*}
\caption{\label{tbl:few_vertices_II}The number of triangulations of the $2$-sphere that belong to the classes 
defined in Section~\ref{subsec:roundness} for small numbers of vertices $n$.}
\begin{ruledtabular}
\begin{tabular}{*{11}{r}}
 $n$ & all & stacked & flag & flag$_{+1}$ & flag$_{+1+1}$ & flag$_{+2}$ & flag$_{\,\geq3}$ & sc$^*$ & sc$^*_{+1}$ & sc$^*_{\,\geq2}$ \\
\hline \noalign{\smallskip}
  4 &      1 &     1 &    0 &    0 &    0 &    0 &      0 &    0 &    0 &     0 \\
  5 &      1 &     1 &    0 &    0 &    0 &    0 &      0 &    0 &    0 &     0 \\
  6 &      2 &     1 &    1 &    0 &    0 &    0 &      0 &    0 &    0 &     0 \\
  7 &      5 &     3 &    1 &    1 &    0 &    0 &      0 &    0 &    0 &     0 \\
  8 &     14 &     7 &    2 &    1 &    3 &    1 &      0 &    0 &    0 &     0 \\
  9 &     50 &    24 &    4 &    3 &    5 &    2 &     11 &    1 &    0 &     0 \\
 10 &    233 &    93 &   10 &   13 &   19 &    7 &     86 &    2 &    3 &     0 \\
 11 &   1249 &    434 &   25 &   47 &   78 &   30 &    590 &    9 &   16 &    20 \\
 12 &   7595 &  2110 &   87 &  217 &  354 &  127 &   4301 &   43 &   99 &   257 \\
 13 &  49566 & 11002 &  313 & 1041 & 1799 &  608 &  30938 &  212 &  650 &  3003 \\
 14 & 339722 & 58713 & 1357 & 5288 & 9780 & 3016 & 224839 & 1115 & 3997 & 31617
\end{tabular}
\end{ruledtabular}
\end{table*}

\begin{table*}
\caption{\label{tbl:distribution_V_data}The number of simple $3$-polytopes in \texttt{V-268402} that belong to the various classes 
defined in Section~\ref{subsec:roundness}, where $n$ is the number of faces. The final row is the sum of entries in the column.}
\begin{ruledtabular}
\begin{tabular}{*{11}{r}}
 $n$ & all & stacked$^*$ & flag$^*$ & flag$^*_{+1}$ & flag$^*_{+1+1}$ & flag$^*_{+2}$ & flag$^*_{\,\geq3}$ & sc & sc$_{+1}$ & sc$_{\,\geq2}$ \\
\hline \noalign{\smallskip}
  4  &  0  &  0  &  0  &  0  &  0  &  0  &  0  &  0  &  0  &  0 \\
  5  &  14  &  14  &  0  &  0  &  0  &  0  &  0  &  0  &  0  &  0 \\
  6  &  94  &  71  &  23  &  0  &  0  &  0  &  0  &  0  &  0  &  0 \\
  7  &  451  &  188  &  99  &  164  &  0  &  0  &  0  &  0  &  0  &  0 \\
  8  &  1556  &  348  &  311  &  453  &  312  &  132  &  0  &  0  &  0  &  0 \\
  9  &  4063  &  385  &  812  &  1177  &  697  &  188  &  788  &  16  &  0  &  0 \\
 10  &  8208  &  389  &  1371  &  2350  &  1661  &  327  &  2027  &  29  &  54  &  0 \\
 11  &  14012  &  252  &  2219  &  3723  &  3001  &  569  &  4058  &  39  &  76  &  75 \\
 12  &  20650  &  152  &  3037  &  5408  &  4516  &  659  &  6580  &  43  &  89  &  166 \\
 13  &  26615  &  64  &  3351  &  6899  &  5983  &  759  &  9153  &  45  &  103  &  258 \\
 14  &  30987  &  30  &  3589  &  7568  &  7252  &  792  &  11263  &  45  &  117  &  331 \\
 15  &  32304  &  8  &  3338  &  7456  &  7615  &  791  &  12663  &  38  &  76  &  319 \\
 16  &  30721  &  1  &  2772  &  6636  &  7402  &  640  &  12908  &  11  &  69  &  282 \\
 17  &  27189  &  0  &  2366  &  5363  &  6297  &  560  &  12271  &  12  &  60  &  260 \\
 18  &  22195  &  0  &  1648  &  4317  &  5268  &  389  &  10345  &  14  &  37  &  177 \\
 19  &  17018  &  0  &  1112  &  3145  &  3906  &  302  &  8398  &  6  &  27  &  122 \\
 20  &  12090  &  0  &  742  &  2103  &  2680  &  157  &  6281  &  9  &  15  &  103 \\
 21  &  8139  &  0  &  456  &  1317  &  1768  &  110  &  4429  &  4  &  9  &  46 \\
 22  &  5182  &  0  &  266  &  772  &  1059  &  72  &  2970  &  3  &  7  &  33 \\
 23  &  3090  &  0  &  157  &  431  &  628  &  37  &  1821  &  1  &  3  &  12 \\
 24  &  1794  &  0  &  69  &  219  &  355  &  19  &  1125  &  0  &  0  &  7 \\
 25  &  998  &  0  &  40  &  117  &  192  &  13  &  629  &  0  &  0  &  7 \\
 26  &  542  &  0  &  10  &  57  &  92  &  7  &  373  &  0  &  0  &  3 \\
 27  &  265  &  0  &  12  &  18  &  40  &  4  &  187  &  0  &  0  &  4 \\
 28  &  120  &  0  &  8  &  10  &  19  &  2  &  80  &  0  &  1  &  0 \\
 29  &  63  &  0  &  1  &  7  &  11  &  0  &  44  &  0  &  0  &  0 \\
 30  &  28  &  0  &  0  &  3  &  5  &  0  &  20  &  0  &  0  &  0 \\
 31  &  9  &  0  &  1  &  1  &  0  &  1  &  6  &  0  &  0  &  0 \\
 32  &  2  &  0  &  0  &  1  &  0  &  0  &  1  &  0  &  0  &  0 \\
 33  &  2  &  0  &  0  &  0  &  0  &  0  &  2  &  0  &  0  &  0 \\
 34  &  1  &  0  &  0  &  0  &  1  &  0  &  0  &  0  &  0  &  0 \\
\hline
    &  268402  &  1902  &  27810  &  59715  &  60760  &  6530  &  108422  &  315  &  743  &  2205
\end{tabular}
\end{ruledtabular}
\end{table*}

\begin{table*}
\caption{\label{tbl:distribution_R_data}The number of simple $3$-polytopes in \texttt{R-268402} that belong to the various classes 
defined in Section~\ref{subsec:roundness}, where $n$ is the number of faces. The final row is the sum of entries in the column.}
\begin{ruledtabular}
\begin{tabular}{*{11}{r}}
 $n$ & all & stacked$^*$ & flag$^*$ & flag$^*_{+1}$ & flag$^*_{+1+1}$ & flag$^*_{+2}$ & flag$^*_{\,\geq3}$ & sc & sc$_{+1}$ & sc$_{\,\geq2}$ \\
\hline \noalign{\smallskip}
 4  &  124  &  124  &  0  &  0  &  0  &  0  &  0  &  0  &  0  &  0 \\
 5  &  676  &  676  &  0  &  0  &  0  &  0  &  0  &  0  &  0  &  0 \\
 6  &  1998  &  899  &  1099  &  0  &  0  &  0  &  0  &  0  &  0  &  0 \\
 7  &  4054  &  548  &  2021  &  1485  &  0  &  0  &  0  &  0  &  0  &  0 \\
 8  &  7699  &  203  &  4965  &  1680  &  514  &  337  &  0  &  0  &  0  &  0 \\
 9  &  12383  &  45  &  8607  &  2721  &  549  &  119  &  290  &  52  &  0  &  0 \\
 10  &  17610  &  9  &  12579  &  3752  &  712  &  199  &  286  &  52  &  21  &  0 \\
 11  &  22876  &  3  &  16698  &  4742  &  956  &  138  &  265  &  42  &  20  &  12 \\
 12  &  26736  &  1  &  19758  &  5454  &  1062  &  150  &  253  &  40  &  16  &  2 \\
 13  &  29072  &  0  &  21249  &  6164  &  1209  &  145  &  273  &  19  &  8  &  5 \\
 14  &  29201  &  0  &  21211  &  6326  &  1237  &  101  &  288  &  20  &  6  &  12 \\
 15  &  27267  &  0  &  19613  &  6074  &  1222  &  91  &  244  &  13  &  7  &  3 \\
 16  &  23688  &  0  &  16942  &  5335  &  1103  &  55  &  241  &  7  &  4  &  1 \\
 17  &  19618  &  0  &  13824  &  4543  &  990  &  56  &  200  &  4  &  1  &  0 \\
 18  &  15154  &  0  &  10463  &  3637  &  794  &  44  &  208  &  3  &  4  &  1 \\
 19  &  10882  &  0  &  7372  &  2732  &  606  &  24  &  143  &  3  &  2  &  0 \\
 20  &  7649  &  0  &  5215  &  1836  &  481  &  20  &  97  &  0  &  0  &  0 \\
 21  &  4921  &  0  &  3295  &  1232  &  308  &  9  &  76  &  1  &  0  &  0 \\
 22  &  2977  &  0  &  1929  &  804  &  190  &  8  &  46  &  0  &  0  &  0 \\
 23  &  1780  &  0  &  1154  &  442  &  135  &  3  &  46  &  0  &  0  &  0 \\
 24  &  1016  &  0  &  633  &  262  &  97  &  1  &  23  &  0  &  0  &  0 \\
 25  &  539  &  0  &  344  &  134  &  39  &  3  &  19  &  0  &  0  &  0 \\
 26  &  268  &  0  &  165  &  74  &  24  &  0  &  5  &  0  &  0  &  0 \\
 27  &  122  &  0  &  78  &  35  &  5  &  0  &  4  &  0  &  0  &  0 \\
 28  &  54  &  0  &  37  &  12  &  3  &  0  &  2  &  0  &  0  &  0 \\
 29  &  23  &  0  &  14  &  7  &  2  &  0  &  0  &  0  &  0  &  0 \\
 30  &  4  &  0  &  4  &  0  &  0  &  0  &  0  &  0  &  0  &  0 \\
 31  &  10  &  0  &  8  &  2  &  0  &  0  &  0  &  0  &  0  &  0 \\
 32  &  1  &  0  &  1  &  0  &  0  &  0  &  0  &  0  &  0  &  0 \\
 \hline
     &  268402  &  2508  &  189278  &  59485  &  12238  &  1503  &  3009  &  256  &  89  &  36
\end{tabular}
\end{ruledtabular}
\end{table*}

\begin{table*}
\caption{\label{tbl:distribution_E_data}The number of simple $3$-polytopes in \texttt{E-268402} that belong to the various classes 
defined in Section~\ref{subsec:roundness}, where $n$ is the number of faces. The final row is the sum of entries in the column.}
\begin{ruledtabular}
\begin{tabular}{*{11}{r}}
 $n$ & all & stacked$^*$ & flag$^*$ & flag$^*_{+1}$ & flag$^*_{+1+1}$ & flag$^*_{+2}$ & flag$^*_{\,\geq3}$ & sc & sc$_{+1}$ & sc$_{\,\geq2}$ \\
\hline \noalign{\smallskip}
 4  &  372  &  372  &  0  &  0  &  0  &  0  &  0  &  0  &  0  &  0  \\
 5  &  1840  &  1840  &  0  &  0  &  0  &  0  &  0  &  0  &  0  &  0  \\
 6  &  4850  &  2521  &  2329  &  0  &  0  &  0  &  0  &  0  &  0  &  0  \\
 7  &  9362  &  1629  &  4001  &  3732  &  0  &  0  &  0  &  0  &  0  &  0  \\
 8  &  14627  &  674  &  8122  &  3656  &  1364  &  811  &  0  &  0  &  0  & 0  \\
 9  &  19097  &  163  &  11489  &  4891  &  1207  &  414  &  804  &  129  & 0  &  0  \\
 10  &  22550  &  30  &  14264  &  5498  &  1514  &  352  &  716  &  96  & 80  &  0  \\
 11  &  23512  &  3  &  15138  &  5781  &  1577  &  248  &  610  &  80  & 45  &  30  \\
 12  &  24041  &  1  &  15495  &  6111  &  1568  &  209  &  557  &  35  & 43  &  22  \\
 13  &  22445  &  1  &  14302  &  5902  &  1546  &  122  &  505  &  24  & 21  &  22  \\
 14  &  21241  &  0  &  13478  &  5621  &  1436  &  125  &  544  &  16  & 12  &  9  \\
 15  &  18712  &  0  &  11507  &  5151  &  1431  &  103  &  501  &  7  & 9  &  3  \\
 16  &  16585  &  0  &  9927  &  4700  &  1429  &  80  &  426  &  7  & 11  &  5  \\
 17  &  14215  &  0  &  8295  &  4135  &  1298  &  69  &  407  &  5  &  3  & 3  \\
 18  &  11997  &  0  &  6691  &  3667  &  1144  &  55  &  431  &  1  &  5  & 3  \\
 19  &  9898  &  0  &  5342  &  3156  &  1006  &  46  &  345  &  0  &  3  &  0  \\
 20  &  7958  &  0  &  4153  &  2559  &  877  &  33  &  334  &  2  &  0  &  0  \\
 21  &  6438  &  0  &  3223  &  2076  &  794  &  17  &  324  &  2  &  1  &  1  \\
 22  &  5077  &  0  &  2459  &  1695  &  637  &  25  &  257  &  1  &  0  &  3  \\
 23  &  3755  &  0  &  1686  &  1256  &  568  &  13  &  231  &  0  &  1  &  0  \\
 24  &  2925  &  0  &  1318  &  973  &  431  &  12  &  191  &  0  &  0  &  0  \\
 25  &  2136  &  0  &  885  &  726  &  363  &  3  &  159  &  0  &  0  &  0  \\
 26  &  1480  &  0  &  571  &  547  &  229  &  9  &  123  &  0  &  1  &  0  \\
 27  &  1068  &  0  &  382  &  376  &  210  &  3  &  97  &  0  &  0  &  0  \\
 28  &  749  &  0  &  263  &  268  &  140  &  2  &  76  &  0  &  0  &  0  \\
 29  &  547  &  0  &  193  &  204  &  89  &  2  &  59  &  0  &  0  &  0  \\
 30  &  332  &  0  &  103  &  115  &  65  &  0  &  49  &  0  &  0  &  0  \\
 31  &  224  &  0  &  72  &  85  &  40  &  0  &  27  &  0  &  0  &  0  \\
 32  &  149  &  0  &  49  &  39  &  35  &  0  &  26  &  0  &  0  &  0  \\
 33  &  101  &  0  &  30  &  29  &  21  &  1  &  20  &  0  &  0  &  0  \\
 34  &  53  &  0  &  18  &  11  &  13  &  0  &  11  &  0  &  0  &  0  \\
 35  &  26  &  0  &  8  &  10  &  6  &  0  &  2  &  0  &  0  &  0  \\
 36  &  17  &  0  &  3  &  4  &  7  &  0  &  3  &  0  &  0  &  0  \\
 37  &  11  &  0  &  3  &  6  &  1  &  0  &  1  &  0  &  0  &  0  \\
 38  &  9  &  0  &  2  &  2  &  4  &  0  &  1  &  0  &  0  &  0  \\
 39  &  1  &  0  &  0  &  0  &  0  &  0  &  1  &  0  &  0  &  0  \\
 40  &  0  &  0  &  0  &  0  &  0  &  0  &  0  &  0  &  0  &  0  \\
 41  &  1  &  0  &  0  &  1  &  0  &  0  &  0  &  0  &  0  &  0  \\
 42  &  0  &  0  &  0  &  0  &  0  &  0  &  0  &  0  &  0  &  0  \\
 43  &  0  &  0  &  0  &  0  &  0  &  0  &  0  &  0  &  0  &  0  \\
 44  &  0  &  0  &  0  &  0  &  0  &  0  &  0  &  0  &  0  &  0  \\
 45  &  1  &  0  &  0  &  0  &  0  &  0  &  1  &  0  &  0  &  0  \\
 \hline
     &  268402  &  7234  &  155801  &  72983  &  21050  &  2754  &  7839  & 405  &  235  &  101
\end{tabular}
\end{ruledtabular}
\end{table*}

\begin{table*}
\caption{\label{tbl:distribution}Percentages of $k$-round grains for the \texttt{V-268402}, \texttt{E-268402} and \texttt{R-268402} data sets, and percentages of $k$-round grains for the same three data sets and for all simple $3$-polytopes with $n = 13$ and $n = 14$.}
\begin{ruledtabular}
\begin{tabular}{l*{8}{r}}
 & round$_{\,0}$ & round$_{\,\leq 1}$ & round$_{\,\leq 2}$ & round$_{\,\leq 3}$ & round$_{\,\leq 4}$ & round$_{\,\leq 5}$ & round$_{\,\leq 6}$ & sc$_{\,\geq}$ \\
\hline \noalign{\smallskip}
\texttt{V-268402} & 10.36 & 32.61 & 57.71 & 77.17 & 88.85 & 94.75 & 97.32 & 1.22 \\
\texttt{E-268402} & 58.19 & 86.06 & 95.87 & 98.79 & 99.54 & 99.69 & 99.72 & 0.28 \\
\texttt{R-268402} & 70.57 & 92.98 & 98.44 & 99.58 & 99.81 & 99.85 & 99.86 & 0.14 \\[1.5mm]

\texttt{all}$^*_{\,n=13}$      &  0.63 &  2.73 &  7.59 & 15.19 & 25.63 & 37.82 & 51.59 & 7.80 \\
\texttt{V-268402}$_{\,n=13}$ & 12.59 & 38.51 & 63.84 & 81.81 & 91.21 & 95.65 & 97.41 & 1.53 \\
\texttt{E-268402}$_{\,n=13}$ & 63.72 & 90.02 & 97.45 & 99.14 & 99.58 & 99.68 & 99.69 & 0.30 \\
\texttt{R-268402}$_{\,n=13}$ & 73.09 & 94.29 & 98.95 & 99.76 & 99.86 & 99.89 & 99.89 & 0.11 \\[1.5mm]

\texttt{all}$^*_{\,n=14}$      &  0.40 &  1.96 &  5.72 & 12.04 & 20.81 & 31.36 & 43.11 & 10.81 \\
\texttt{V-268402}$_{\,n=14}$ & 11.58 & 36.01 & 61.96 & 80.29 & 90.58 & 95.40 & 97.30 & 1.59 \\
\texttt{E-268402}$_{\,n=14}$ & 63.45 & 89.92 & 97.26 & 99.32 & 99.73 & 99.81 & 99.82 & 0.17 \\
\texttt{R-268402}$_{\,n=14}$ & 72.64 & 94.30 & 98.88 & 99.72 & 99.85 & 99.87 & 99.87 & 0.13
\end{tabular}
\end{ruledtabular}
\end{table*}

\begin{table*}
\caption{\label{tbl:flag_few_vertices}Number of flag triangulations of the $2$-sphere with $n$ vertices and split type $k\!:\!m$.}
\begin{ruledtabular}
\begin{tabular}{r *{6}{D{x}{\times}{-1}}}
 \diagbox{$n$}{$k$} &
 \multicolumn{1}{c}{0} &
 \multicolumn{1}{c}{1} &
 \multicolumn{1}{c}{2} &
 \multicolumn{1}{c}{3} &
 \multicolumn{1}{c}{4} &
 \multicolumn{1}{c}{5} \\
 \hline \noalign{\smallskip}
 6 && 1 x 1\cc1 \\
 7 && 1 x 1\cc2 \\
 8 &&& 2 x 2\cc2 \\
 9 && 1 x 1\cc4 & 3 x 2\cc3 \\
10 && 2 x 1\cc5 & 1 x 2\cc4 & 7 x 3\cc3 \\
11 && 3 x 1\cc6 & 6 x 2\cc5 & 16 x 3\cc4 \\
12 & 1 x 0\cc8 & 9 x 1\cc7 & 15 x 2\cc6 & 16 x 3\cc5 & 46 x 4\cc4 \\
13 && 28 x 1\cc8 & 52 x 2\cc7 & 62 x 3\cc6 & 171 x 4\cc5 \\
14 & 1 x 0\cc10 & 97 x 1\cc9 & 209 x 2\cc8 & 235 x 3\cc7 & 261 x 4\cc6 & 554 x 5\cc5
\end{tabular}
\end{ruledtabular}
\end{table*}

\begin{table*}
\caption{\label{tbl:flag_distribution_V_data}Number of flag$^*$ grains in \texttt{V-268402} with $n$ faces and split type $k\!:\!m$.}
\begin{ruledtabular}
\setlength{\tabcolsep}{-1pt}
\begin{tabular}{r *{10}{D{x}{\times}{-1}}}
 \diagbox{$n$}{$k$} &
 \multicolumn{1}{c}{0} &
 \multicolumn{1}{c}{1} &
 \multicolumn{1}{c}{2} &
 \multicolumn{1}{c}{3} &
 \multicolumn{1}{c}{4} &
 \multicolumn{1}{c}{5} &
 \multicolumn{1}{c}{6} &
 \multicolumn{1}{c}{7} &
 \multicolumn{1}{c}{8} &
 \multicolumn{1}{c}{9} \\
\hline \noalign{\smallskip}
  6 && 23 x 1\cc1 \\ 
  7 && 99 x 1\cc2 \\
  8 &&& 311 x 2\cc2 \\
  9 && 172 x 1\cc4 & 640 x 2\cc3 \\ 
 10 && 264 x 1\cc5 & 407 x 2\cc4 & 700 x 3\cc3 \\
 11 && 536 x 1\cc6 & 711 x 2\cc5 & 972 x 3\cc4 \\ 
 12 & 6 x 0\cc8 & 748 x 1\cc7 & 932 x 2\cc6 & 638 x 3\cc5 & 713 x 4\cc4 \\
 13 && 903 x 1\cc8 & 1134 x 2\cc7 & 642 x 3\cc6 & 672 x 4\cc5 \\ 
 14 & 1 x 0\cc10 & 1060 x 1\cc9 & 1166 x 2\cc8 & 648 x 3\cc7 & 426 x 4\cc6 & 288 x 5\cc5 \\ 
 15 & 1 x 0\cc11 & 969 x 1\cc10 & 1167 x 2\cc9 & 609 x 3\cc8 & 300 x 4\cc7 & 292 x 5\cc6 \\ 
 16 & 1 x 0\cc12 & 814 x 1\cc11 & 1005 x 2\cc10 & 472 x 3\cc9 & 220 x 4\cc8 & 153 x 5\cc7 & 107 x 6\cc6 \\ 
 17 & 1 x 0\cc13 & 688 x 1\cc12 & 923 x 2\cc11 & 384 x 3\cc10 & 186 x 4\cc9 & 103 x 5\cc8 & 81 x 6\cc7 \\
 18 && 534 x 1\cc13 & 647 x 2\cc12 & 269 x 3\cc11 & 100 x 4\cc10 & 48 x 5\cc9 & 29 x 6\cc8 & 21 x 7\cc7 \\ 
 19 && 331 x 1\cc14 & 469 x 2\cc13 & 188 x 3\cc12 & 64 x 4\cc11 & 28 x 5\cc10 & 21 x 6\cc9 & 11 x 7\cc8 \\ 
 20 && 234 x 1\cc15 & 334 x 2\cc14 & 108 x 3\cc13 & 38 x 4\cc12 & 10 x 5\cc11 & 13 x 6\cc10 & 4 x 7\cc9 & 1 x 8\cc8 \\ 
 21 && 144 x 1\cc16 & 200 x 2\cc15 & 66 x 3\cc14 & 29 x 4\cc13 & 8 x 5\cc12 & 4 x 6\cc11 & 3 x 7\cc10 & 2 x 8\cc9 \\
 22 && 82 x 1\cc17 & 113 x 2\cc16 & 49 x 3\cc15 & 15 x 4\cc14 & 5 x 5\cc13 &&& 1 x 8\cc10 & 1 x 9\cc9 \\
 23 && 63 x 1\cc18 & 54 x 2\cc17 & 31 x 3\cc16 & 6 x 4\cc15 & 2 x 5\cc14 & 1 x 6\cc13 \\ 
 24 && 21 x 1\cc19 & 37 x 2\cc18 & 6 x 3\cc17 & 4 x 4\cc16 &&&& 1 x 8\cc12 \\ 
 25 && 14 x 1\cc20 & 19 x 2\cc19 & 3 x 3\cc18 & 4 x 4\cc17 \\ 
 26 && 5 x 1\cc21 & 5 x 2\cc20 \\
 27 && 1 x 1\cc22 & 9 x 2\cc21 & 1 x 3\cc20 & 1 x 4\cc19 \\ 
 28 && 1 x 1\cc23 & 5 x 2\cc22 & 1 x 3\cc21 & 1 x 4\cc20 \\
 29 &&& 1 x 2\cc23 \\
 30 \\
 31 && 1 x 1\cc26
\end{tabular}
\end{ruledtabular}
\end{table*}

\begin{table*}
\caption{\label{tbl:flag_distribution_R_data}Number of fllag$^*$ grains in \texttt{R-268402} with $n$ faces and split type $k\!:\!m$.}
\begin{ruledtabular}
\begin{tabular}{r *{9}{D{x}{\times}{-1}}}
 \diagbox{$n$}{$k$} &
 \multicolumn{1}{c}{0} &
 \multicolumn{1}{c}{1} &
 \multicolumn{1}{c}{2} &
 \multicolumn{1}{c}{3} &
 \multicolumn{1}{c}{4} &
 \multicolumn{1}{c}{5} &
 \multicolumn{1}{c}{6} &
 \multicolumn{1}{c}{7} &
 \multicolumn{1}{c}{8} \\
\hline \noalign{\smallskip}
  6 && 1099 x 1\cc1 \\
  7 && 2021 x 1\cc2 \\
  8 &&& 4965 x 2\cc2 \\
  9 && 3632 x 1\cc4 & 4975 x 2\cc3 \\
 10 && 4895 x 1\cc5 & 3878 x 2\cc4 & 3806 x 3\cc3 \\
 11 && 7676 x 1\cc6 & 4843 x 2\cc5 & 4179 x 3\cc4 \\
 12 & 184 x 0\cc8 & 9577 x 1\cc7 & 5555 x 2\cc6 & 2617 x 3\cc5 & 1825 x 4\cc4 \\
 13 && 11358 x 1\cc8 & 6030 x 2\cc7 & 2216 x 3\cc6 & 1645 x 4\cc5 \\
 14 & 102 x 0\cc10 &  11957 x 1\cc9 & 5964 x 2\cc8 & 1816 x 3\cc7 & 867 x 4\cc6 & 505 x 5\cc5 \\
 15 & 82 x 0\cc11 & 11673 x 1\cc10 & 5421 x 2\cc9 & 1390 x 3\cc8 & 626 x 4\cc7 & 421 x  5\cc6 \\
 16 & 58 x 0\cc12 & 10511 x 1\cc11 & 4581 x 2\cc10 & 1099 x 3\cc9 & 387 x 4\cc8 & 195 x 5\cc7 & 111 x 6\cc6 \\
 17 & 66 x 0\cc13 & 8793 x 1\cc12 & 3766 x 2\cc11 & 775 x 3\cc10 & 238 x 4\cc9 & 101 x 5\cc8 & 85 x 6\cc7 \\
 18 & 38 x 0\cc14 & 6802 x 1\cc13 & 2846 x 2\cc12 & 541 x 3\cc11 & 141 x 4\cc10 & 57 x 5\cc9 & 20 x 6\cc8 & 18 x 7\cc7 \\
 19 & 36 x 0\cc15 & 4930 x 1\cc14 & 1937 x 2\cc13 & 329 x 3\cc12 & 85 x 4\cc11 & 31 x 5\cc10 & 12 x 6\cc9 & 12 x 7\cc8 \\
 20 & 16 x 0\cc16 & 3556 x 1\cc15 & 1353 x 2\cc14 & 206 x 3\cc13 & 58 x 4\cc12 & 17 x 5\cc11 & 5 x 6\cc10 & 2 x 7\cc9 & 2 x 8\cc8 \\
 21 & 12 x 0\cc17 & 2238 x 1\cc16 & 892 x 2\cc15 & 123 x 3\cc14 & 20 x 4\cc13 & 5 x 5\cc12 & 2 x 6\cc11 & 1 x 7\cc10 & 2 x 8\cc9 \\
 22 & 5 x 0\cc18 & 1324 x 1\cc17 & 500 x 2\cc16 & 77 x 3\cc15 & 18 x 4\cc14 & 5 x 5\cc13 \\
 23 & 5 x 0\cc19 & 799 x 1\cc18 & 299 x 2\cc17 & 44 x 3\cc16 & 7 x 4\cc15 \\
 24 & 1 x 0\cc20 & 462 x 1\cc19 & 152 x 2\cc18 & 17 x 3\cc17 & 1 x 4\cc16 \\
 25 & 3 x 0\cc21 & 246 x 1\cc20 & 82 x 2\cc19 & 12 x 3\cc18 & 1 x 4\cc17 \\
 26 && 127 x 1\cc21 & 37 x 2\cc20 & 1 x 3\cc19 \\
 27 & 1 x 0\cc23 & 52 x 1\cc22 & 25 x 2\cc21 \\
 28 && 22 x 1\cc23 & 14 x 2\cc22 && 1 x 4\cc20 \\
 29 && 10 x 1\cc24 & 4 x 2\cc23 \\
 30 && 4 x 1\cc25 \\
 31 && 5 x 1\cc26 & 3 x 2\cc25 \\
 32 &&& 1 x 2\cc26
\end{tabular}
\end{ruledtabular}
\end{table*}

\begin{table*}
\caption{\label{tbl:flag_distribution_E_data}Number of flag$^*$ grains in \texttt{E-268402} with $n$ faces and split type $k\!:\!m$.}
\begin{ruledtabular}
\begin{tabular}{r *{9}{D{x}{\times}{-1}}}
 \diagbox{$n$}{$k$} &
 \multicolumn{1}{c}{0} &
 \multicolumn{1}{c}{1} &
 \multicolumn{1}{c}{2} &
 \multicolumn{1}{c}{3} &
 \multicolumn{1}{c}{4} &
 \multicolumn{1}{c}{5} &
 \multicolumn{1}{c}{6} &
 \multicolumn{1}{c}{7} &
 \multicolumn{1}{c}{8} \\
\hline \noalign{\smallskip}
  6 && 2329 x 1\cc1 \\
  7 && 4001 x 1\cc2 \\
  8 &&& 8122 x 2\cc2  \\
  9 && 5024 x 1\cc4 & 6465 x 2\cc3 \\
 10 && 5941 x 1\cc5 & 4110 x 2\cc4 & 4213 x 3\cc3 \\
 11 && 7347 x 1\cc6 & 4226 x 2\cc5 & 3565 x 3\cc4 \\
 12 & 187 x 0\cc8 & 7739 x 1\cc7 & 4279 x 2\cc6 & 2001 x 3\cc5 & 1289 x 4\cc4 \\
 13 && 7894 x 1\cc8 & 3993 x 2\cc7 & 1403 x 3\cc6 & 1012 x 4\cc5 \\
 14 & 69 x 0\cc10 & 7855 x 1\cc9 & 3692 x 2\cc8 & 1091 x 3\cc7 & 493 x 4\cc6 & 278 x 5\cc5 \\
 15 & 42 x 0\cc11 & 6964 x 1\cc10 & 3159 x 2\cc9 & 801 x 3\cc8 & 321 x 4\cc7 & 220 x 5\cc6 \\
 16 & 38 x 0\cc12 & 6178 x 1\cc11 & 2742 x 2\cc10 & 626 x 3\cc9 & 200 x 4\cc8 & 98 x 5\cc7 & 45 x 6\cc6 \\
 17 & 27 x 0\cc13 & 5278 x 1\cc12 & 2322 x 2\cc11 & 456 x 3\cc10 & 121 x 4\cc9 & 54 x 5\cc8 & 37 x 6\cc7 \\
 18 & 28 x 0\cc14 & 4315 x 1\cc13 & 1858 x 2\cc12 & 349 x 3\cc11 & 92 x 4\cc10 & 28 x 5\cc9 & 13 x 6\cc8 & 8 x 7\cc7 \\
 19 & 13 x 0\cc15 & 3513 x 1\cc14 & 1480 x 2\cc13 & 256 x 3\cc12 & 53 x 4\cc11 & 19 x 5\cc10 & 5 x 6\cc9 & 3 x 7\cc8 \\
 20 & 19 x 0\cc16 & 2740 x 1\cc15 & 1155 x 2\cc14 & 179 x 3\cc13 & 40 x 4\cc12 & 11 x 5\cc11 & 6 x 6\cc10 & 3 x 7\cc9 \\
 21 & 12 x 0\cc17 & 2127 x 1\cc16 & 911 x 2\cc15 & 144 x 3\cc14 & 18 x 4\cc13 & 8 x 5\cc12 && 1 x 7\cc10 & 2 x 8\cc9 \\
 22 & 5 x 0\cc18 & 1639 x 1\cc17 & 683 x 2\cc16 & 104 x 3\cc15 & 21 x 4\cc14 & 5 x 5\cc13 & 2 x 6\cc12 \\
 23 & 1 x 0\cc19 & 1083 x 1\cc18 & 517 x 2\cc17 & 68 x 3\cc16 & 12 x 4\cc15 & 5 x 5\cc14 \\
 24 & 3 x 0\cc20 & 851 x 1\cc19 & 401 x 2\cc18 & 56 x 3\cc17 & 6 x 4\cc16 & 1 x 5\cc15 \\
 25 & 3 x 0\cc21 & 597 x 1\cc20 & 250 x 2\cc19 & 31 x 3\cc18 & 3 x 4\cc17 && 1 x 6\cc15 \\
 26 & 1 x 0\cc22 & 382 x 1\cc21 & 169 x 2\cc20 & 16 x 3\cc19 & 3 x 4\cc18 \\
 27 & 1 x 0\cc23 & 263 x 1\cc22 & 105 x 2\cc21 & 11 x 3\cc20 & 1 x 4\cc19 & 1 x 5\cc18 \\
 28 && 168 x 1\cc23 & 84 x 2\cc22 & 11 x 3\cc21 \\
 29 && 125 x 1\cc24 & 60 x 2\cc23 & 8 x 3\cc22 \\
 30 && 75 x 1\cc25 & 24 x 2\cc24 & 4 x 3\cc23 \\
 31 && 47 x 1\cc26 & 23 x 2\cc25 & 1 x 3\cc24 && 1 x 5\cc22 \\
 32 && 35 x 1\cc27 & 13 x 2\cc26 & 1 x 3\cc25 \\
 33 && 22 x 1\cc28 & 7 x 2\cc27 & 1 x 3\cc26 \\
 34 && 10 x 1\cc29 & 7 x 2\cc28 & 1 x 3\cc27 \\
 35 && 4 x 1\cc30 & 4 x 2\cc29 \\
 36 && 2 x 1\cc31 && 1 x 3\cc29 \\
 37 && 1 x 1\cc32 & 2 x 2\cc31 \\
 38 &&& 2 x 2\cc32 \\
\end{tabular}
\end{ruledtabular}
\end{table*}

\end{document}